%% file: main.tex
\documentclass[acmtog,screen,nonacm]{acmart}
\usepackage{hyperref} 
\usepackage{hyperxmp}

\usepackage{booktabs} 
\usepackage{subcaption}
\usepackage{overpic}

\usepackage{algorithm}
\usepackage{algorithmic}
\usepackage{multirow}

\acmJournal{TOG}
\input{notations}

\usepackage{color}
\usepackage{soul}
\usepackage{xcolor}

\begin{document}

\title{{\em Patch-Grid}: An Efficient and Feature-Preserving Neural Implicit Surface Representation}

\author{Guying Lin}
\authornote{Both authors contributed equally to this research.}
\email{guyingl@andrew.cmu.edu}

\affiliation{%
  \institution{The University of Hong Kong}
  \city{Hong Kong}
  \country{China}
}

\author{Lei Yang}
\authornotemark[1]
\email{l.yang@transgp.hk}
\authornote{Corresponding authors.}

\affiliation{%
  \institution{The University of Hong Kong}
  \city{Hong Kong}
  \country{China}
}

\author{Congyi Zhang}
\email{zhcy@outlook.com}

\affiliation{%
  \institution{The University of Hong Kong}
  \city{Hong Kong}
  \country{China}
}

\author{Hao Pan}
\email{haopan@microsoft.com}

\affiliation{%
  \institution{Microsoft Research Asia}
  \city{Beijing}
  \country{China}
}

\author{Yuhan Ping}
\email{csyhping@connect.hku.hk}

\affiliation{%
  \institution{The University of Hong Kong}
  \city{Hong Kong}
  \country{China}
}

\author{Guodong Wei}
\email{g.d.wei.china@gmail.com}

\affiliation{%
  \institution{The University of Hong Kong}
  \city{Hong Kong}
  \country{China}
}

\author{Taku Komura}
\email{taku@cs.hku.hk}

\affiliation{%
  \institution{The University of Hong Kong}
  \city{Hong Kong}
  \country{China}
}

\author{John Keyser}

\email{keyser@cse.tamu.edu}

\affiliation{%
  \institution{Texas A\&M University}
  \city{College station}
  \country{United States of America}
}

\author{Wenping Wang}
\authornotemark[2]
\email{wenping@tamu.edu}

\affiliation{%
  \institution{Texas A\&M University}
  \city{College station}
  \country{United States of America}
}

\input{sections/abstract.tex}


\begin{teaserfigure}
\centering
  \includegraphics[width=1\textwidth]{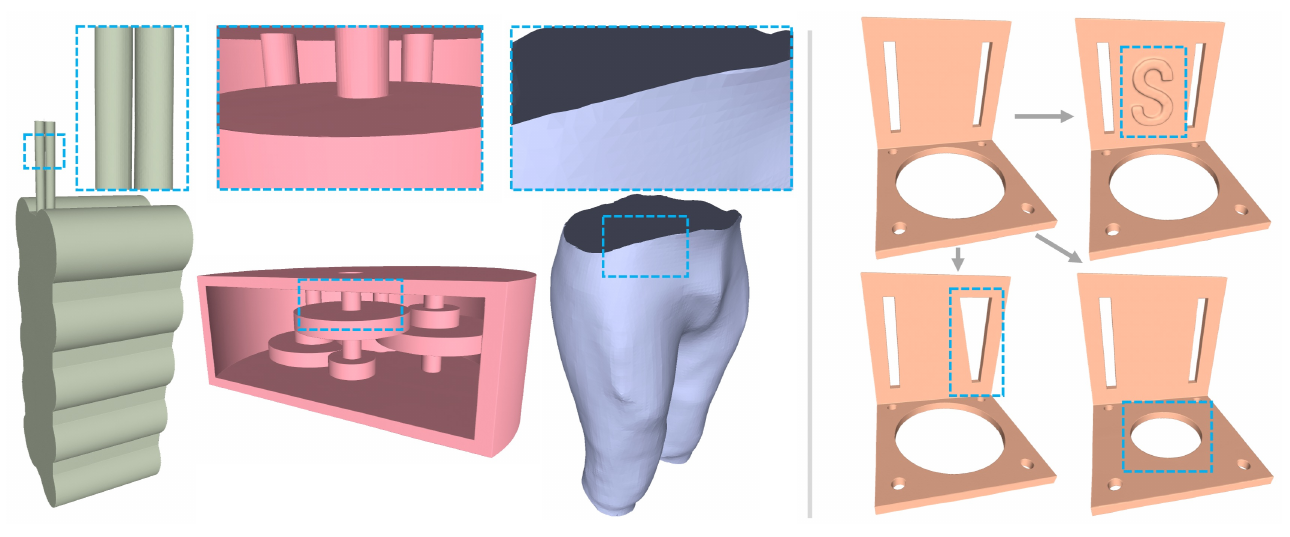}
\put(-370,-9){(a) Shape fitting}
\put(-120,-9){(b) Shape editing}
  \caption{
  We propose \textit{Patch-Grid}, a method capable of (a) efficiently modeling diverse geometric features, such as thin geometric features, sharp features, and open surface boundaries (from left to right); (b) \textit{Patch-Grid} also allows updating a thus learned implicit representation to incorporate bump textures (top right) or reshape a geometric feature (bottom) within 2 seconds.}
  \Description{A set of models represented by our neural implicit representation.}
  \label{fig:teaser}
\end{teaserfigure}

\maketitle

\input{sections/introduction}

\input{sections/related_works}

\input{sections/method}

\input{sections/global_local_blending}

\input{sections/results}
\input{sections/limitation}

\input{sections/conclusions}

\bibliographystyle{ACM-Reference-Format}
\bibliography{ref}

\end{document}

%% file: notations.tex
\newcommand{\loss}{\mathcal{L}}

\usepackage[capitalize]{cleveref}
\crefname{section}{Sec.}{Secs.}
\Crefname{section}{Section}{Sections}
\Crefname{table}{Table}{Tables}
\crefname{table}{Tab.}{Tabs.}

%% file: sections/abstract.tex
\begin{abstract}

Neural implicit representations are increasingly used to depict 3D shapes owing to their inherent smoothness and compactness, contrasting with traditional discrete representations. Yet, the multilayer perceptron (MLP) based neural representation, because of its smooth nature, rounds sharp corners or edges, rendering it unsuitable for representing objects with sharp features like CAD models. Moreover, neural implicit representations need long training times to fit 3D shapes.
While previous works address these issues separately, we present a unified neural implicit representation called {\em Patch-Grid}, which efficiently fits complex shapes, preserves sharp features delineating different patches, and can also represent surfaces with open boundaries and thin geometric features. 

{\em Patch-Grid} learns a signed distance field (SDF) to approximate an encompassing surface patch of the shape with a learnable patch feature volume. To form sharp edges and corners in a CAD model, \textit{Patch-Grid} merges the learned SDFs via the constructive solid geometry (CSG) approach. 
Core to the merging process is a novel {\em merge grid} design that organizes different patch feature volumes in a common octree structure. This design choice ensures robust merging of multiple learned SDFs by confining the CSG operations to localized regions. Additionally, it drastically reduces the complexity of the CSG operations in each merging cell, allowing the proposed method to be trained in seconds to fit a complex shape at high fidelity.

Experimental results demonstrate that the proposed \textit{Patch-Grid} representation is capable of accurately reconstructing shapes with complex sharp features, open boundaries, and thin geometric elements, achieving state-of-the-art reconstruction quality with high computational efficiency within seconds.

\end{abstract}

%% file: sections/introduction.tex
\section{Introduction}\label{sec:intro}

The implicit surface representation typically defines a shape as the zero-level set of some function, such as the signed distance function of the shape. The implicit representation is widely used for shape modeling~\cite{turk2002modelling,ohtake2005multi,mitchell2015non} and downstream engineering applications like simulation~\cite{nTopoWhitePaper}. 
Recent years have seen a surge in research on {\em neural implicit representations}~\cite{Park2019DeepSDF,Gropp2020IGR,tancik2020fourier,martel2021acorn} where a deep neural network is used to encode the implicit function in question. Neural implicit representations are inherently smooth and can represent complex shape details more compactly than traditional discrete representations, e.g.\ point clouds and polygonal meshes~\cite{Sitzmann2020SIREN,takikawa2021neural}.

There are two main challenges with current neural implicit surface representations. First, they struggle to represent various geometric features, such as sharp geometric edges, surfaces with open boundaries, and narrow gaps formed by surfaces in spatial proximity; see the examples in Fig.~\ref{fig:teaser}. Although two exemplar shapes possess sharp geometric features and narrow gaps as shown in the zoom-in views, no existing method can handle these geometric features simultaneously, leading to less satisfactory modeling results. Furthermore, open surfaces, which find important applications in modeling cloth and representing clothed humans, are known to be challenging to model with implicit surface representations. Therefore, a framework capable of faithfully representing these geometric features in a shape can significantly extend the application of neural implicit surface representations.

Second, learning a neural implicit representation to fit a given shape accurately often takes an excessively long time, from minutes to around an hour, precluding its application to interactive design. Several methods~\cite{saragadam2022miner, Wang_2022} have proposed to address this efficiency issue. While these methods can improve training efficiency through a hierarchical design or a pretrained shape space, they struggle to balance high quality and efficiency.
In contrast to these previous methods, \textit{InstantNGP}~\cite{muller2022instant} is designed with algorithmic improvements and customized CUDA kernels, establishing state-of-the-art performance in both efficiency and fitting accuracy. 

To tackle these two challenges, we present \textit{Patch-Grid}, a compositional framework for modeling neural implicit surfaces with two distinctive advantages: 1) {\bf Versatile representation}: it can represent sharp geometric features, open surface boundaries, and thin geometric features (such as slender tubes or narrow gaps), which are challenging or impossible for current neural surface representations, as shown in Fig.~\ref{fig:teaser}. 2) \textbf{Efficiency:}
\textit{Patch-Grid} is faster to train than most existing neural representations that achieve state-of-the-art performance. It typically takes about 5 seconds to train \textit{Patch-Grid} to fit a given shape with state-of-the-art fitting accuracy and less than 2 seconds to complete a local shape update of an existing surface represented by \textit{Patch-Grid} to the same level of accuracy. This enables interactive modeling and editing of neural implicit surfaces. 

The problem we address is formulated as follows. Given a boundary representation (B-Rep) of a 3D shape, which defines the 3D surface shape as a set of surface patches, we aim to convert this B-rep representation into a neural implicit representation, by representing each surface patch as a zero-level set of a neural implicit function. 
Specifically, each surface patch is tightly contained in a bounding volume that comprises a regular grid of cubic cells with a learnable feature vector assigned to each grid point. 

The collection of these grid cells is referred to as the {\em patch volume} and the collection of feature vectors defined on it is referred to as the {\em patch feature volume} of the surface patch.
For any point $\mathbf{x}$ inside a cubic cell, a Multi-Layer Perceptron (MLP) decoder with one hidden layer
is used to map the feature vector at $\mathbf{x}$, which is obtained via trilinear interpolation from feature vectors at the corners of the cell, to a signed distance value, as shown in Fig.~\ref{fig:pipeline}. 

We observe that the global CSG approach adopted by \textit{NH-Rep}~\cite{Guo2022NHRep} for the same task often leads to failure cases in highly concave and narrow regions (cf. Fig.~\ref{fig:nhrep_failure}). This is because merging zero-level sets of multiple surface patches requires careful coordination to avoid undesired interference (cf. Fig.~\ref{fig:clique}). Since the intersection of two or more surface patches forming a sharp geometric feature (edges or corners) is only a local operation, we propose to use a {\em merge grid} for robustly modeling sharp features of a given shape in a local manner. Specifically, the merge grid uses an octree structure to adaptively subdivide the spatial domain into cubic cells so that each leaf cell ideally contains no more than \textit{one} sharp feature to simplify the task. This simplifies the task by circumventing the difficulty in merging learned SDFs of multiple patches and trimming off the \textit{extraneous zero-level sets} (c.f. Fig.~\ref{fig:clique}) into various sharp features globally, as encountered in~\cite{Guo2022NHRep}. By adopting a divide-and-conquer strategy for merging learned SDFs of multiple patches in a local region, our approach achieves robust and superior performance for modeling various geometric features, as shown in Fig.~\ref{fig:teaser}.

Besides sharp features, our local approach is also capable of modeling \textbf{thin geometric features} and \textbf{open surface boundaries}. 
In a general sense, a slender tube or a narrow gap/slit is considered a thin geometric feature (e.g. columns 1, 2, and 4 of Fig.~\ref{fig:our_show}). By adopting a local approach, \textit{Patch-Grid} robustly models {\em thin geometric features} by placing spatially proximal but disjoint surface patches in different patch volumes which may overlap. In this way, the distance fields induced by the different surface patches do not interfere with each other, thereby enabling the modeling of narrow gaps or thin solids formed by these surface patches.
Furthermore, we model a boundary curve of an open surface patch as the result of a trimming operation. This involves constructing a trimming surface followed by a trimming operation, which can be performed locally to produce accurate results.

\textbf{Efficiency.}
Even without using a customized CUDA implementation, the proposed approach that uses patch-level feature vectors and a merge grid for modeling geometric features can achieve high training efficiency. Typically, our method takes about 5 seconds to fit a given shape from scratch and supports local shape updates at an interactive rate (within 2 seconds). The increased efficiency of \textit{Patch-Grid} stems from two key aspects. Firstly, \textit{Patch-Grid} decomposes a given shape into patches and independently learns a patch feature volume for each patch, reducing the overall learning complexity. Secondly, we propose an \textit{adaptive merge grid} to simplify the CSG operations among several surface patches that form a sharp feature. With this adaptive merge grid, the training objective is drastically simplified and thus improves the overall training efficiency.
Moreover, Patch-Grid can leverage a pretrained shallow MLP to significantly reduce the training time for learning the patch feature volumes.
We show that the proposed representation can facilitate local updates of individual surface patches, enabling interactive shape editing tasks. We will release our code upon the acceptance of the paper to enable future works to build on our work.

In summary, the contributions of this paper are:

\begin{itemize}
    \item \textit{Versatile representation}. We present \textit{Patch-Grid}, a compositional neural implicit representation capable of modeling a variety of geometric features that are challenging for previous methods, such as sharp or thin geometric features and open boundaries; 
    
    \item \textit{Robustness}. A novel \textit{merge grid} as illustrated in Fig. \ref{fig:clique} is proposed that adaptively partitions the spatial domain and locally composes neural implicit surface patches to faithfully model a target surface shape with sharp features (i.e. edges and corners) in a more robust manner than the existing global approach~\cite{Guo2022NHRep}.
    
    \item \textit{Superior efficiency}.  Combining the design choices of \textit{Patch feature volume} and \textit{Merge Grid}, our method \textit{Patch-Grid}, implemented solely in PyTorch, achieves state-of-the-art fitting quality within a few seconds, outperforming most existing methods.
    Additionally, \textit{Patch-Grid} supports local shape updates at an interactive rate (within 2 secs).

\end{itemize}

%% file: sections/related_works.tex
\begin{figure*}[t!]
   \centering
   \begin{overpic}
   [width=1.\linewidth]{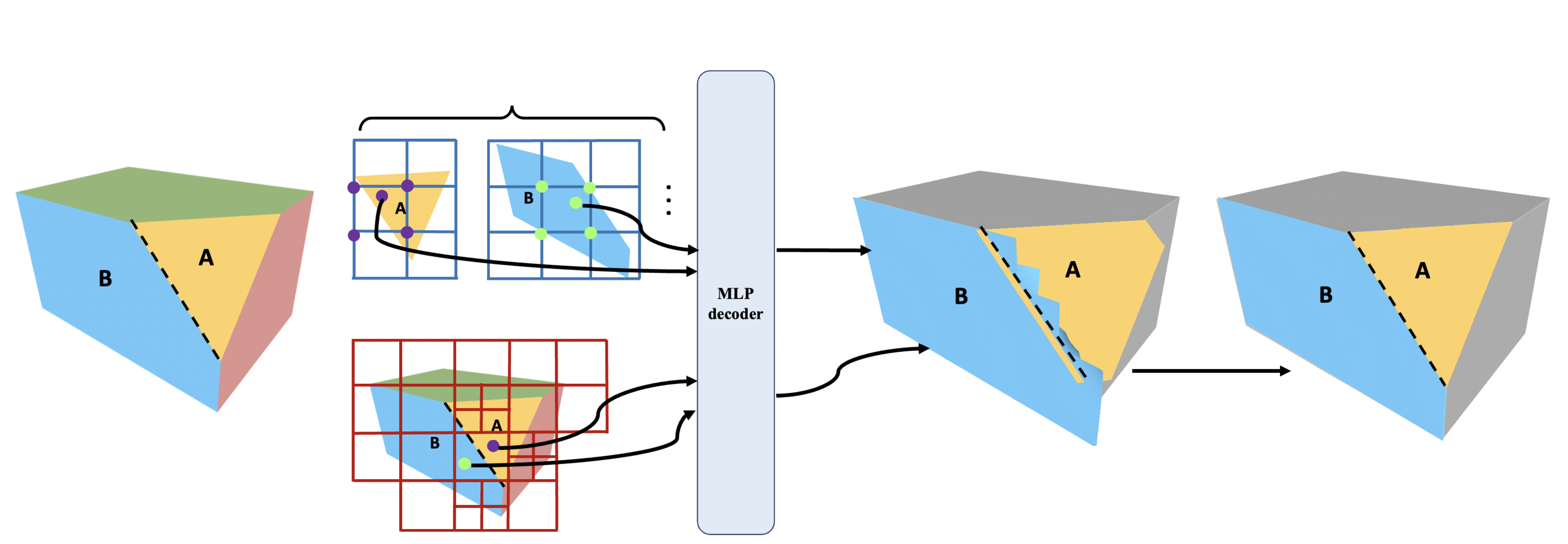}
   \put(5,32){(a) Input shape}
   \put(26,32){(b) \textit{Patch-Grid} representation}
   \put(23,29){\textbf{Patch feature volumes $FV$}}
    \put(23,14){\textbf{Merge grid $\mathcal{G}$}}
   \put(46,12){$f$}
   \put(50,17){$\loss_{\mathrm{patch}} $}
   \put(50,7.5){$\loss_{\mathrm{merge}} $}
   \put(73,8){Composition}
   \put(56,32){(c) Individual patches}
   \put(80,32){(d) Composed result}
   \end{overpic}
   \caption{\textbf{Overview.} (a) Input to \textit{Patch-Grid} is a B-Rep model of a 3D shape $\mathcal{S}$; 
   (b) \textit{Patch-Grid} encloses each surface patch of the given shape with a patch volume. We assign a learnable feature vector to each grid point of the patch volume and convert this volume into a patch feature volume $FV$, which defines a continuous feature field $F(\mathbf{x})$ via trilinear interpolation within each cell. Then, a shared MLP decoder $f$ is used to map $F(\mathbf{x})$ to $\mathbb{R}$, representing the SDF of the target surface patch. 
   As there exists an extended zero-level set beyond the fitting surface (called \textit{extraneous zero-level set} in the following), patches are robustly trimmed off through CSG merging operations to derive a neural implicit field that reproduces the SDF of the given shape.
   To this end, we construct a \textit{merge grid} $\mathcal{G}$ and enforce a merge loss within each merge grid cell. The merge grid confines the CSG operations required to merge multiple learned SDFs to localized regions. 
   (c) We design two types of loss terms to enable accurate fitting of individual patches ($\loss_{\mathrm{patch}}$) and model the relationship between connected patches ($\loss_{\mathrm{merge}}$), respectively;
   (d) Finally, the individual neural surface patches are composed into the final surface shape that reconstructs the target shape. It takes on average 5 seconds to fit a 3D CAD model. 
   }
   
   \label{fig:pipeline}
\end{figure*}

\section{Related works}\label{sec:relatedworks}

\subsection{Constructive solid geometry and implicit functions}

Implicit representations employ a (continuous) function to determine whether a given query point, specified by its x-y-z coordinates, is located inside or outside of a shape. A more complicated shape can be composed by combining several implicit functions. Specifically, constructive solid geometry (CSG) is a widely used technique for modeling complex geometries by organizing a set of implicit functions as a tree through recursive application of Boolean operations \cite{requicha1977constructive} or more general transformations like R-functions \cite{Pasko1995FRep,Shapiro2007RFunction}. Compared with standard boundary representation (B-Rep), CSG representations based on continuous implicit functions are more compact, can handle flexible topologies, and allow for shape recovery at arbitrary resolutions. These advantages make it a prominent choice for simulation, inverse optimization, and manufacturing \cite{nTopoWhitePaper}.

In recent years, machine learning approaches have been applied to CSG modeling. These methods either recover CSG trees from unstructured point clouds using neural networks \cite{sharma2018csgnet, kania2020ucsgnet, ren2021csgstump, yu2023d2csg}, or perform CSG operations on neural SDFs \cite{Guo2022NHRep, marschner2023csgnsdf}. 
Our method is similar to the latter which applies CSG operations on neural SDFs to compose complex geometry. This drastically extends the CSG modeling capability beyond simple geometric primitives, such as spheres, cones, boxes, or convex shapes.
Our work differs from previous works with the use of localized patch-grids to efficiently convert any B-Rep solid into a feature-preserving implicit representation. 
In the following, we expand in detail on the relationships of our approach and various neural implicit functions.

\subsection{Neural implicit representations}

\paragraph{Global neural implicit functions}

Neural implicit representation is a novel approach that turns traditional explicit discrete representations (e.g., point clouds, polygon meshes, or voxels) into the iso-surface of some continuously defined differentiable fields represented through a neural network \cite{Park2019DeepSDF,Mescheder2019OccNet, Chen2019Decoder}. To improve the performance when overfitting the network to complex 3D shapes, several techniques have been proposed including exploring the optimal hyper-parameters \cite{Davies2020Overfit}, different training strategies \cite{Duan2020CurSDF}, positional encoding \cite{Tancik2020PE}, and sinusoidal activation \cite{Sitzmann2020SIREN}. Another line of work focuses on extending the capability of neural implicit representations to general shapes like open surfaces using a neural unsigned distance field \cite{Chibane2020UDF} or modified SDF \cite{Chen20223PSDF}. However, these prior approaches cannot faithfully recover sharp features or cannot scale between fine-grained shapes and large scenes.

\paragraph{Hierarchical neural implicit functions}

A class of methods based on hierarchical representation has been proposed to achieve more compact representations, reconstruct highly detailed geometry, and adaptively scale to different levels of detail (LODs). These methods can be divided into two branches. The first branch focuses on hierarchical designs within the network architecture~\cite{lindell2022bacon, chibane2020implicit}. For example, Lindell et al.~\shortcite{lindell2022bacon} introduce a multiplicative filter network~\cite{fathony2020multiplicative} to limit the bandwidth of each layer's output, enabling the network to learn a multi-resolution decomposition of the output. The second branch involves using explicit representations, such as octrees or hash tables~\cite{martel2021acorn,takikawa2021neural,tatarchenko2017octree,muller2022instant}. 
\textit{NGLOD}~\cite{takikawa2021neural} and \textit{ACORN}~\cite{martel2021acorn} are two examples. \textit{NGLOD} maps the point coordinates to octree-organized features and decodes these features into an SDF, while \textit{ACORN}~\cite{martel2021acorn} utilizes an octree-like coordinate network to learn an adaptive decomposition of the 3D occupancy fields, enabling the network to fit 3D shapes faster and more accurately. While these hierarchical methods have made significant advancements in terms of spatial and computational efficiency, as well as LOD applications in neural surface representation, they still lack the ability to express diverse geometric features that include sharp features and open boundaries.

While our work utilizes an adaptive grid, the purpose of this adaptive grid is different from that in the previous works which build the hierarchical feature volumes on top of the octree data structure. The adaptive grid in our work is complementary to the state-of-the-art methods. It is used to confine CSG operations to localized grid cells, thereby avoiding undesired interference between spatially close but unrelated patches and ensuring the validity of the merged results.

\paragraph{Patch-based neural implicit functions}

Representing the entire shape using a set of shape primitives is a classic point of view in traditional explicit geometry processing \cite{Schnabel2007RANSAC,Li2011Globfit,Nan2017PolyFit}. Those patch-wise representations decompose complex models into multiple patches with parametric shape priors to extract high-level shape structures for downstream applications like shape completion or semantic editing.
Motivated by the simplicity of patches, neural implicit representations adopted similar ideas to decompose shapes, allowing learning of locally controllable models \cite{Genova2020LDIF}, generalizable parts \cite{Tretschk2020PatchNets}, or a semantic compositional parametric model \cite{Zhang2022DMM}. However, it is not easy to achieve high representation accuracy by using patch-based implicit representations, due to the difficulty in stitching together or trimming patches. 

Focusing on neural implicit modeling of CAD shapes with sharp features, Guo et al.~\shortcite{Guo2022NHRep} give a global CSG-based solid entity representation that is based on patch-wise halfspaces. It first adopts a top-down constructive method to build the global CSG tree that ensures the order of Boolean operations yields theoretically sound results. Then it learns patch-wise halfspace-based representations on all sample points in the domain, thus coordinating them in the whole space. A similar work [Reddy et al., 2021] uses the global superposition of multiple learned 2D SDFs to reconstruct fonts while preserving their sharp corners and edges.
As we will show in experiments, in contrast to our local approach to CSG construction for sharp feature modeling, the global approach~\cite{Guo2022NHRep} has difficulty precisely representing complex 3D shapes with many concave patches. Moreover, our approach handles more general surface types than solid CAD shapes, including open surfaces, self-intersecting surfaces, and non-orientable surfaces.

%% file: sections/method.tex
\section{Our method}\label{sec:method}

\subsection{Overview}

We present the overview of \textit{Patch-Grid} in Fig.~\ref{fig:pipeline}. Input to our method is a shape $\mathcal{S} = \bigcup_{p=1}^K S_p$ composed of a collection of surface patches $S_p$ along with the type of connection between adjacent patches.
Our goal is to represent the shape $\mathcal{S}$ as a neural implicit surface $\mathcal{Z}$ composed of a collection of neural signed distance fields. We denote the zero-level set of a neural signed distance field as $Z_p$ which contains a target surface patch $S_p$. 
To facilitate fast training and effective modeling, we introduce a patch volume that tightly encloses a surface patch $S_p$ and defines the extent of $Z_p$ as shown in Fig.~\ref{fig:pipeline}(b). Furthermore, we impose a regular grid structure on the patch volume of $S_p$ and assign a feature vector at each grid point. Then the collection of all these feature vectors is referred to as the {\em patch feature volume}. Here, the grid resolution is determined by the local shape complexity, as will be explained later. Given the patch feature volume and a query coordinate relative to the patch volume, we use an MLP to map the interpolated feature vector at the query point to a signed distance value, thus approximating the SDF of the surface patch $S_p$ (Sec.~\ref{sec:patch_grid}).
We use a local approach based on another grid structure, named the \textit{merge grid} $\mathcal{G}$, to effectively assemble $\{Z_p\}_{p=1}^K$ into the final neural representation $\mathcal{Z}$. The merge grid adaptively subdivides the spatial domain around the sharp geometric features to simplify the adjacency graph of surface patches enclosed in the individual merge grid cells (Sec.~\ref{sec:merge_grid}). A patch volume and the merge grid of a 3D shape are shown in Fig.~~\ref{fig:patch_merge_grid}.
The overall training process is presented in Sec.~\ref{sec:training_loss}. In addition, we also present a fast adapting strategy for local shape update in Sec.~\ref{sec:local_shape_update}.

\subsection{Patch Feature Volumes}\label{sec:patch_grid}

\begin{figure}
    \centering
    \begin{overpic}
    [width=1.\linewidth]{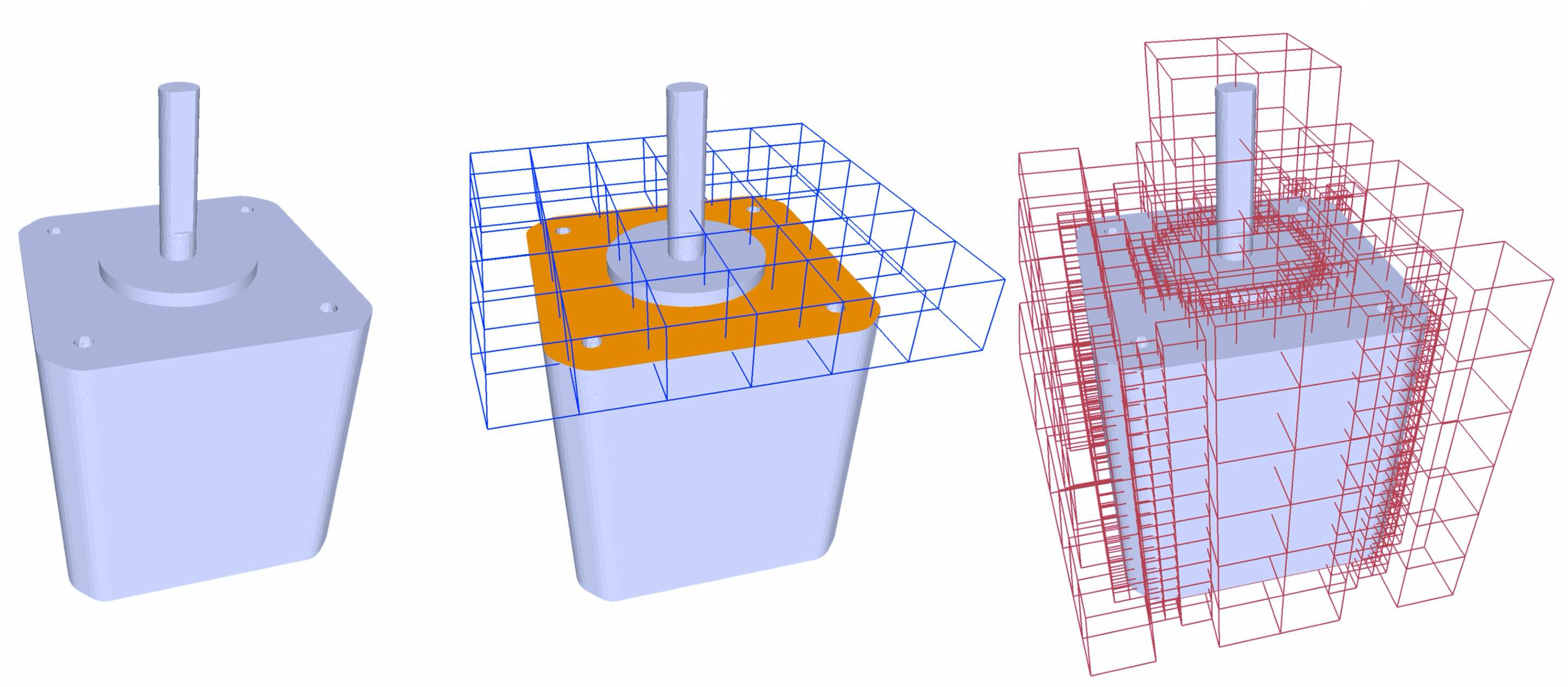}
    \put(7,-3){(a) GT}
    \put(29,-3){(b) A patch volume}
    \put(67,-3){(c) The merge grid}
    \end{overpic}
    \vspace{2mm}
    \caption{Given a 3D shape (a), we show (b) a patch volume for the patch in orange and (c) the merge gird.}
    \label{fig:patch_merge_grid}
\end{figure}

A \textit{patch feature volume} is constructed as follows.
First, for each surface patch $S_p$ of a given shape, we determine its axis-aligned rectangular bounding box and partition the bounding box into a grid of regular rectangular cells whose size is determined adaptively according to the shape complexity of the patch. Then, we prune the cells to keep only those that enclose part of the surface patch $S_p$, so that the union of these remaining nonempty cells forms a tight bounding volume of $S_p$.  We call these cells a {\em patch volume} denoted by $V_p$, as shown in Fig.~\ref{fig:patch_merge_grid}(b). 

Next, for each patch volume, we introduce a learnable {\em feature volume}, $FV_p = \{ \mathbf{f}_{(i,j,k)}^p \}$, where feature vectors $\mathbf{f}_{(i,j,k)}^p \in \mathbb{R}^D$ are assigned to the grid points of the patch volume, indexed by $(i, j, k)$. A feature field $F_p(\mathbf{x}): V_p \rightarrow \mathbb{R}^D$ is defined over the patch volume $V_P$ by trilinear interpolation of the feature vectors at the 8 corner points of the cell containing the spatial query point $\mathbf{x} \in \mathbb{R}^3$.
The grid resolution is determined adaptively according to the shape complexity of individual patches. Specifically, it is based on the averaged shape diameter function~\cite{shapira2008consistent}, which is a reasonable approximation of the local feature size~\cite{amenta1998surface}.

Given the continuous feature vector field $F_p(\mathbf{x})$, we use an MLP decoder $f$ to represent surface patch $S_p$ as part of the zero-level set $Z_p$ of the following function, 
\begin{equation}\label{eq:embed_surface}
    S_p \subset Z_p = \{ \mathbf{x} | f(F_p(\mathbf{x})) = 0 \} .
\end{equation}
The surface shape thus generated is independent of the absolute position of the query point but depends solely on the interpolated features. The use of the patch feature volume allows the decoder $f$ to be pretrained on a shape dataset to learn a local shape prior, as we will explain in Sec.~\ref{sec:local_shape_update}.

Representing a surface patch locally by the patch volume is a key design that distinguishes our method from the global approach used in \textit{NH-Rep}~\cite{Guo2022NHRep}. In this way, each learned implicit function of a surface patch is only defined within the patch volume as shown in Fig.~\ref{fig:pipeline}(c). This effectively trims off the extraneous zero-level set of a learned implicit function outside the patch volume, thus avoiding undesired interference between extraneous zero-level sets faced by the global approach.

For the remaining extraneous zero-level sets within the patch volumes $V_p$, we need to merge $Z_p$ with its neighboring patches following a sequence of pre-defined Boolean operations, i.e., a CSG tree, to form $S_p$, as will be represented next. 

\begin{figure}
    \centering
    \begin{overpic}
    [width=1.\linewidth]{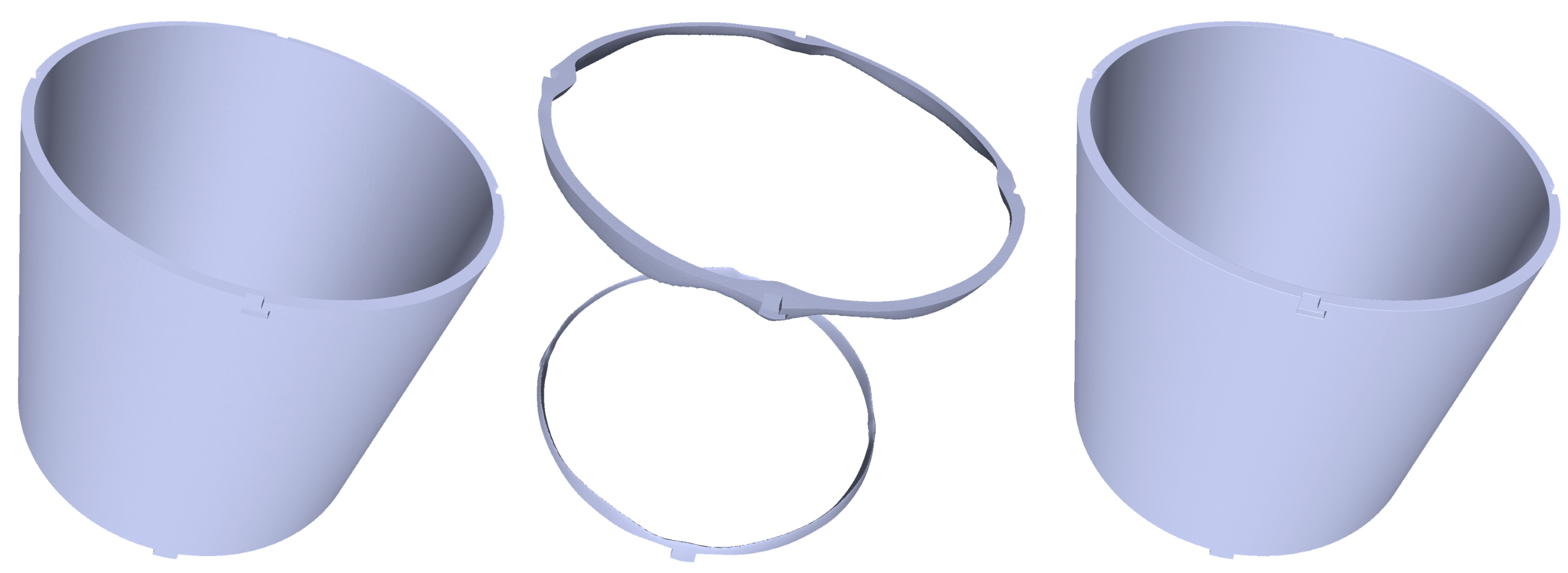}
    \put(7,-5){(a) GT}
    \put(38,-5){(b) NH-Rep}
    \put(71,-5){(c) Patch-Grid}
    \end{overpic}
    \vspace{2mm}
    \caption{(a) A challenging case with highly concave regions at the top and bottom of the cylindrical shell. (b) \textit{NH-Rep} fails in this challenging case due to the difficulty in merging the learned implicit functions in a global manner. (c) Our \textit{Patch-Grid} performs robustly due to the localized approach adopted.}
    \label{fig:nhrep_failure}
\end{figure}

\subsection{Merge Grid and Merging Constraints}\label{sec:merge_grid}

\begin{figure}
    \centering
    \begin{overpic}
    [width=0.95\linewidth]{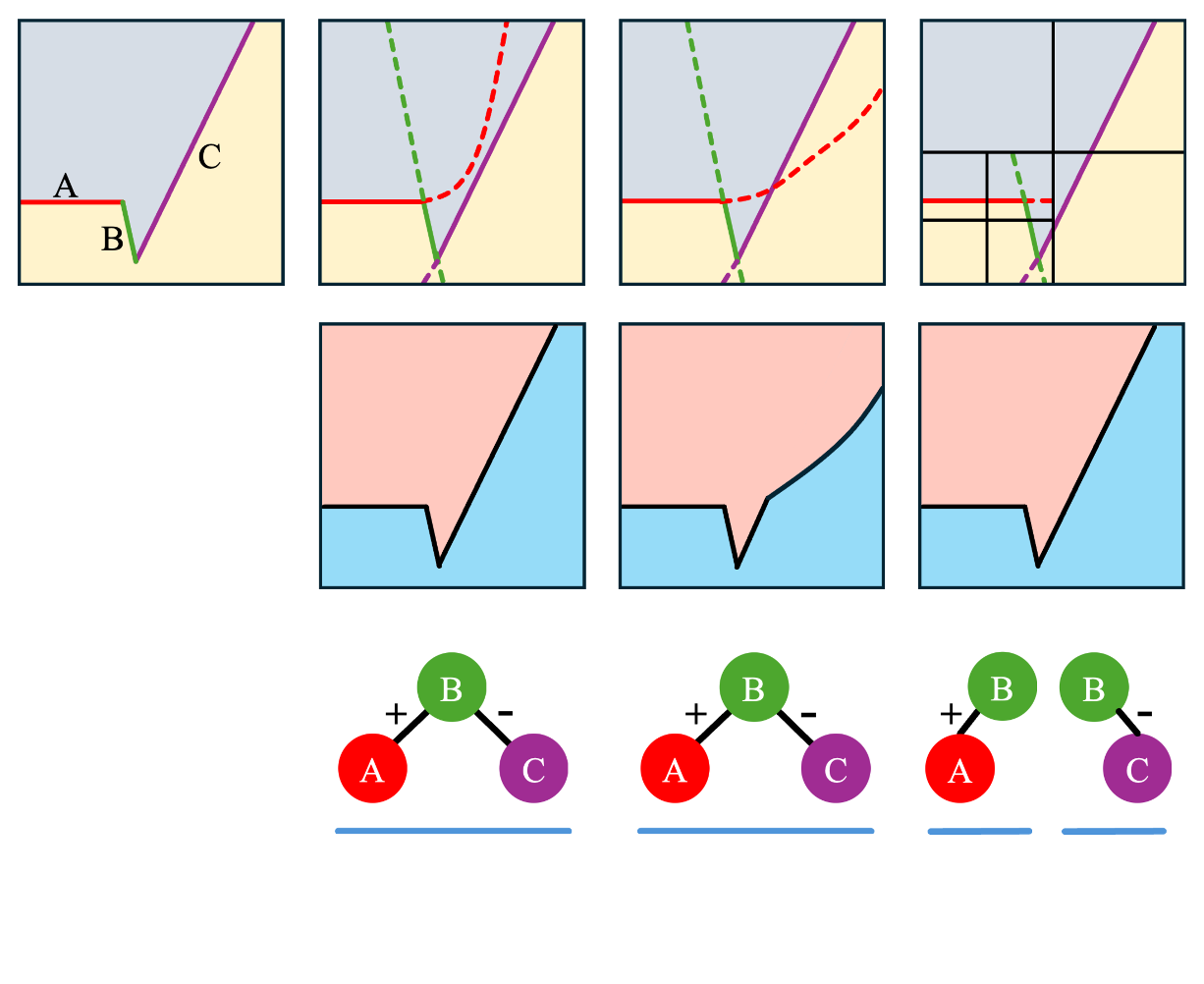}
    \put(3,76){$C_0$}
    \put(28,76){$C_0$}
    \put(53,76){$C_0$}

    \put(9,-2.5){(a)}
    \put(35,-2.5){(b)}
    \put(59,-2.5){(c)}
    \put(85,-2.5){(d)}    
    \put(78,76){$C_{00}$}
    \put(76.5,61){\tiny{$C_{033}$}}
    \put(76.5,66.5){\tiny{$C_{030}$}}
    \put(82.5,66.5){\tiny{$C_{031}$}}
    \put(82.5,61){\tiny{$C_{032}$}}
    \put(92,76){$C_{01}$}
    \put(92,60){$C_{02}$}

    \put(36,10){\tiny$C_0$}
    \put(61,10){\tiny$C_0$}

    \put(79,10){\tiny$C_{031}$}
    \put(90,10){\tiny$C_{032}$}
    \end{overpic}
    \caption{(a) The ground truth shape; (b) An ideal case of the extraneous zero-level set of curve $A$ learned by \textit{Patch-Grid}; (c) A failure case caused by undesired interference between the zero-level sets of curves $A$ and $C$; (d) Applying the merge grid $\mathcal{G}$ converts the original CSG operations to localized regions (i.e., cell $C_{031}$). 
    The first row depicts 2D illustrations of the curves. The second row shows the merged zero-level sets in black. The third row shows the CSG operations under different circumstances, and applying \textit{merge grid} $\mathcal{G}$ simplifies the CSG operations in each cell and significantly reduces the burden for learning coordinated extraneous zero-level sets.}
    \label{fig:clique}
\end{figure}

\paragraph{Patch connectivity and Boolean operations}
 We adhere to the definition used in \textit{NH-Rep} \cite{Guo2022NHRep}, where the connectivity between any two adjacent patches is determined as \textit{concave}, \textit{convex}, or \textit{smooth}, based on the dihedral angles of patches adjacent to a common boundary curve.
Assuming a point $b\in \Gamma$ on a boundary curve $\Gamma$, the dihedral angle $\alpha_{b} \in [0,2\pi]$ is formed by patch tangent directions pointing away from the boundary curve and excluding the outward normal vector. 
If $\alpha_{b}< \pi-\epsilon, \forall b\in \Gamma$, the boundary curve $\Gamma$ is classified as \textit{convex}; if $\alpha_{b}> \pi+\epsilon, \forall b\in \Gamma$, $\Gamma$ is classified as \textit{concave}; otherwise, $\Gamma$ is \textit{smooth}. We follow Guo et al.~\shortcite{Guo2022NHRep} in using $\epsilon=0.35$ as the tolerance.

For hybrid boundary curves that simultaneously exhibit convex and concave features at different points, we adopt the patch decomposition algorithm from \textit{NH-Rep} \cite{Guo2022NHRep} to divide the boundary curve into segments of singular connectivity types, and draw \textit{smooth}-type curves within the adjacent patches to subdivide them and maintain consistent topology.

To create the boundary edge shared by two adjacent patches, we employ a Boolean operation ($\textsc{max}$ for intersection or $\textsc{min}$ for union) based on whether the edge is \textit{convex} or \textit{concave}; the \textit{smooth} type can be regarded as either \textit{convex} or \textit{concave} without special treatment. The adjacency of two connected patches forming a \textit{convex} edge is denoted as $+$, indicating the $\textsc{max}$ operation. Conversely, if the adjacency is \textit{concave}, it is denoted by $-$, indicating the $\textsc{min}$ operation. Consequently, a CSG tree, i.e., a sequence of Boolean operations, assembles the surface patches into the desired shape.

In contrast to the global approach of constructing the CSG tree~\cite{Guo2022NHRep}, which is susceptible to robustness issues (refer to Fig.~\ref{fig:nhrep_failure} for an example), we adopt a local approach based on an adaptive \textit{merge grid} $\mathcal{G}$ to enhance the robustness of our learned neural implicit representation.

\paragraph{Global approach lacks robustness. }

We will use a simple 2D example in Fig.~\ref{fig:clique} to illustrate the robustness issue encountered by the global approach. 
Consider a 2D solid shape (shaded in pale yellow) bounded by three curve segments $A$, $B$, and $C$, with two sharp geometric features, i.e., corners in 2D. 

Because the SDF of curves $A$, $B$, and $C$ will be merged into an entire signed distance field according to the Boolean operations in the CSG tree, there are certain requirements for the extended zero-level set of each curve. Here, we refer to the naturally extended part of each curve's zero-level sets, excluding the curve itself, as the \textit{extraneous zero-level set}, which is shown by dashed curves. 

Due to the lack of SDF supervision in the extended region, the learning progress may lead to extended SDF as shown in either (b) or (c). Since curves $A$ and $C$ are not directly connected, supervising the extraneous zero-level set of $A$ to avoid undesired interference with patch $C$ (c), thus leading to an ideal result (b), relies on the supervision imposed on the entire SDF after merging. While the illustrated case is rather simple and thus easy to handle by a neural network, a more complicated 3D shape can lead to the failure of the global approach as shown in Fig.~\ref{fig:nhrep_failure}. 
In contrast, our localized approach first models each surface patch (or curves in this example) within its bounding volume to constrain the extent of the extraneous zero-level set of the patch. Then, a merge grid is applied to subdivide the spatial domain, ensuring that each merge grid cell contains the patch relationship in the simplest possible form as explained next.

\paragraph{Adaptive merge grid. }
We adopt a divide-and-conquer strategy by subdividing the spatial domain with an adaptive merge grid. Let us continue our explanation with the aforementioned 2D example. By subdividing the spatial domain (denoted $C_0$) in Fig.~\ref{fig:clique}(a), each subdivided cell in Fig.~\ref{fig:clique}(d) contains a simpler geometry -- at most one sharp corner is enclosed in each cell. The benefit is two-fold. First, the curve segment $C$ is now defined only in cell $C_{01}$, $C_{02}$, and $C_{032}$ and so is its extraneous zero-level set. This circumvents the demanding task of globally coordinating extraneous zero-level sets of multiple patches. Second, the construction of the sharp corner formed by $B$ and $C$ (in $C_{032}$) now involves only these two relevant patches; curve segment $A$ is trimmed off in $C_{031}$ avoiding any potential interference from other unconnected patches (e.g.\ as it did to $C$ in Fig.~\ref{fig:clique}(c)). This localized approach drastically improves the robustness of CSG operations compared to the global approach.

\begin{figure}
    \centering
    \begin{overpic}
    [width=1.\linewidth]{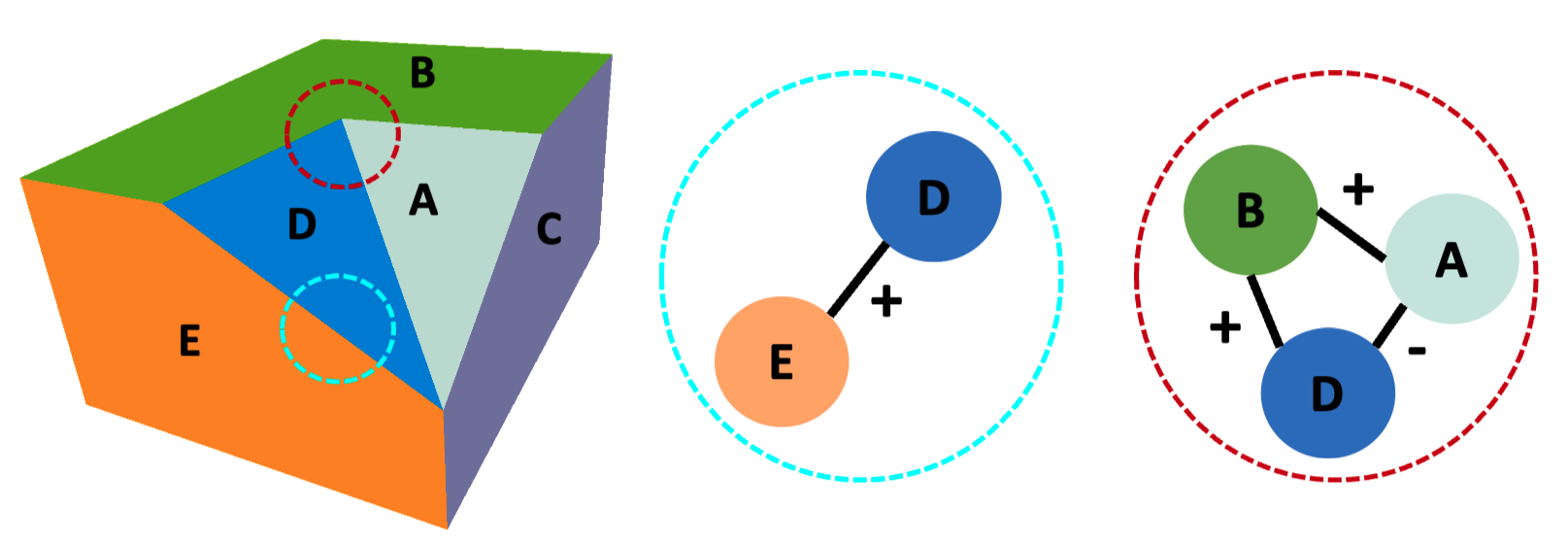}
    \put(18,0){(a)}
    \put(48,0){(b)}
    \put(75,0){(c)}
    \end{overpic}
    \caption{Sharp geometric features in a 3D shape. (a) A 3D example; (b) The adjacency graph of an edge feature; (c) The adjacency graph of a corner feature.}
    \label{fig:sharp_features}
\end{figure}

Now, we elaborate on this subdividing scheme for 3D shapes. Consider each patch as a node. An undirected \textit{adjacency graph} can be defined for the patches bounded in each domain. Then, a sharp edge formed by two adjacent patches corresponds to a link in the adjacency graph and a corner feature at which multiple patches intersect corresponds to a complete subgraph of the adjacency graph, as shown in Fig.~\ref{fig:sharp_features}. 

We initialize the merge grid as an octree grid and subdivide it based on two criteria: 1) whether this cell contains a patch feature volume with a finer resolution, and 2) whether the adjacency graph of the cell is a complete graph (i.e., a \textit{clique}). The second criterion translates into the situation where the CSG operations are defined between all pairs of patches in the grid cell, therefore the cell cannot be further simplified by subdivision.
The second criterion can be achieved ideally through subdivision, but for computational practicality we impose a maximum depth constraint on the octree subdivision.  Thus, there may be cells whose graphs are not cliques at the maximum depth. 
That said, the presented octree subdivision can already significantly reduce the complexity of the graph in a cell, ensuring the robustness of the proposed method as shown empirically.
The pseudo-code for constructing the adaptive merge grid is provided in Alg.~\ref{alg:octree_sub}.

There may exist more than two connected components in the adjacency graph of a leaf cell of the octree merge grid.
Such cases can occur for disjoint patches with close proximity, or for modeling self-intersecting shapes. To handle such cases uniformly, drawing on ideas of how they can be meshed~\cite{LiBarbic18_SelfIntersectionImmersion}, we simply create $Q$ copies of the cell, with each copy representing one connected component within the cell, and assign the duplicated cells for merge loss computation according to their corresponding patch identity.

\paragraph{Merge loss design. }

Once the merge grid has been adaptively subdivided, for each connected component at the non-empty leaf node, we utilize the Boolean tree construction algorithm described in Guo et al.~\shortcite{Guo2022NHRep} to build a CSG tree. The Boolean operations of the CSG tree impose merging constraints on the learned implicit functions. The objective is to ensure that the zero-level set of the merged implicit function, denoted as $M$, accurately reproduces the target geometry.

We now introduce how the patch-wise merge loss $\loss^p_{\mathrm{merge}}$ of surface patch $S_p$ is computed. 
First, assume that a merge cell $C_i$ contains two patches $S_p$ and $S_q$. We compute the merge loss $\loss^i_{\mathrm{merge}}$ in this cell as follows:
\begin{equation}\label{eq:merge_loss}
    \loss^i_{\mathrm{merge}} = \frac{1}{N_x}\sum_{\mathbf{x} \in C_i}|M(\mathbf{x})|,
\end{equation}
where $\mathbf{x}$ are sampled from $S_{\{p,q\}}$ in cell $C_i$ and $N_x$ is the number of sample points.
Then, we assign this merge loss to the involved patches $S_{\{p,q\}}$. 
Note that a cell in the patch volume (e.g.\ $C_0$ in Fig.~\ref{fig:clique}(d)) can be subdivided into smaller, nested merge cells (e.g.\ $C_{031}$ or $C_{032}$ inside $C_0$). We define the patch-wise merge loss as follows to balance the patch loss and the merge loss:
\begin{equation}
    \loss^p_{\mathrm{merge}} = \frac{1}{L}\sum_{l=1}^{L}(\frac{1}{J}\sum_{j=1}^{J}{\loss^j_{\mathrm{merge}}}).
\end{equation}\label{eq:merge_loss_2}
Here, we denote as $\{FC_l\}_{l=1}^L$ the collection of cells in $S_p$ containing merge cells and $\{C^l_j\}_{j=1}^J$ is the collection of merge cells in $FC_l$.

\begin{algorithm}[t]
\caption{Construction of Merge Grid}\label{alg:octree_sub}
\begin{algorithmic}[1]

\REQUIRE~~\\
Surface shape $\mathcal{S}$\\
Initial merge grid $\mathcal{G}$\\
Maximal depth $d_{max}$;\\
\ENSURE~~\\
Merge grid $\mathcal{G}$;\\
\WHILE{$\#$subdivisible leaf cells > 0}
    \FOR{$C \in \mathcal{G}$}
        \IF{$C.sub == False$}
            \STATE \textit{continue};
        \ELSIF{$\mathcal{S} \bigcap C \neq \varnothing$ and $C.d < d_{max}$}
            \FOR{every connected graph $g$ in $C$}
                \IF{$g$ is not a {\bf clique} graph}
                    \STATE  $M \gets subdivide(C)$; \\
                    \STATE  $C'.sub \gets True, \forall C' \in M$; \\
                    \STATE  \textit{break}; \\
                \ENDIF
            \ENDFOR
            \STATE  $C.sub \gets False$; \\
            \COMMENT {/*clique cell*/} \\
        \ELSE
            \STATE  $C.sub \gets False$ \\
            \COMMENT {/*empty cell or cell reaches $d_{max}$*/} \\
        \ENDIF
    \ENDFOR
\ENDWHILE
\end{algorithmic}
\end{algorithm}

\subsection{Learning Patch-Grid Representation}\label{sec:training_loss}

The patch feature volumes $\{FV_p\}_{p=1}^K$ and the weights of the shared decoder $f$ are trainable parameters and optimized to fit the target shape $\mathcal{S}$. We adopt a surface fitting loss $\loss_{\mathrm{surface}}$ and a surface normal fitting loss $\loss_{\mathrm{normal}}$ as the data terms for the fitting loss. Besides, we also adopt a pseudo SDF loss $\loss_{\mathrm{SDF}}$ and a regularization term $\loss_{\mathrm{eikonal}}$ proposed by Gropp et al.~\shortcite{Gropp2020IGR} to improve the reconstruction quality.
Hence, the loss function for learning each patch is defined as
\begin{align}\label{eq:patch_loss}
    \loss_{\mathrm{patch}} &= \lambda_{\mathrm{surface}}\loss_{\mathrm{surface}} + 
    \lambda_{\mathrm{normal}}\loss_{\mathrm{normal}} \\ \nonumber
    &+
    \lambda_{\mathrm{SDF}}\loss_{\mathrm{SDF}} + 
    \lambda_{\mathrm{Eikonal}}\loss_{\mathrm{Eikonal}},
\end{align}
where 
\begin{align}
    \loss_{\mathrm{surface}} &= \frac{1}{ |S_p|}\sum_{\mathbf{x}\in S_p} \big| f(F_p(\mathbf{x})) \big|, \\
    \loss_{\mathrm{normal}} &= \frac{1}{ |S_p|}\sum_{\mathbf{x} \in S_p}
    \big\| \nabla_{\mathbf{x}} f(F_p(\mathbf{x})) - \mathbf{n}(\mathbf{x}) \big\|_2, \\
    \loss_{\mathrm{SDF}} &= \frac{1}{ |S_p|}\sum_{\mathbf{x}\in S_p} \big| f(F_p(\mathbf{x}+d\mathbf{n})) - d\big|,\\
    \loss_{\mathrm{Eikonal}} &= \frac{1}{|\Omega_p|} \sum_{\mathbf{x} \in \Omega_p} \big| \|\nabla_{\mathbf{x}} f(F_p(\mathbf{x}))\| - 1 \big|.
\end{align}
Here, $\mathbf{n}(\mathbf{x})$ denotes the unit normal vector at $\mathbf{x}$; $d$ is a small offset distance within a range of $(0, 0.1)$ of the grid cell size; and $\Omega_p$ denotes the set of point samples from the patch volume of surface patch $S_p$.

The final loss of the optimization problem takes the following form:
\begin{equation}\label{eq:total_loss}
    \loss_{\mathrm{total}} = \frac{1}{K} \sum_{p=1}^K \loss^p_{\mathrm{patch}} + \lambda_{\mathrm{merge}}\frac{1}{K} \sum_{p=1}^K \loss^p_{\mathrm{merge}}.
\end{equation}

\subsection{Fast Local Shape Updating with Pretrained Decoder}\label{sec:local_shape_update}

Editing is an important part of 3D shape modeling. In CAD modeling, users commonly adjust the size of a component or deform its geometry to edit an existing 3D model. Therefore, it would be desirable if the learned representation can be updated at an interactive rate as the CAD model is edited. Existing methods~\cite{Guo2022NHRep,martel2021acorn,takikawa2021neural} must overfit the edited shape from scratch, which takes too long for interactive editing. 
For all these prior methods, a change to a surface patch will inevitably propagate to the rest of the patches in proximity due to using a shared feature volume to model these spatially close surface patches. \textit{Patch-Grid} circumvents this limitation by modeling each surface patch with a separate patch feature volume, allowing for independent manipulation and efficient update of a surface patch.

We leverage the compositional nature of \textit{Patch-Grid} and develop a new strategy to enable a fast (i.e.\ less than 2 seconds) local update. The overall idea is to perform a test-time optimization that only optimizes the patch feature volumes of the edited patches while fixing all other patch feature volumes and the shared MLP decoder $f$. This local updating strategy thus reduces the computational time significantly which allows for interactive local shape updates.

To this end, we pretrain the shared MLP decoder $f$ with various surface patches. During the pretraining stage, given a patch $S_p$ from a shape, we construct its patch volume $V_p$ and randomly initialize the patch feature volume $FV_p$ as described earlier. The decoder $f$ along with the patch feature volume $FV_p$ are trained to reconstruct the corresponding surface patch $S_p$ during the pretraining process using Eq.~\ref{eq:patch_loss}. This process exposes the decoder $f$ to a diverse collection of surface patches and thus instills to decoder $f$ a local shape prior. 
We will explain the pretraining details and demonstrate its benefit later in Sec.~\ref{sec:results}.

%% file: sections/global_local_blending.tex
\begin{figure}
    \centering
    \begin{overpic}
    [width=1.\linewidth]{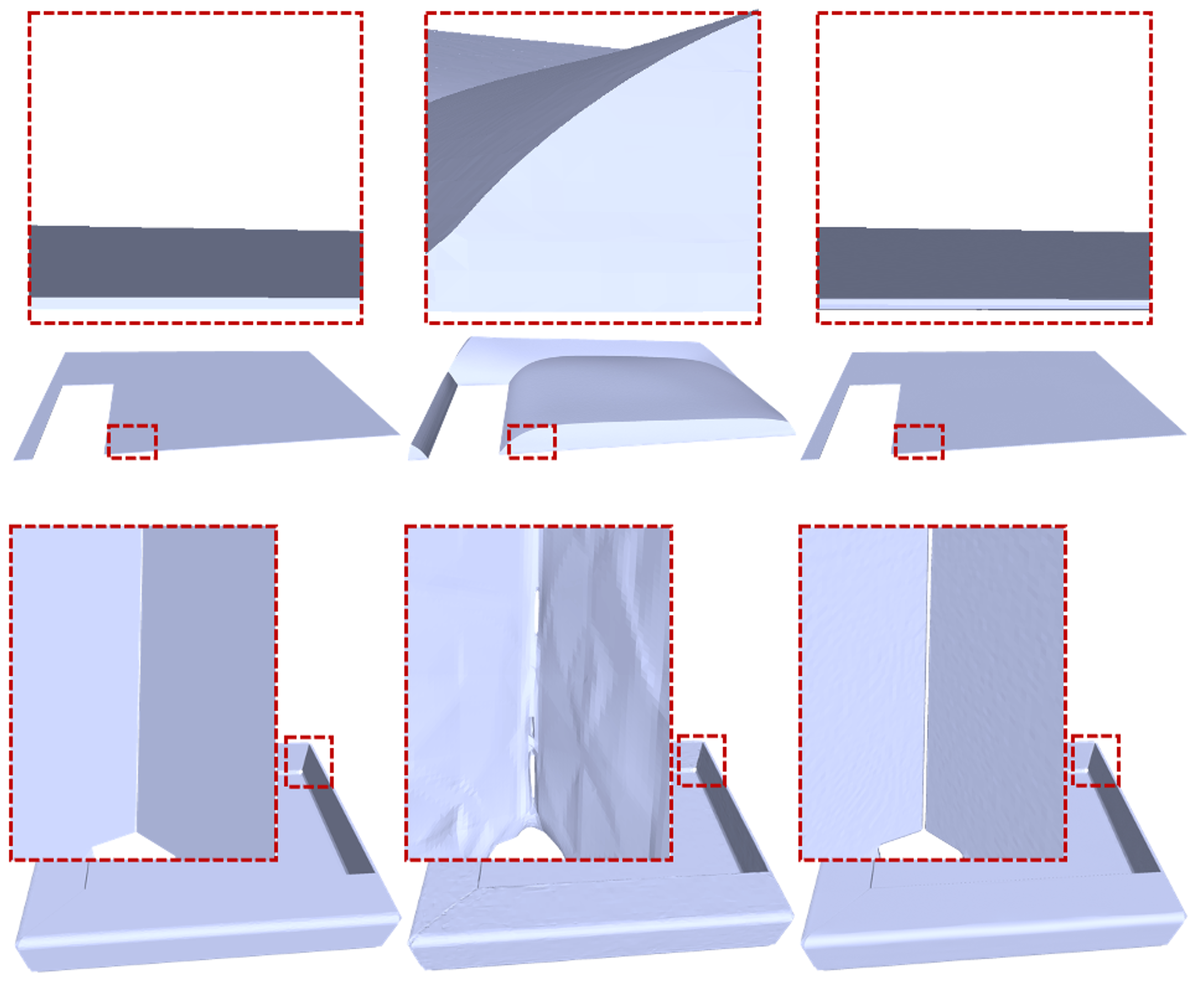}
    \put(9,40){(a) GT}
    \put(40,40){(b) NH-Rep}
    \put(72,40){(c) Patch-Grid}
    \put(9,-3){(d) GT}
    \put(40,-3){(e) NGLOD}
    \put(72,-3){(f) Patch-Grid}
    \end{overpic}
    \caption{Modeling of thin structures. Due to their global scheme, \textit{NGLOD} and \textit{NH-Rep} struggle to represent thin structures; however, our localized method robustly and faithfully addresses these challenging cases.}
    \label{fig:thin_motivation}
\end{figure} 

\section{Modeling diverse geometric features}\label{sec:global_local_blending}

In this section, we first briefly describe how the mesh is extracted from the learned compositional SDFs and then elaborate on how the proposed \textit{Patch-Grid} representation can be applied to modeling diverse geometric features, such as narrow slits or thin solids (which are together termed here as thin geometric features) or open surface boundaries, and enabling global distance queries via a local-global blending scheme.

\textbf{Mesh extraction. } Patch-Grid encloses the given shape composed of multiple surface patches with patch volumes. When two or more patches intersect, the overlapped grid cells in the relevant patch volumes subdivide into a merge grid. To extract the mesh representation, we perform the MarchingCubes algorithm to mesh the zero-level sets in the patch volume cells which contain only a single surface patch. For patch volume cells containing intersecting patches, we only mesh the zero-level sets of the patches in the merge cells. In this way, we ensure the validity of the obtained mesh surface by extracting it from the cells where the patch loss and the merge loss are enabled.

\begin{figure}
    \centering
    \begin{overpic}
    [width=1.\linewidth]{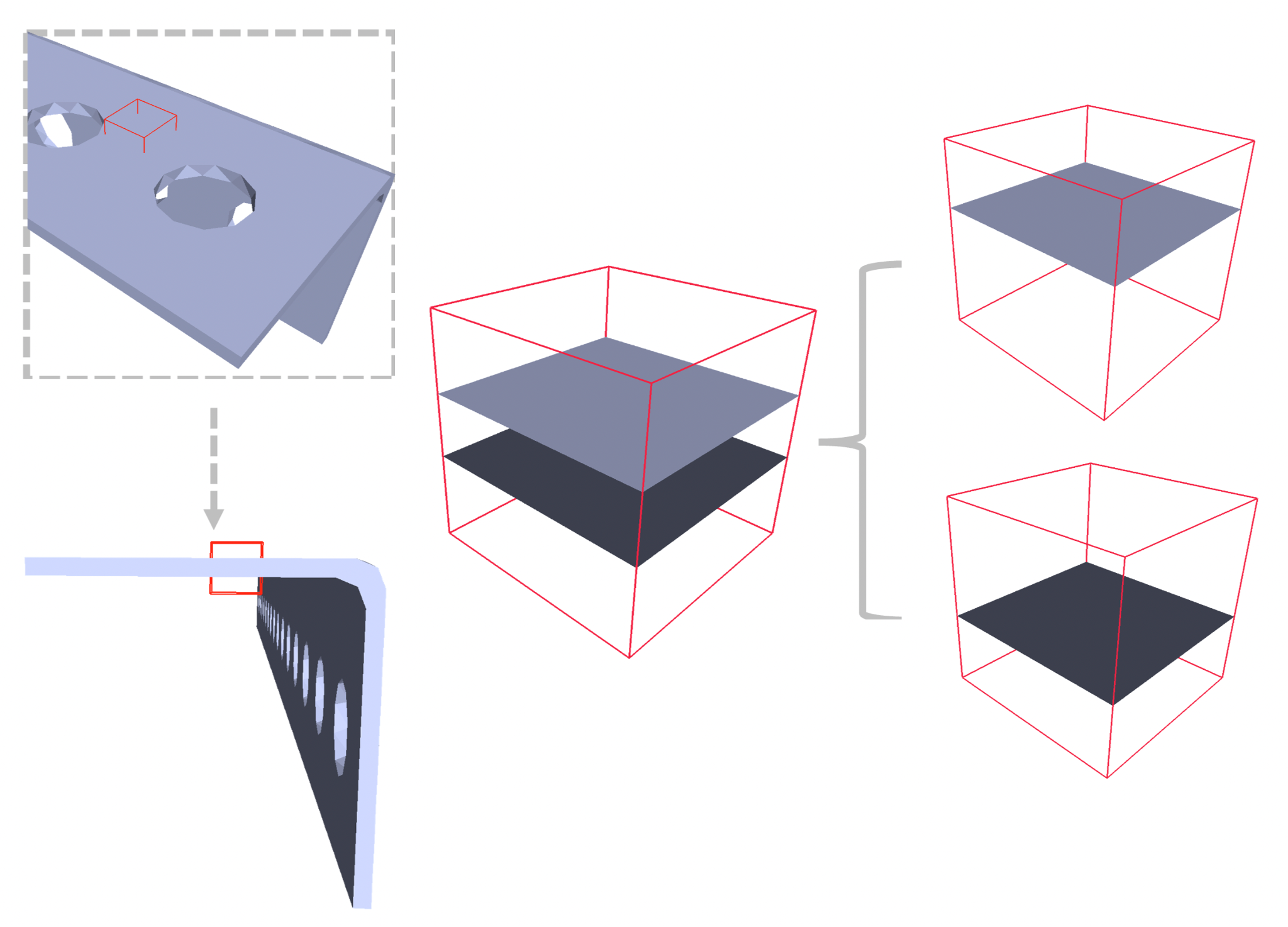}
    \end{overpic}
    \caption{Processing thin geometric features. Each of the two surface patches bounding the thin solid is modeled by a separate independent neural signed distance field.}
    \label{fig:thin_feature}
\end{figure}

\subsection{Modeling of Thin Geometric Features}

It is challenging for a global approach like \textit{NH-Rep}~\cite{Guo2022NHRep} or \textit{NGLOD}~\cite{takikawa2021neural} to fit a neural implicit representation to a 3D shape with thin geometric features, such as the thin solid shown in Fig.~\ref{fig:thin_motivation} (upper row) and the very thin gap shown in Fig.~\ref{fig:thin_motivation} (bottom row).
On the one hand, insufficient samples around these thin features, coupled with the inherent difficulty that the MLP faces in handling sharp changes in SDF, accounts for the failures encountered in such regions by \textit{NGLOD}~\cite{takikawa2021neural}. 

On the other hand, \textit{NH-Rep}~\cite{Guo2022NHRep}, while being a patch-based representation, is limited by the global approach it adopts. As described earlier, the need for managing global interaction between all pairs of patches makes it prone to fail when modeling thin features, as shown in the top of Fig.~\ref{fig:thin_motivation}. 

In contrast to previous methods, our \textit{Patch-Grid} \textit{locally} represents each composing surface patch with a bounding patch volume and achieves robust results in regions of thin geometric features; see our results in Fig.~\ref{fig:thin_motivation}. 
We demonstrate how our local representation benefits the modeling of thin geometric features with Fig.~\ref{fig:thin_feature} where a thin solid is bounded by two spatially close yet disconnected surface patches.
First, the two surface patches are separately represented by their respective patch volumes, and so are their signed distance fields. 
Therefore, learning the two \textit{individual} signed distance fields avoids excessively dense sampling around the thin solid as required by \textit{NGLOD}.
On the other hand, while it adopts a similar patch-based representation, \textit{NH-Rep} defines the patches in the global domain and can only evaluate the geometry through the CSG tree in a global manner that struggles to disentangle the two disconnected but almost overlapping patches. 
In contrast, our \textit{Patch-Grid} enables local evaluation of a surface patch within its patch volume. If two patches are not connected (see Fig.~\ref{fig:thin_feature}), each surface patch can be individually extracted without going through the CSG tree. Hence, our local approach exhibits high flexibility in modeling these thin geometric features.

\begin{figure}
    \centering
    \begin{overpic}
    [width=\linewidth]{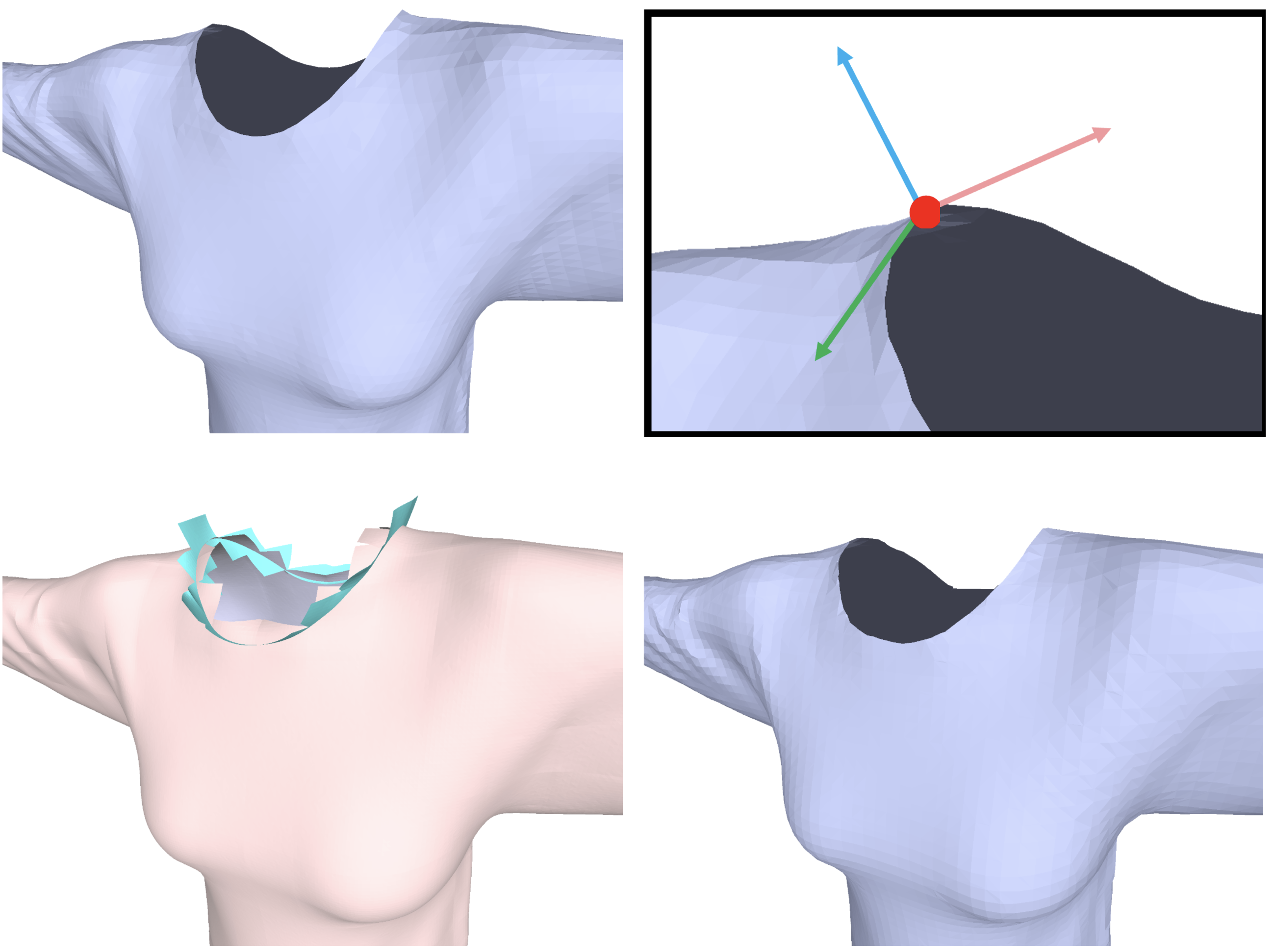}
    \put(12,-3){(c) Individual patches}
    \put(61,-3){(d) Final result} 
    \put(22,37){(a) GT}
    \put(58,37){(b) Training supervision}
    \put(83,60){$N_s$}
    \put(63,65){$N_n$}
    \put(61,50){$N_t$}
    
    \end{overpic}
    \caption{Modeling open boundary surfaces. To learn the trimming patch, the normal supervision $N_s$ is obtained by computing the cross product of the surface normal $N_n$ and the tangent vector of the boundary $N_t$.}
    \label{fig:trimming_open}
\end{figure}

\subsection{Modeling of Open Surface Boundaries}

The modeling of an open surface boundary can be considered as a trimming process, where the open surface patch is formed by trimming away an extended surface at the boundary curve using a trimming surface. 
An example is illustrated in \ref{fig:trimming_open}. We denote a given open surface as $S_o$ and its boundaries as $\partial S_o$. 

We follow the method presented earlier to train a neural surface patch $Z_o$ that reconstructs $S_o$ with some extraneous part (see \ref{eq:embed_surface}). 
To represent the trimming surface as a zero-level set of a neural implicit function, we need to provide sample points in the zero-level set and gradient directions at these points. Specifically, we first sample from the boundary curve $\partial S_o$ a set of points $\mathbf{x}_b$ at which the zero-level set of the trimming surface passes through.
Then, we compute at each of these sampled points a vector $N_s$ that represents the gradient direction of the target implicit functions. $N_s$ is computed as the cross product of the surface normal $N_n$ at this point and the tangent $N_t$ along the boundary curve $\partial S_o$ as shown in \ref{fig:trimming_open}(b). SDF values and gradients of this virtual cutting surface are computed from $\mathbf{x}_b$ and $N_s(\mathbf{x}_b)$. 
We adopt the same training loss, \ref{eq:patch_loss}, to obtain a learned implicit function serving as the trimming patch. Its zero-level set is then used to cut off the extraneous part of the zero-level set $Z_o$ to form a clean boundary curve. 

We implement this trimming operation as a Boolean max operation, assuming that the trimming surface and the corresponding open surface form a sharp convex feature at the boundary curve. Hence, we can adopt the same pipeline as described before to model this virtual sharp edge. During mesh extraction, we simply mask out the sample triangles lying on the trimming surface. We show several results of the modeled open surfaces along with zoom-in views at the surface boundaries in Fig.~\ref{fig:open_surfaces}.

\subsection{Enabling global distance query} \label{sec:global_local_blend_detail}

\begin{figure}
    \centering
    \subcaptionbox[width=0.22\linewidth]{GT shape}{
    \includegraphics[width=0.22\linewidth]{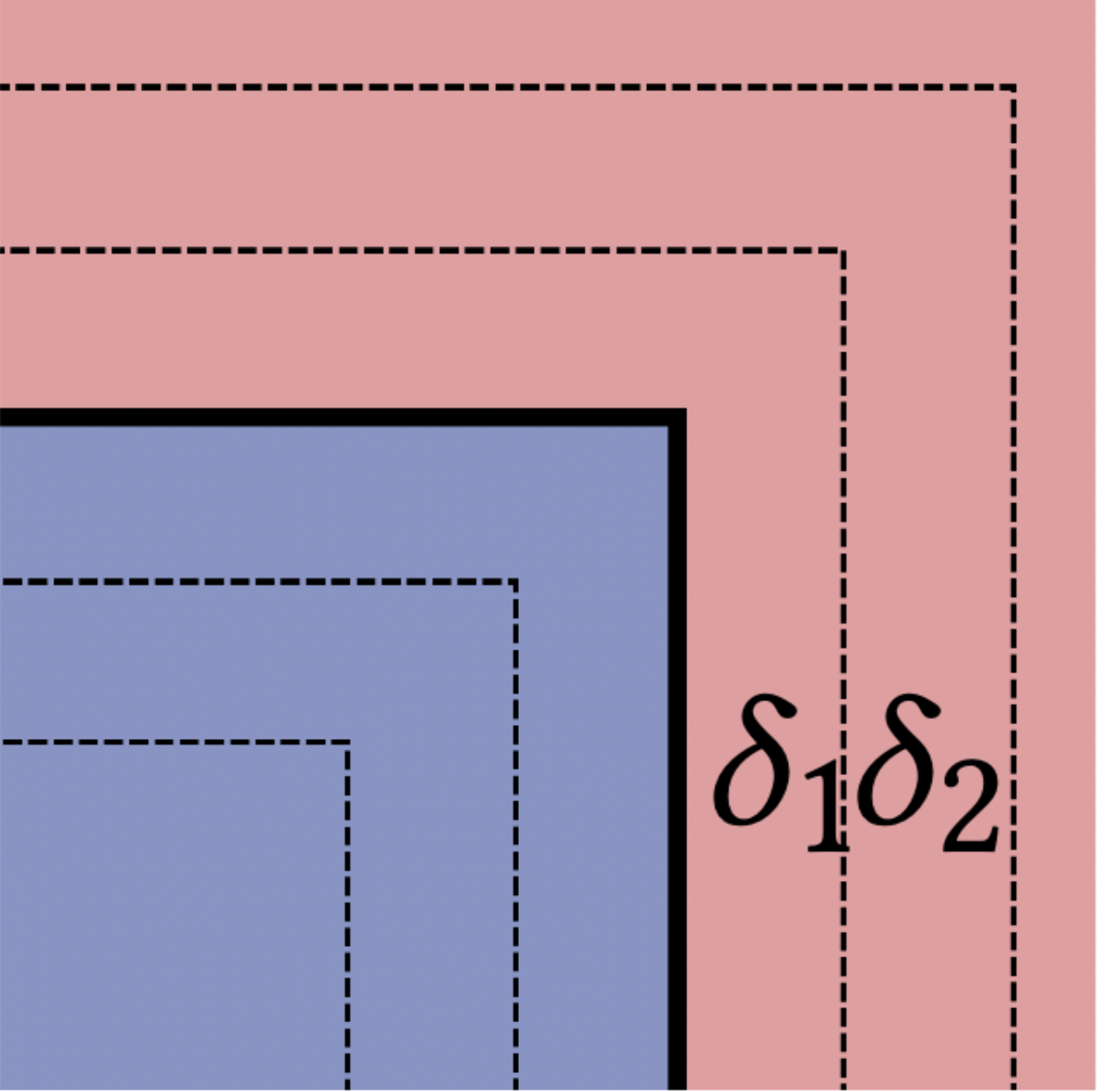}
    }
    \subcaptionbox[width=0.22\linewidth]{Local SDF}{
    \includegraphics[width=0.22\linewidth]{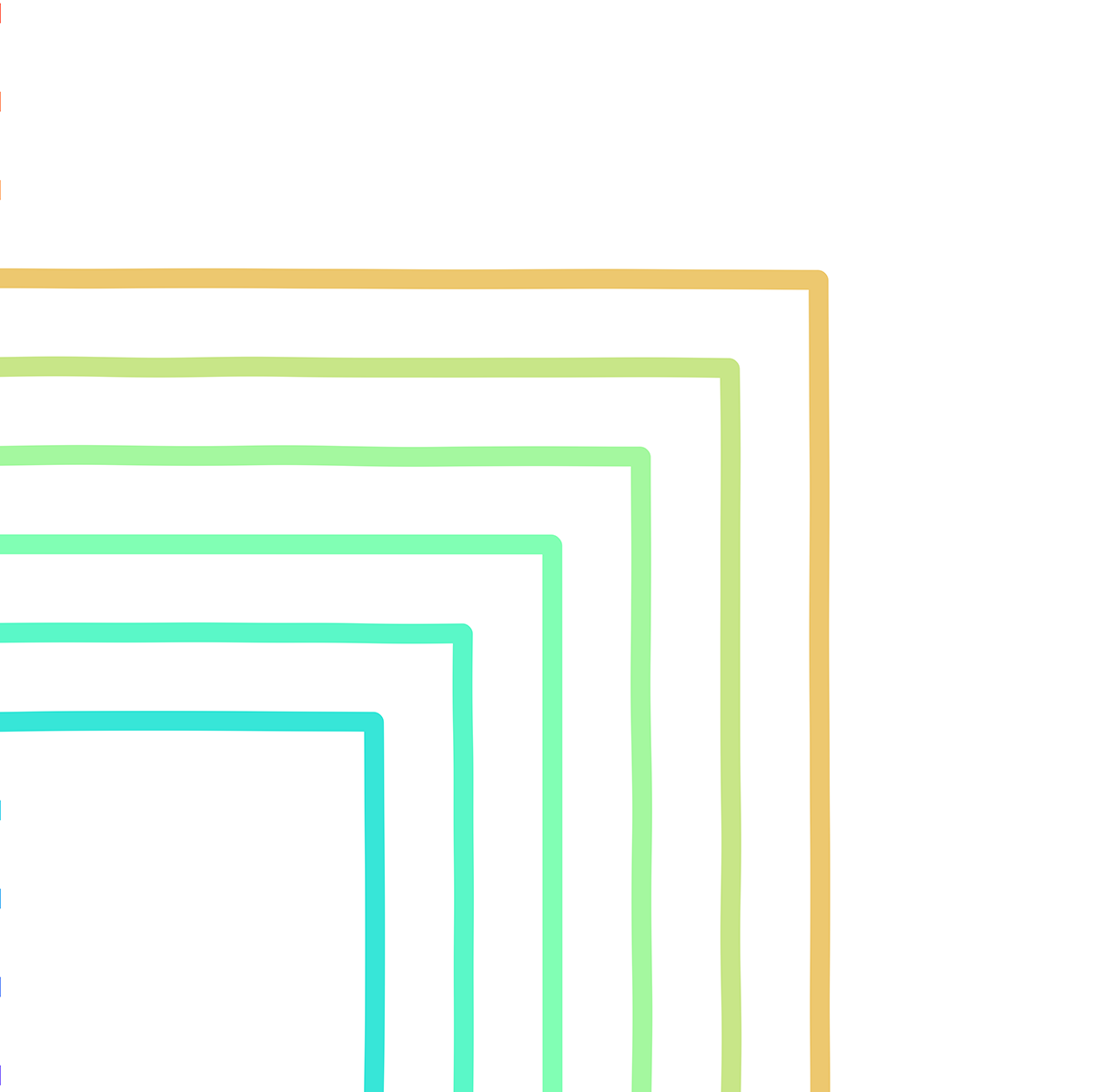}
    }
    \subcaptionbox[width=0.22\linewidth]{Global SDF}{
    \includegraphics[width=0.22\linewidth]{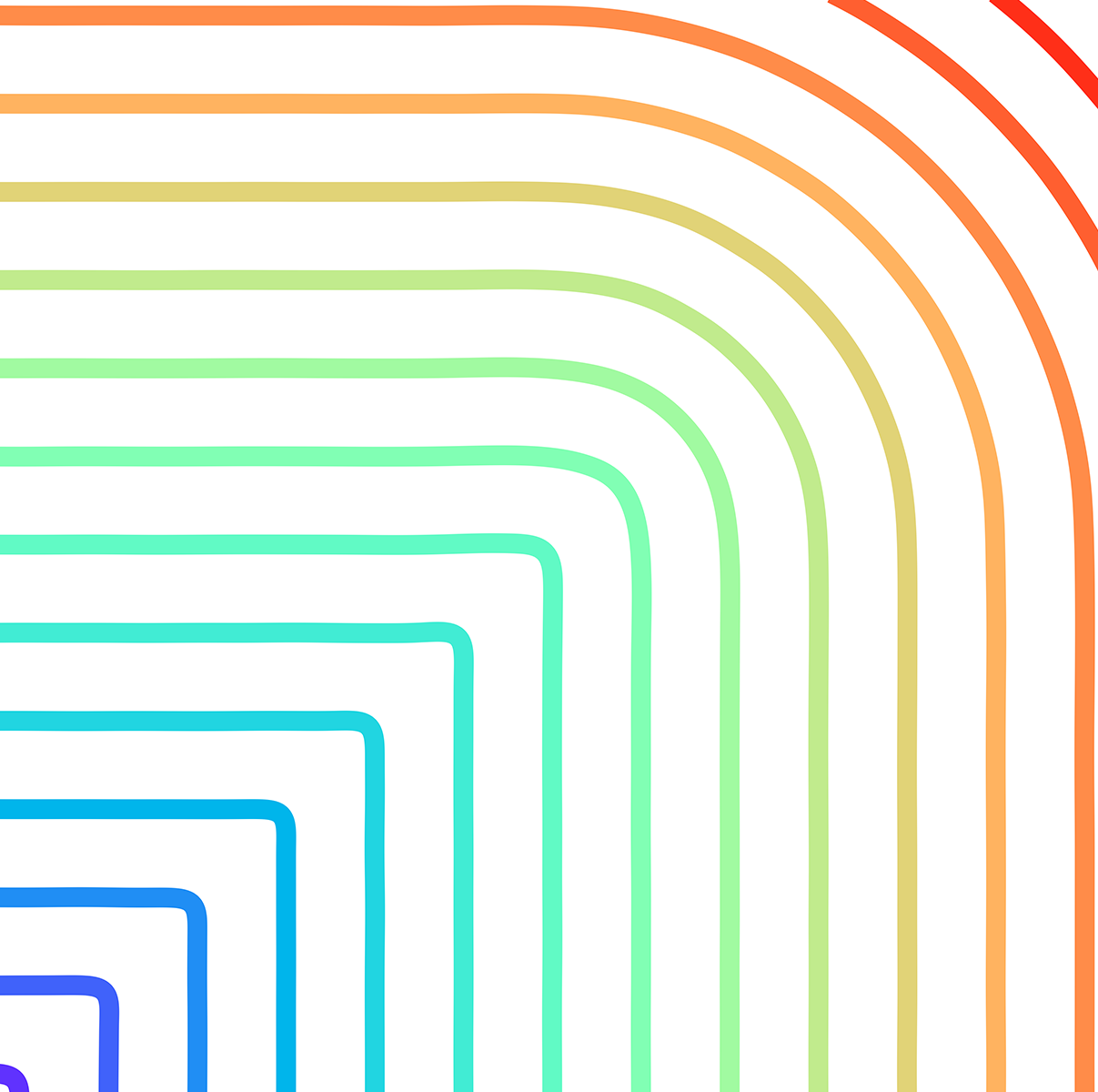}
    }
    \subcaptionbox[width=0.22\linewidth]{Blended SDF}{
    \includegraphics[width=0.22\linewidth]{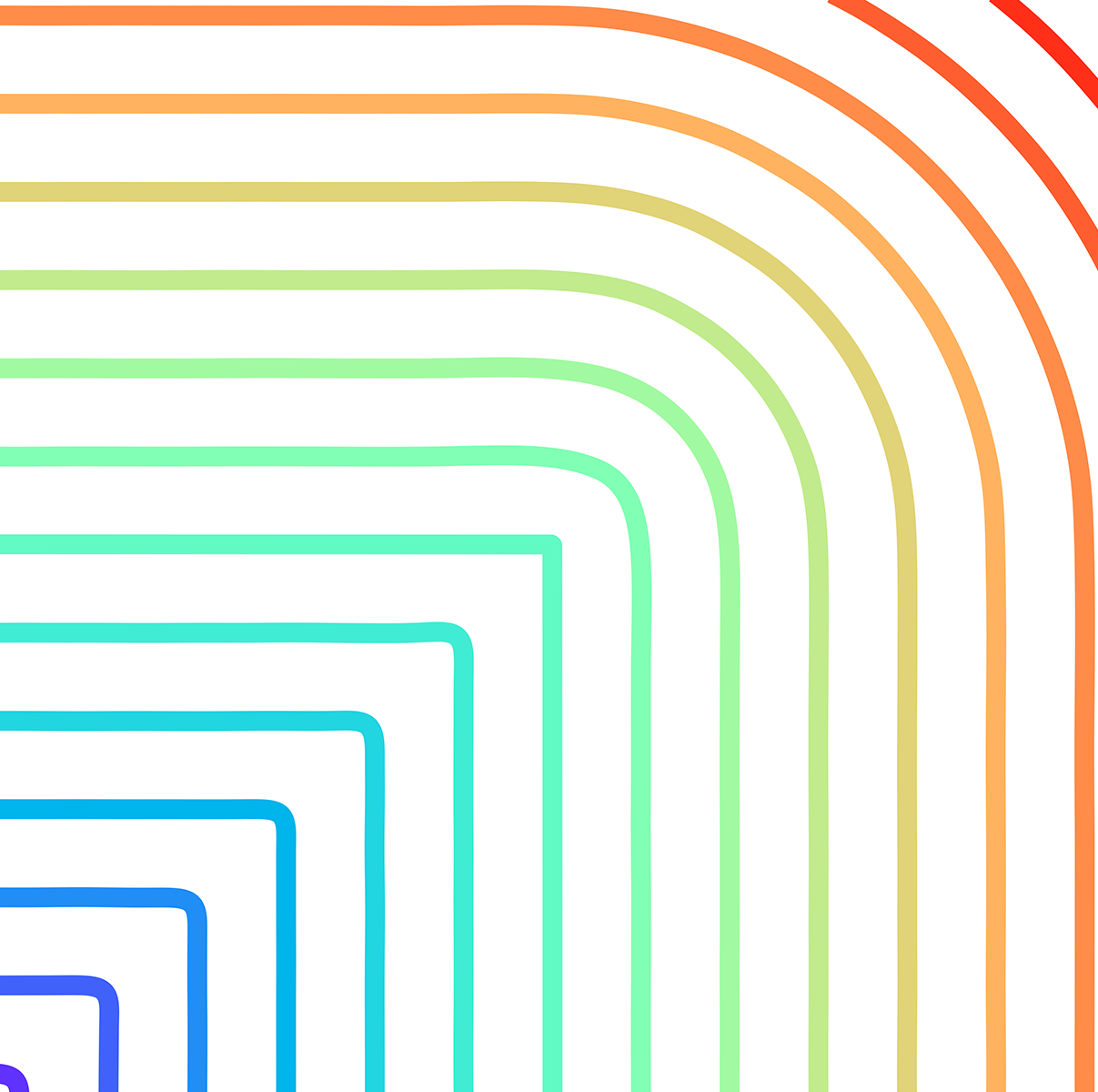}
    }
    \caption{Given the blending region around the GT shape (a), the local SDF (b) and the global SDF (c) are blended to generate the blended SDF (d) that well preserves the sharp corner at its zero-level set and is globally defined in the whole domain.
    }
    \label{fig:2d_blending_eg}
\end{figure}

CAD applications often require a global signed distance function to support inquiring whether a given point is inside or outside of a CAD model. To address this issue, we extend our local patch-based surface representation to a \textit{feature-preserving} global signed distance field $F_G$ for CAD models by blending the local patch-based representation $F_L$ with an ordinary global signed distance field $F_O$. 
A 2D example is shown in \ref{fig:2d_blending_eg} where the GT shape contains a sharp corner. The patch-based representation $F_L$ is defined only in a region near the GT surface, while the learned global field $F_O$ smoothly approximates the sharp corner in the GT shape. In our implementation, we obtain the global field $F_0$ by following the strategy outlined by Lin et al.~\shortcite{lin2024optimal}. In this paper, we utilize a vanilla 8-layered MLP with each layer consisting of 512 neurons, equipped with positional encoding as introduced in ~\cite{mildenhall2020nerf,tancik2020fourier}. A blended field obtained by the following method can retain the sharp feature in the GT shape as well as enable a global distance query.

\begin{figure}
    \centering
    \includegraphics[width=0.75\linewidth]{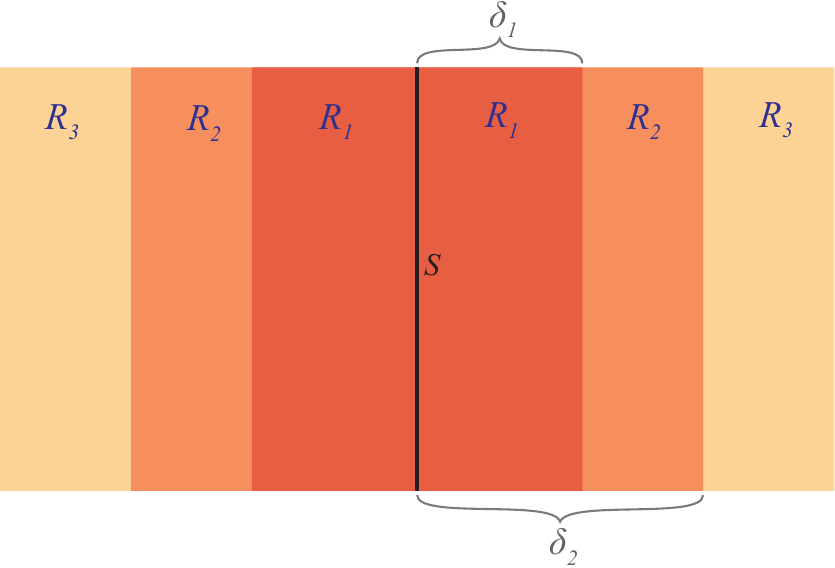}
     \caption{We demonstrate the blending strategy in a 2D case.}
    \label{fig:global_local_demo}
\end{figure}

\begin{figure}
    \includegraphics[width=0.5\linewidth]{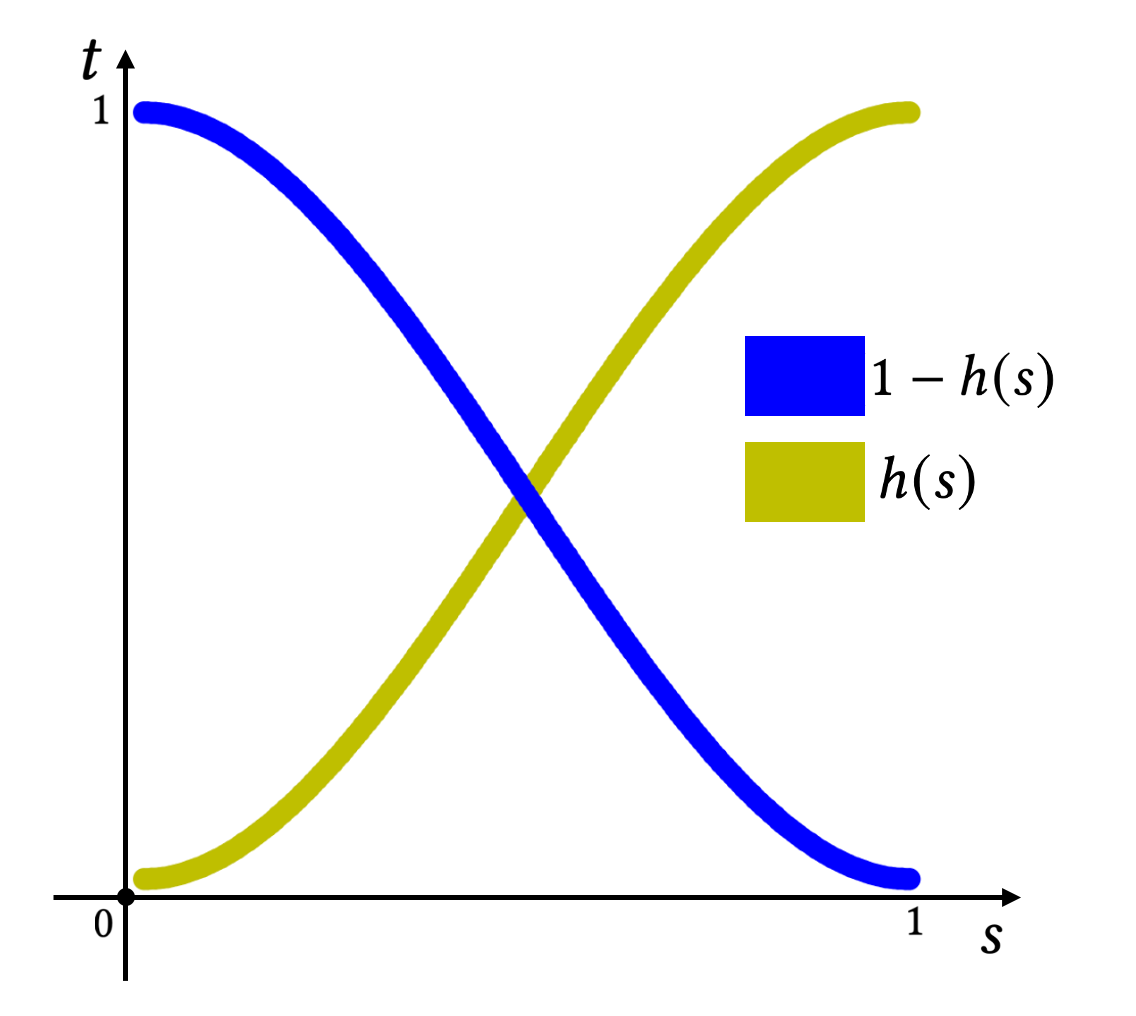}
     \caption{Weight functions for smoothly blending the local and global SDFs.}
    \label{fig:blend_func}
\end{figure}

With $F_L$ being a sufficiently accurate distance value for points near the surface $S$, we define three regions as follows: the region $R_1 = \{ \mathbf{x}|~|F_L(\mathbf{x})| < \delta_1 \}$, $R_2 = \{ \mathbf{x}| \delta_1 \leq |F_L(\mathbf{x})| \leq \delta_2 \}$, and $R_3 = D\setminus (R_1 \cup R_2)$, where $D$ is the entire bounding domain ($[-1,1]^3$) of a given shape.
Then, we design the global feature-preserving distance function $F_G$. Inside the region $R_1$, we set $F_P = F_L$, because the local distance function $F_L$ is accurate and feature-preserving in $R_1$. In our implementation, with all shape bounding boxes as $[-1,1]^3$, we choose $\delta_1=0.001$ in the definition of $R_1$. 
In $R_3$, since $F_L$ becomes inaccurate and unstable because of the lack of supervision far from $S$, we set $F_G=F_O$. Similarly, we choose $\delta_2=0.03$ in the definition of $R_3$.

To make a smooth interpolation between $F_L$ and $F_O$ in $R_2$, we first define the weight functions $w_0(s)$ and $w_1(s)$ so that 
\begin{equation}
F_G(\mathbf{x}) = w_0 F_L(\mathbf{x}) + w_1 F_O(\mathbf{x}), ~ \mathbf{x} \in R_2.
\label{eqn:blend}
\end{equation}
Let us denote $d(\mathbf{x}) =|F_L(\mathbf{x})| \in [\delta_1, \delta_2], \mathbf{x} \in R_2$ and $s(\mathbf{x}) = (d(\mathbf{x}) - \delta_1)/(\delta_2-\delta_1)$ is a linear mapping that maps $d(\mathbf{x}) \in [\delta_1, \delta_2]$ to $s(\mathbf{x})\in [0, 1]$. We define the weight functions by 
\[
w_0(s) = (1 - h(s(\mathbf{x})), w_1(s) = h(s(\mathbf{x}))
\]
where $h(s) = 3s^2 - 2s^3$, for $s\in [0, 1]$; see \ref{fig:blend_func}.

Clearly, $h(0) = h'(0) = h'(1) = 0$ and $h(1)=1$, from which one can verify that $F_G(\mathbf{x}) = F_L(\mathbf{x})$ and $\nabla F_G(\mathbf{x}) = \nabla F_L(\mathbf{x})$ along the inner boundaries of $R_2$ (i.e.\ where $|F_L(\mathbf{x})| = \delta_1$), and that $F_G(\mathbf{x}) = F_O(\mathbf{x})$ and $\nabla F_G(\mathbf{x}) = \nabla F_O(\mathbf{x})$ along the outer boundaries of $R_2$ (i.e.\ where $|F_L(\mathbf{x})| = \delta_2$). That is, $F_G$ thus defined is an $C^1$-continuous extension of $F_L$ in $R_2\cup R_3$. 
To summarize, the distance function $F_G(\mathbf{x})$ is a {\em smooth} blending of $F_L(\mathbf{x})$ and $F_O(\mathbf{x})$ and is globally defined on $D$. In particular, the zero-level set of $F_G$ agrees with that of $F_L(\mathbf{x})$, so it preserves the sharp features of the original target surface.

\begin{figure}
    \centering
    \subcaptionbox[width=0.28\linewidth]{}{
    \includegraphics[width=0.28\linewidth]{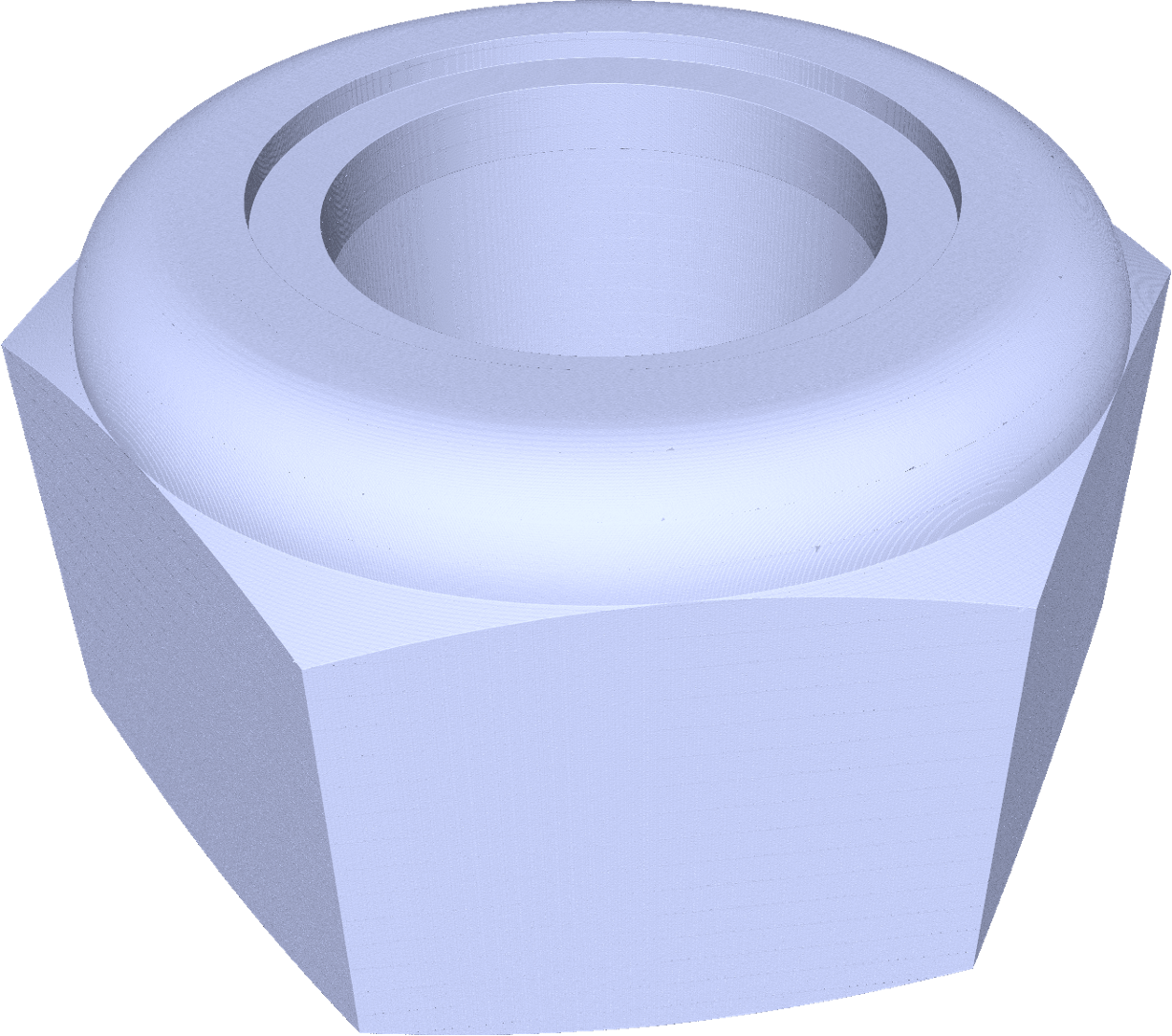}
    }
    \subcaptionbox[width=0.32\linewidth]{}{
    \includegraphics[width=0.32\linewidth]{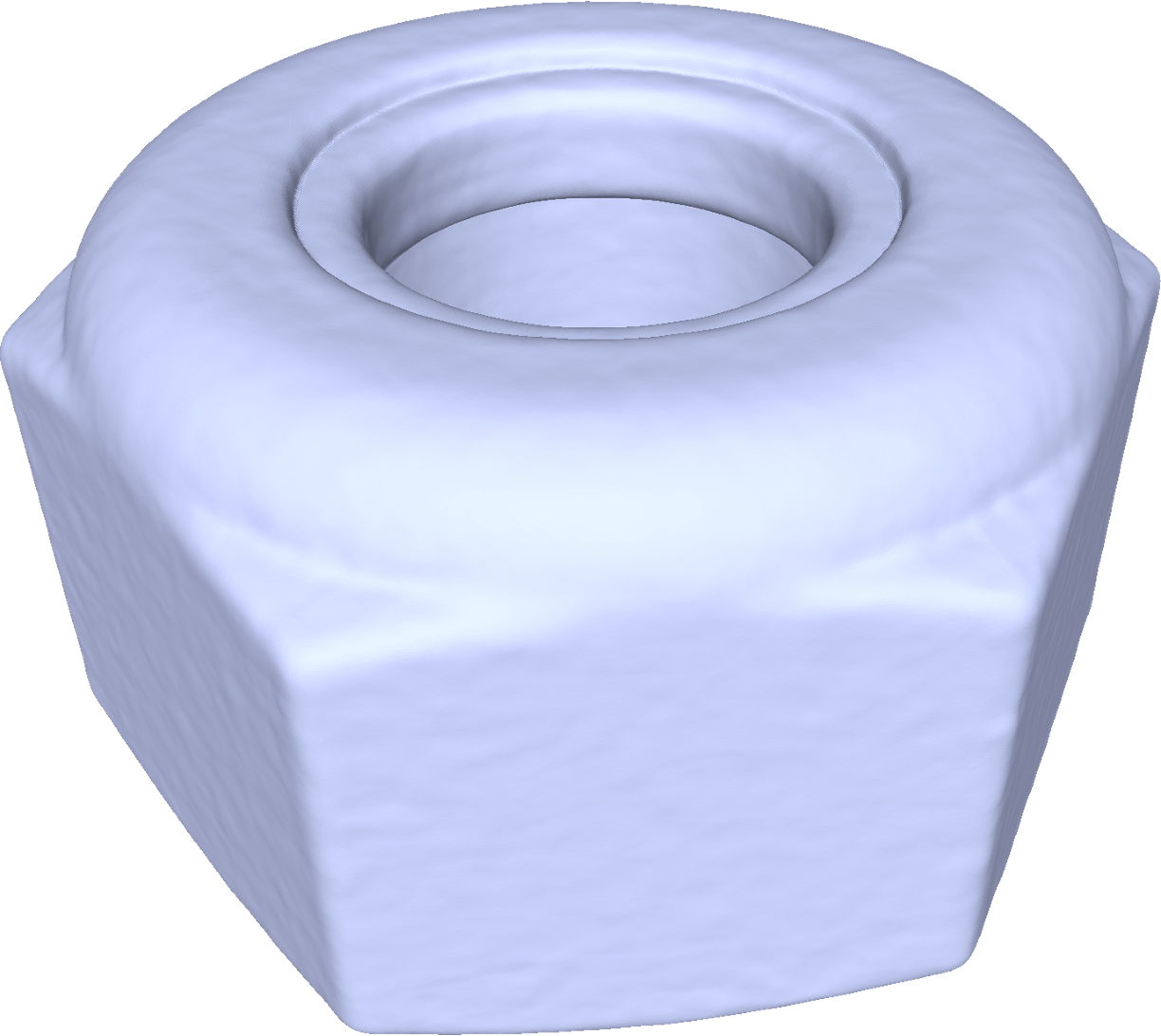}
    }
    \subcaptionbox[width=0.25\linewidth]{}{
    \includegraphics[width=0.25\linewidth]{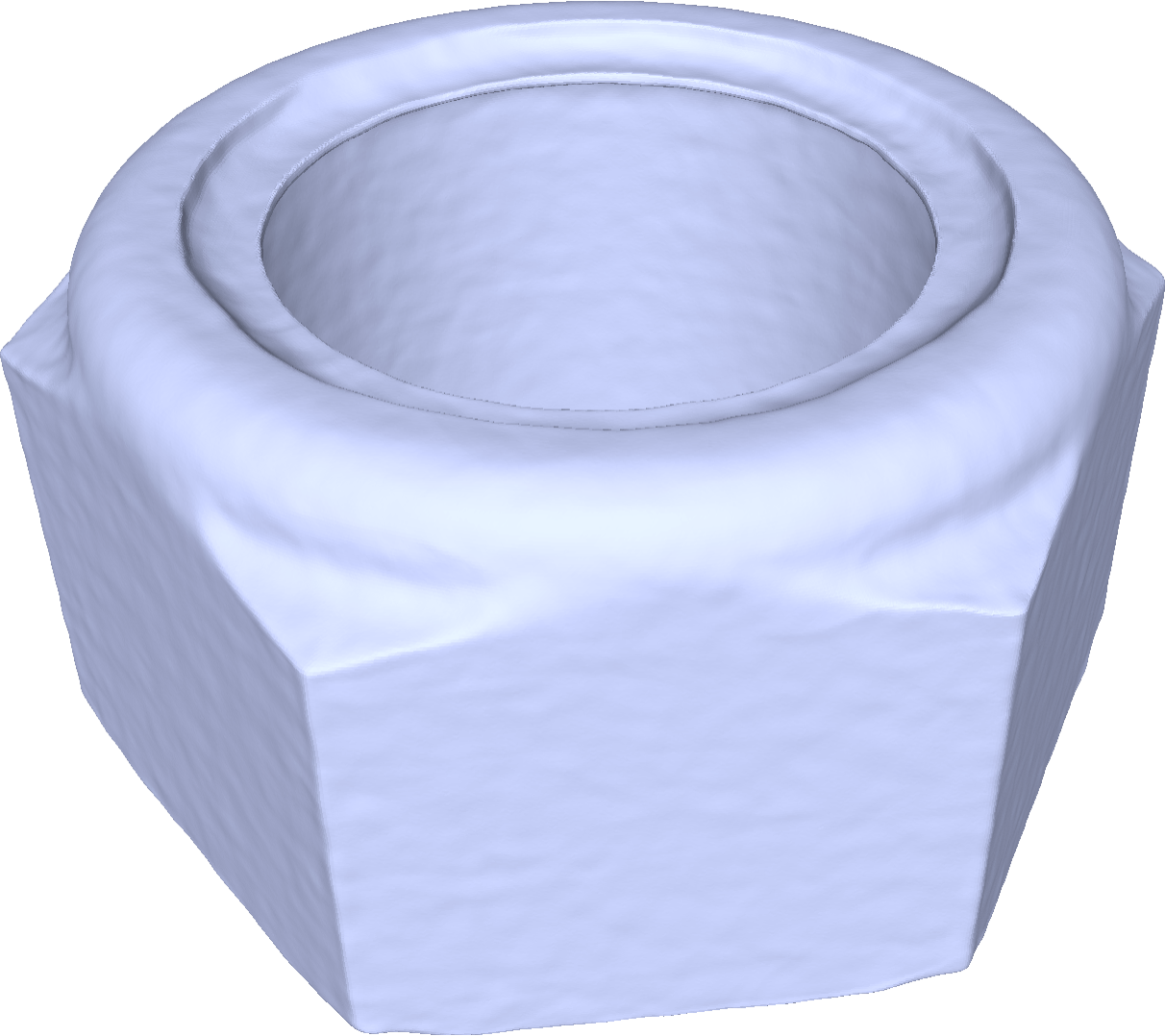}
    }
    \caption{
    The zero-level set of the blended field (a) and its outward (b) and inward (c) offsets are shown.
    }
    \label{fig:offset}
\end{figure}

Finally, we show a 3D result of our blending strategy in \ref{fig:offset}(a) and demonstrate an offset application of our global distance function $F_G$ in \ref{fig:offset}(b, c).

%% file: sections/results.tex
\section{Experimental results}\label{sec:results}

\begin{figure*}
\centering
  \includegraphics[width=1.\linewidth]{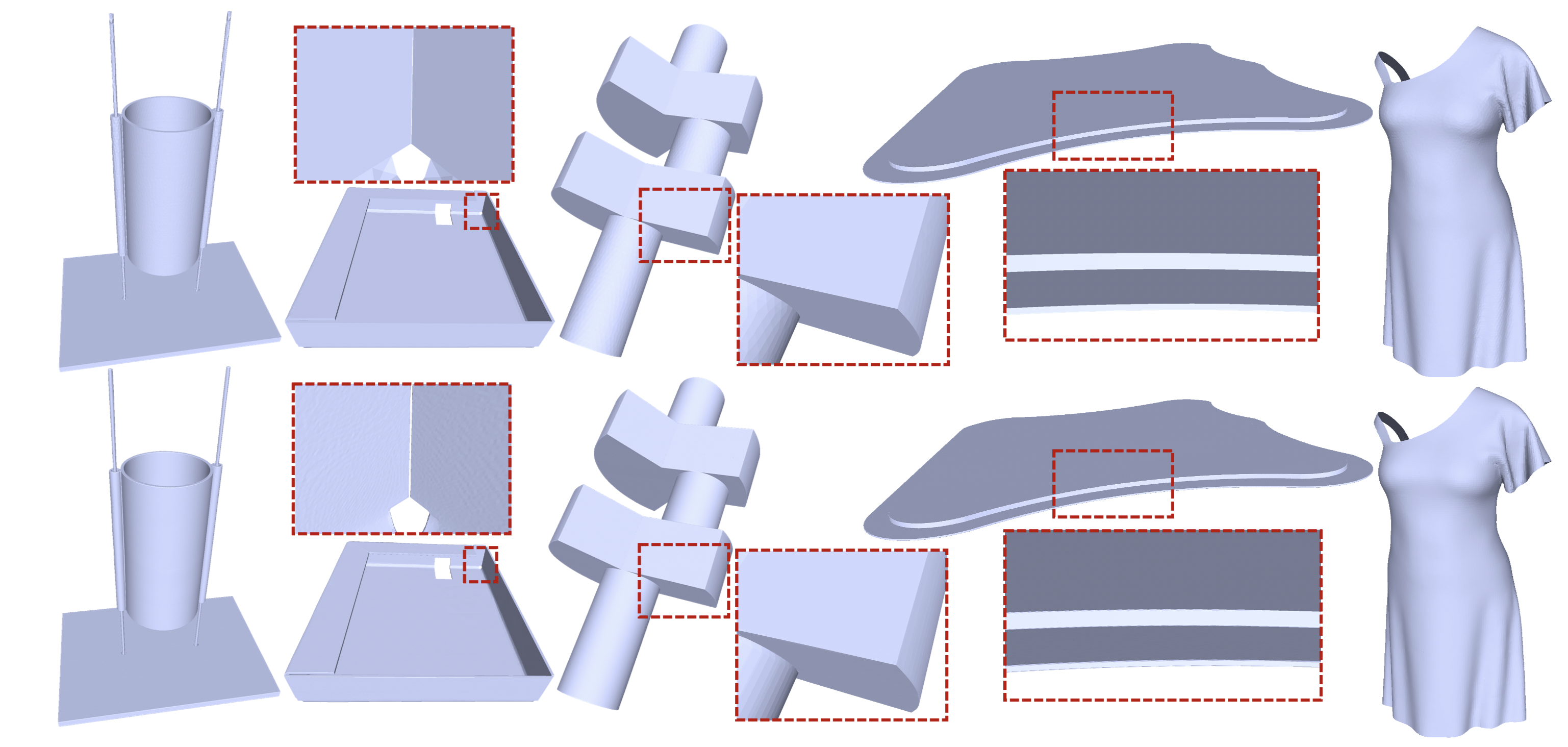}
    \put(-510,180){GT}
    \put(-520,60){Patch-Grid}
    \put(-460,-10){(a) }
    \put(-380,-10){(b)}
    \put(-280,-10){(c)}
    \put(-150,-10){(d)}
    \put(-40,-10){(e)}
  \caption{Our proposed approach, \textit{Patch-Grid}, is capable of representing shapes that possess various types of features and demonstrates exceptional accuracy and robustness. (a, b): thin structures; (c, d): sharp boundary edges and corners; (e): clean boundary curves in the open surfaces.}
  \label{fig:our_show}
\end{figure*}

\begin{figure*}
\centering
    \vspace*{-25px} 
    \hspace*{60px} 
    \includegraphics[width=0.8\linewidth]{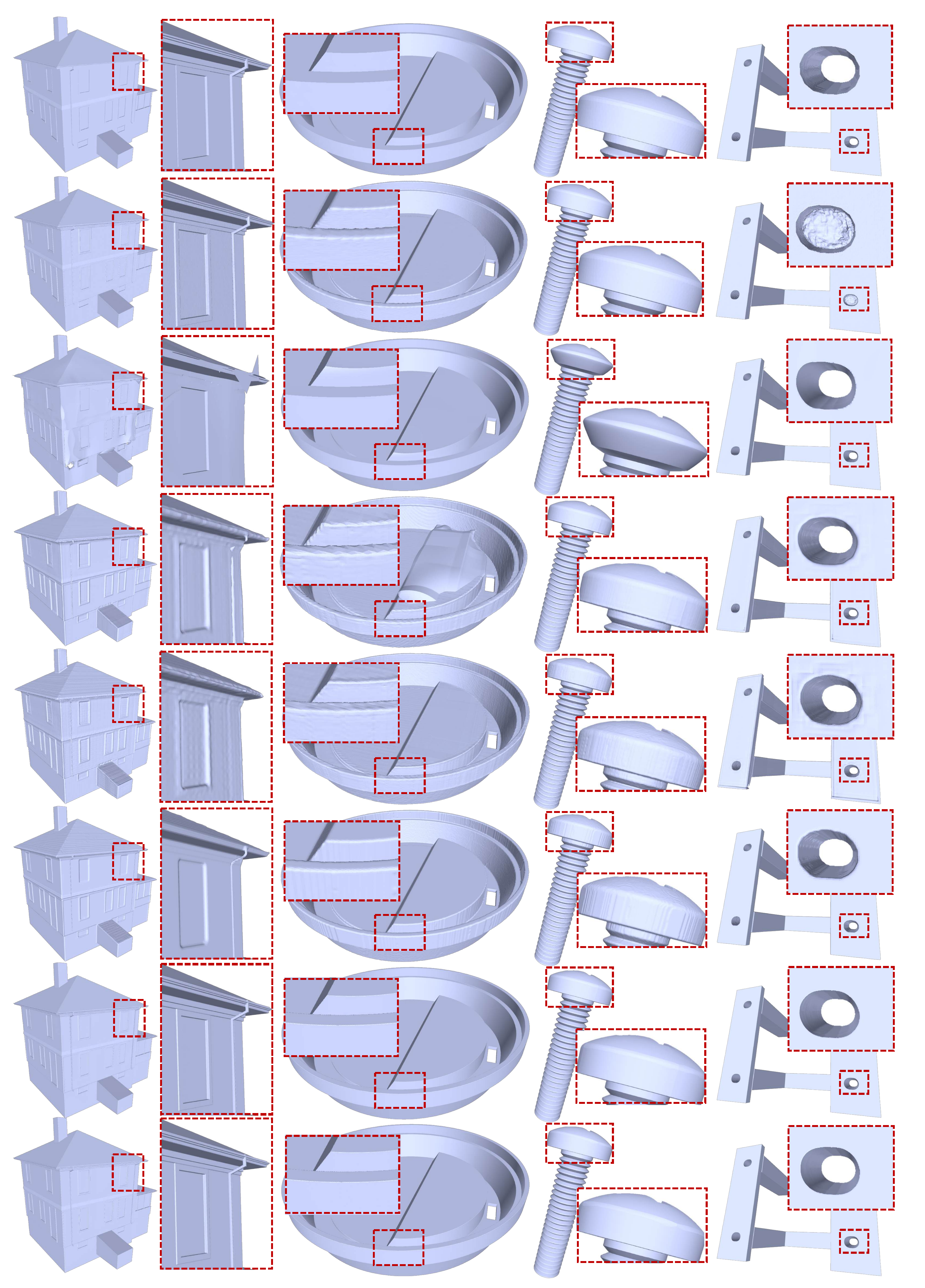}
    \put(-485, 523){(a) GT}
    \put(-485, 454){(b) InstantNGP}
    \put(-485, 385){(c) NH-Rep }
    \put(-485, 316){(d) DualOctreeGNN}
    \put(-485, 247){(e) DualOctreeGNN-FT}
    \put(-485, 178){(f) MINER}
    \put(-485,109){(j) Patch-Grid}
    \put(-485, 40){(k) Patch-Grid-TS}
  \caption{Comparison with other baseline methods. \textit{Patch-Grid}, \textit{Patch-Grid-TS}, and \textit{NH-Rep}~\cite{Guo2022NHRep}, which adopt a patch-based representation, can faithfully capture \textbf{sharp geometric features} presented in the shapes, while \textit{InstantNGP}~\cite{muller2022instant}, \textit{DualOctreeGNN} and its variant \textit{DualOctreeGNN-FT}~\cite{Wang_2022}, and \textit{MINER}~\cite{saragadam2022miner} fail to do so, as shown in the zoom-in views of column 2. 
  Only \textit{Patch-Grid}, \textit{Patch-Grid-TS} and \textit{InstantNGP} can capture the \textbf{thin structure} (the pipe of the house, column 1). 
  Qualitative results, along with the quantitative results in Tab.~\ref{tab:quantitative}, demonstrate the robustness of our method as compared to prior methods,  all of which experience varying degrees of failure.}
  \label{fig:big_compare}
\end{figure*}

\begin{figure*}
    \centering
    \includegraphics[width=0.88\linewidth, trim=500 30 40 40, clip]{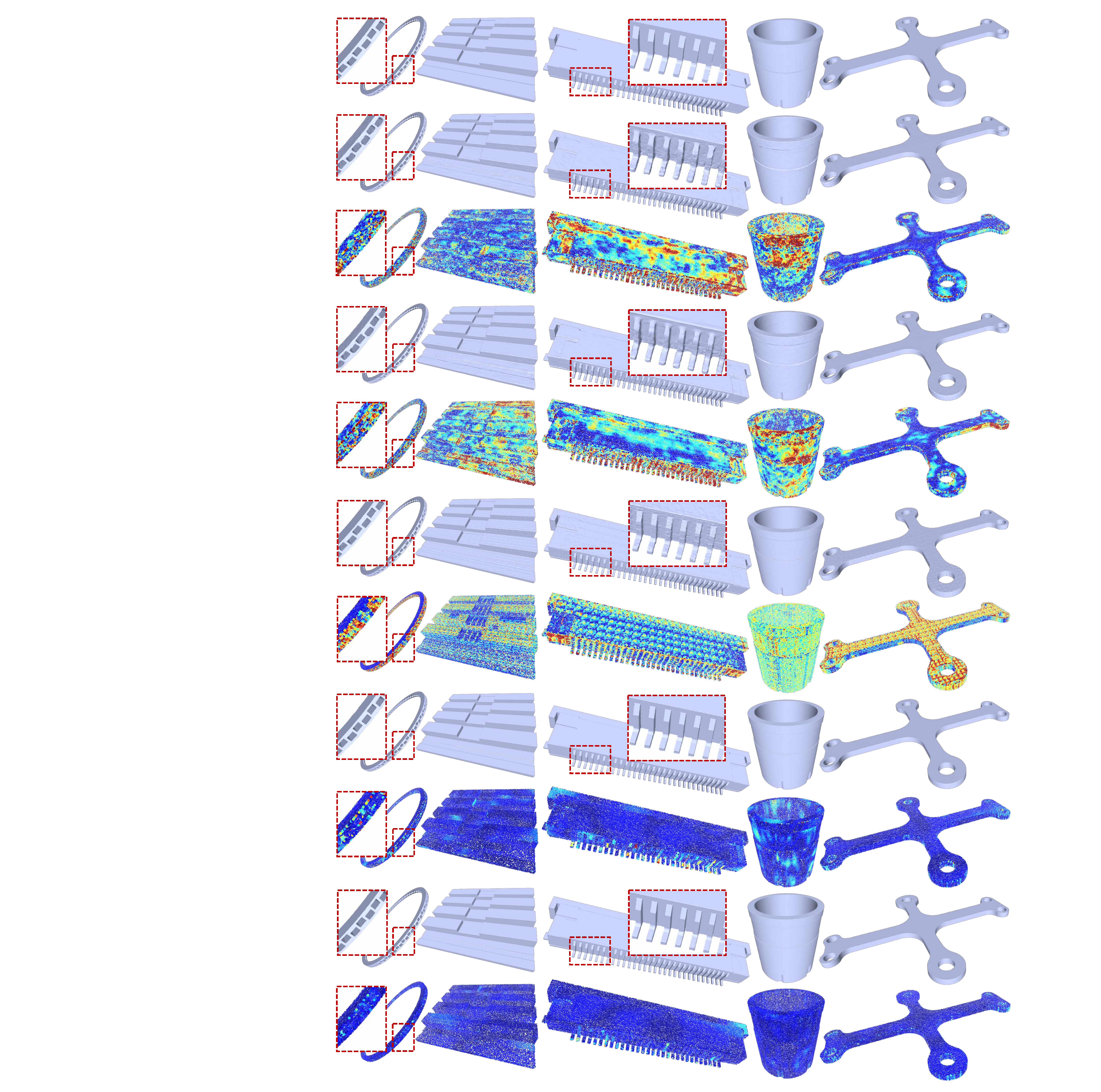}
    \put(-470, 525){(a) GT}
    \put(-470, 470){(b) NGLOD}
    \put(-470, 380){(c) NGLOD-SP}
    \put(-470, 280){(d) ACORN}
    \put(-470, 185){(e) Patch-Grid}
    \put(-470, 85){(f) Patch-Grid-TS}
    \caption{
    Comparison with \textit{NGLOD}~\cite{takikawa2021neural}, \textit{NGLOD-SP}, and \textit{ACORN}~\cite{martel2021acorn}. 
    Extracted meshes from the learned zero-level sets (upper) and their corresponding error maps (lower) are shown.
    A warmer color indicates a higher error, with the error being clipped by 0.001. \textit{Patch-Grid} and \textit{Patch-Grid-TS} outperform the comparing methods significantly regarding the fitting accuracy, especially around the sharp features of the presented shapes.}
    \label{fig:small_compare}
\end{figure*}

\subsection{Implementation details} \label{sec:implementation_details}

\paragraph{Data preprocessing and preparation}
We test our method primarily on the CAD models randomly sampled from the ABC dataset~\cite{Koch2019ABC}. All the shapes are segmented in advance and normalized to $[-0.8,0.8]^3$. For each segmented surface patch of a given shape, we first construct its patch volume $V_p$. The resolution of the patch volume is determined by the average shape diameter~\cite{shapira2008consistent} of this surface patch computed with \textit{LIBIGL}~\cite{libigl}. Specifically, we stipulate that the size of the grid cell in $V_p$ should not exceed 2.5 times the average shape diameter of the patch. 
We observe that the resolution of a patch feature volume ranges from $2^3$ to $2^6$ depending on the local feature size. 
We construct the merge grid $\mathcal{G}$ as described in Sec.~\ref{sec:method} and derive the merge constraints from the merge grid cells accordingly. We empirically set the maximal depth of the merge grid octree to 8.

To prepare the training samples, for each patch, we uniformly sample 1,000 surface points and 1,000 spatial points in each cell of the corresponding feature volume volumes, forming the sampling pool. Subsequently, in each training iteration, we randomly sample 10,000 surface points and 10,000 spatial points from the prepared sampling pool.

\paragraph{Training details}
We set the balance weights for the training loss terms as follows:
$\lambda_{\mathrm{surface}}=200$, $\lambda_{\mathrm{normal}}=50$, $\lambda_{\mathrm{SDF}}=50$, $\lambda_{\mathrm{eikonal}}=5$, $\lambda_{\mathrm{merge}}=400$. 
We implemented the proposed method with PyTorch. We adopt the ADAM~\cite{kingma14adam} optimizer with default hyperparameters. The feature volumes are comprised of 256-dimensional latent codes. The decoder is implemented as a 1-layer MLP whose hidden layer has 128 neurons. The activation function used for each layer is the \textit{SoftPlus} function ($\beta$ = 100 as suggested by Atzmon and Lipman~\shortcite{atzmon2020sald}), with the exception that no activation function is applied at the last layer. 

For shape fitting, there are two training approaches: training from scratch and training with a pretrained fixed decoder. To train from scratch, we train the neural network (both the decoder and the patch feature volumes) for 500 iterations. The initial learning rate is 0.001, which is then reduced by a factor of 0.3 at the 470th and 485th iterations. To train with a pretrained decoder, we only update the patch feature volumes for 300 iterations. The initial learning rate is 0.001, and it is decayed by a factor of 0.3 at the 270th and 285th iterations.
  
In practice, it is often necessary to edit a 3D shape and update its corresponding implicit representation as well. Typically, the editing involves only a few patches. Therefore, to reuse the previously learned feature volumes while updating only the changed ones, we adopt the test-time optimization scheme similar to Park et al.~\shortcite{Park2019DeepSDF}. 
For this shape updating application, our results were obtained with 80 training iterations. Similar to the fitting application, the learning rate is initialized at 0.001 and decayed by a factor of 0.3 at the 65th and 72nd iterations.

\textbf{Pretraining the decoder. }
To speed up the shape fitting process and to improve the overall quality of the shape updating results, one can use a pretrained decoder. 

During the pertaining stage, we collected 1,849 patches from 25 CAD shapes. Each patch was trained under its patch resolution as explained earlier. The batch size was set to 240k to fully utilize the GPU memory.
The decoder was trained with 200k iterations (or 20 hours on our hardware), and the learning rate was decayed at 150k and 175k iterations, by a factor of 0.3. The training objective was the patch fitting loss $\loss_{\mathrm{patch}}$; the merging loss term, $\loss_{\mathrm{merge}}$, was not involved. All other parameters were kept the same as our default training setting. 

All results produced by our \textit{Patch-Grid} and the comparing methods were obtained on a desktop with an NVIDIA RTX4090 graphical card and an Intel i9 13900kf CPU. More implementation details can be found in our codes. 

\subsection{Evaluation metrics}

To evaluate fitting accuracy, we use the following metrics.
1) \textbf{Symmetric Chamfer distance} (CD) quantifies the average reconstruction quality of a given shape. 
2) \textbf{Hausdorff distance} (HD) measures the maximum reconstruction error. 
3) \textbf{F-score based on CD} is a statistical measure that provides an overall assessment of the reconstruction quality. Specifically, the F-score is computed as the percentage of points with a reconstruction error smaller than $0.001$ throughout the paper.  
4) \textbf{Intersection over Union} (IoU) measures the degree of overlap between the reconstructed shape and the ground-truth shape.
5) \textbf{Normal consistency} (NC) assesses the similarity between the normal vectors at the reconstructed surface and those at the ground-truth mesh surface. 

6) \textbf{Sharp feature error} (SFE) evaluates how accurately the learned field reconstructs the sharp features present in the ground-truth mesh. 
7) \textbf{Field error} (FE) measures the overall volumetric quality of the learned field. Points are uniformly sampled within the domain of definition at a density of $2^7$ per unit length, and the error is computed between the SDF values of the sampled points and the corresponding ground-truth values in the learned field.

\subsection{Results and discussions}

\subsubsection{Shape fitting}
Our method can model 3D surface shapes with various geometry features, e.g.\ sharp features, narrow gaps, or open boundaries, at high fidelity as is shown in Fig.~\ref{fig:teaser}(a) and Fig.~\ref{fig:our_show}. We report the quantitative results in Table~\ref{tab:quantitative}, produced by two variants of our approach (\textit{Patch-Grid} and \textit{Patch-Grid-TS}) and the comparing methods. \textit{Patch-Grid} denotes training our approach with a fixed, pretrained decoder as described in the previous subsection, while \textit{Patch-Grid-TS} denotes Patch-Grid \underline{\textbf{T}}rained from \underline{\textbf{S}}cratch.

\textbf{Baselines. }
We compare our results to those produced by the baseline methods.
The baseline methods are: \textit{InstantNGP}~\cite{muller2022instant}, \textit{NH-Rep}~\cite{Guo2022NHRep}, \textit{DualOctreeGNN}~\cite{Wang_2022}, \textit{MINER}~\cite{saragadam2022miner}, \textit{ACORN}~\cite{martel2021acorn}, and \textit{NGLOD}~\cite{takikawa2021neural}. 

Specifically, \textit{DualOctreeGNN} predicts a hierarchical feature volume based on an octree from an input point cloud via a single forward pass. It then uses a learned MLP to map the feature volume to the SDF. 

We also consider its variant, denoted \textit{DualOctreeGNN-FT}, which is fine-tuned for 2000 iterations for each shape.
Note that \textit{NGLOD} has a variant using SoftPlus to replace ReLU as the activation function, we denote this variant \textit{NGLOD-SP}.

\textbf{Datasets. }
First, a set of 100 shapes from the ABC dataset is established for comparison between our method and \textit{InstantNGP}, \textit{NH-Rep}, \textit{DualOctreeGNN}, \textit{DualOctreeGNN-FT}, and \textit{MINER} that can fit a given shape within minutes. 
Since \textit{NGLOD}, \textit{NGLOD-SP}, and \textit{ACORN} require a prolonged training time, we establish a second group of 10 shapes randomly sampled from the ABC dataset for comparison.

Finally, we randomly sampled 8 shapes, each having over 100 patches, to validate the robustness of our method in handling complex shapes.

Our approach, in both its variants (\textit{Patch-Grid} and \textit{Patch-Grid-TS}), can achieve significantly better reconstruction quality than the comparing baselines as shown in Tab.~\ref{tab:quantitative}.
Acknowledging the severe failure cases in \textit{NH-Rep} due to its robustness issues as discussed previously, we have excluded these shapes from the quantitative comparison. Nevertheless, our approach still performs favorably compared to \textit{NH-Rep}. We attribute this performance gain to the use of the adaptive merge grid, which circumvents the difficulty in managing the extended zero-level sets of the learned patches to satisfy the global CSG constraints. 

The training time averages around 5 seconds for 300 training iterations and is at least 7 times faster than most of our baselines, as reported in Tab.~\ref{tab:quantitative}.
For qualitative comparison, we show several results produced by \textit{Patch-Grid}, \textit{Patch-Grid-TS}, and the other methods in Fig.~\ref{fig:big_compare} and Fig.~\ref{fig:small_compare}.

\begin{table*}[htbp]
\centering
\caption{Quantitative evaluation of different neural implicit representations in terms of the symmetric Chamfer distance (CD), Hausdorff distance (HD), F-score (CD < 0.001), Intersection over Union(IoU), normal consistency (NC), sharp feature error (SFE), and field error (FE). $\uparrow$ indicates the higher the better, while $\downarrow$ indicates the lower the better. CD, HD, NC, SFE, and FE are presented in units of $\times10^{-4}$, $\times10^{-3}$,  $\times10^{-2}$, $\times10^{-5}$, and $\times10^{-4}$, respectively. We report the training time in the unit of seconds. We do not report the SFE and FE metrics based on SDF for ACORN and MINER since they predict occupancy fields. For our metrics of FE, we also report in parentheses the FE of the blended field, which is blended according to the method described in Section~\ref{sec:global_local_blend_detail}.}\label{tab:quantitative}
\begin{tabular}{c|c c c c c c c c }
\toprule
 Metrics & CD $\downarrow$ & HD $\downarrow$ & F-score $\uparrow$ & IoU $\uparrow$ & NC $\uparrow$ & SFE $\downarrow$ & FE $\downarrow$  & Time $\downarrow$ \\ \hline
\\[-1em]

NH-Rep & 3.18 & 4.47 & 95.32 & 98.57 & 99.71 & 40.2 & 55.9 & 185  \\
\\[-1em]

InstantNGP & 1.85 & 9.93 & 99.60 & 95.89 & 99.65 & 11.1 & \textbf{7.97} & 101  \\
\\[-1em]

MINER & 5.55 & 6.82 & 80.00 & 91.72 & 99.06 & \# & \# &  37.3\\
\\[-1em]
DualOctreeGNN & 22.5  & 45.8 & 89.91 & 97.42 & 98.93 & 122 & 328 & \textbf{0.03} \\
\\[-1em]

DualOctreeGNN-FT & 6.94 & 13.3 & 76.88 & 97.26 & 99.21 & 164 & 451 & 731  \\
\\[-1em]

Patch-Grid & 1.00 & \textbf{2.47} & \textbf{99.78} & \textbf{99.61} & \textbf{99.83} & 9.89 & 67.27(3.58) & 5.44 \\
\\[-1em]
Patch-Grid-TS & \textbf{0.93} & 2.70 & 99.71 & 99.54 & 99.70 & \textbf{8.87} & 59.4(3.59) & 9.81  \\

\hline
\\[-1em]

NGLOD* & 4.17 & 6.09 & 92.46 & 97.18 & 97.93 & 79.4 & 7.78 & 2296 \\
\\[-1em]
NGLOD-SP* & 4.05 & 8.34 & 94.22 & 97.55 & 97.91 & 80.3 & \textbf{7.65} & 2296 \\
\\[-1em]
ACORN* & 4.23 & 3.08 & 96.44 & 98.11 & 98.31 & \# & \# & 7200 \\
\\[-1em]

Patch-Grid* & 1.03 & \textbf{2.82} & \textbf{99.82} & 98.60 & \textbf{99.15} & 12.2 &  42.1(5.69) & \textbf{6.56}  \\
\\[-1em]

Patch-Grid-TS* & \textbf{0.80} & 3.13 & 99.81 & \textbf{98.79} & 99.05 & \textbf{8.19} &  31.6(5.67) & 10.9 \\
\bottomrule
\end{tabular}
\end{table*}

\textit{InstantNGP} largely reduces the training time by utilizing a small neural network augmented by a multiresolution hash table of trainable feature vectors and achieves high-quality geometric reconstruction within two minutes. However, it often fails to model elongated and slender tubes as shown in the fourth column of Fig. \ref{fig:big_compare}(b). This failure can be alleviated or avoided with extra training time. Since such geometric features (i.e., elongated and slender tubes) are frequently observed in the ABC dataset, the averaged reconstruction quality of \textit{InstantNGP} is inferior to ours, as shown in Tab.~\ref{tab:quantitative}. 
While \textit{InstantNGP} can capture the sharp geometric features well, our method slightly outperforms it.

Compared to \textit{NH-Rep}, which also uses a patch-wise representation, \textit{Patch-Grid} and \textit{Patch-Grid-TS} can robustly model geometric features presented in shapes as shown in the first and third columns of Fig.~\ref{fig:big_compare}. \textit{NH-Rep} adopts a global approach to assembling the learned implicit functions into the target shape, rendering it prone to failure due to the undesirable interference between the extraneous zero-level sets of the individual learned implicit patches. In contrast, \textit{Patch-Grid} consistently produces robust, high-quality results even in these challenging cases, validating the superiority of the proposed local approach.

Although \textit{DualOctreeGNN} can predict a reasonable SDF with a single forward pass instantly, the reconstruction often exhibits wavy artifacts (see the second shape of Fig.~\ref{fig:big_compare}(d, e)), accounting for the poor performance in Tab.~\ref{tab:quantitative}.
While these wavy artifacts can be alleviated by fine-tuning as shown by the same shape in Fig.~\ref{fig:big_compare}(e), both \textit{DualOctreeGNN} and its fine-tuned variant fail to fit the thin structures (the pipes of the house) as shown in the zoom-in view of the first shape in Fig.~\ref{fig:big_compare}(d, e).

\textit{MINER} employs a Laplacian pyramid to apply sparse multiscale decomposition of the 3D shape for modeling. Although it can fit the shape within one minute, it inherits the drawbacks of common implicit neural representations, being unable to reconstruct thin structures and sharp features, as shown in the zoom-in views of the first and second shapes of Fig.~\ref{fig:big_compare}(f). Meanwhile, grid-like patterns are observable in the zoom-in view of the third and fourth shapes of Fig.~\ref{fig:big_compare}(f).

Next, we compare our results with \textit{NGLOD}, \textit{NGLOD-SP}, and \textit{ACORN}. \textit{NGLOD} yields less satisfactory results when modeling thin structures, often resulting in pronounced artifacts which can be seen in the first and second columns of Fig. \ref{fig:small_compare}(b). \textit{ACORN} is another hierarchical network architecture that uses an octree-like, multiscale block-coordinate decomposition to adaptively represent a given shape. We trained ACORN for 2 hours for each given shape to obtain satisfactory fitting results. As shown in Fig.~\ref{fig:small_compare}(f), ACORN's results exhibit noticeable grid-like artifacts, which can be seen from the accompanied error map.

Similar to NH-Rep, our method merges multiple SDFs of surface patches into a merged implicit field. Hence, directly querying the distance value at a spatial point may return an invalid distance. Nevertheless, we show that this limitation can be fixed by the local-global blended field described in Sec.~\ref{sec:global_local_blending}, which exhibits high accuracy as shown in the parenthesis under the FE metric in Tab.~\ref{tab:quantitative}.

It is worth noting that Fig.~\ref{fig:big_compare} and Fig.~\ref{fig:small_compare} show not only complex shapes with more than 100 patches for comparison but also relatively ``simple'' shapes with fewer patches which turn out to be challenging because of sharp features and thin structures (see the fourth column of Fig.~\ref{fig:big_compare} and the fourth and fifth columns of Fig.~\ref{fig:small_compare}). 

Our method demonstrates significantly better performance than methods equipped with hierarchical feature grids even on the shapes with fewer patches. 

Meanwhile, although NH-Rep employs a similar patch-based representation to ours and utilizes CSG operations to decompose these challenging geometric features to reduce the learning burden for the network, it also introduces mutual extension constraints between every pair of patches (Sec.~\ref{sec:merge_grid}). If there are $N$ patches, these constraints are in the order of $\mathcal{O}(N^2)$. For shapes with only 10 patches, this might result in hundreds of patch pair constraints. In contrast, our algorithm introduces a merge grid, allowing the network to learn patch extensions with only the simplest and most necessary constraints, resulting in patch pair constraints in the order of $\mathcal{O}(N)$. Therefore, even compared to NH-Rep, our method provides advantages in fitting ``simple'' shapes.

We also report the finest spatial resolution each method accesses and the number of trainable parameters of each method in Tab.~\ref{tab:resolution}, where we chose to balance the two factors for fair comparison. Note that among the hierarchical methods our method requires the least spatial resolution (same as NGLOD) and incurs only a moderate size of training parameters. 
\begin{table}[]
    \centering
    \begin{tabular}{c|c r r}
        \toprule
            & Finest Res. & \#Param. \\
        \hline
        NH-Rep* & n.a. & 0.1 M  \\
        \hline
        InstantNGP & $2^{11}$ & 12.2 M  \\ 
        MINER & $2^{10}$ & 3.2 M  \\
        DualOctreeGNN & $2^8$ & 1.9 M  \\
        NGLOD (LOD=5) & $\mathbf{2^6}$ & \textbf{0.3 M}  \\
        ACORN & $2^{10}$ & 17.0 M  \\
        Ours & $\mathbf{2^6}$ & 0.8 M  \\
        \bottomrule
    \end{tabular}
    \caption{
The finest resolution (Finest Res.) of each method can access and the number of training parameters (\#Param.) of each method. *: NH-Rep is not a hierarchical approach.
    }
    \label{tab:resolution}
\end{table}

In summary, we compare our method, Patch-Grid, to a line of methods, such as InstantNGP, that can fit a given shape within five minutes on a pool of 100 shapes. Our method is also compared to NGLOD, its variant, and ACORN on a pool of 10 shapes due to the prolonged training time required by the baselines. Both comparisons show that Patch-Grid achieves state-of-the-art reconstruction quality at a remarkably high efficiency, i.e.\ around 5 seconds.

\begin{figure}
    \centering
    \includegraphics[width=1\linewidth]{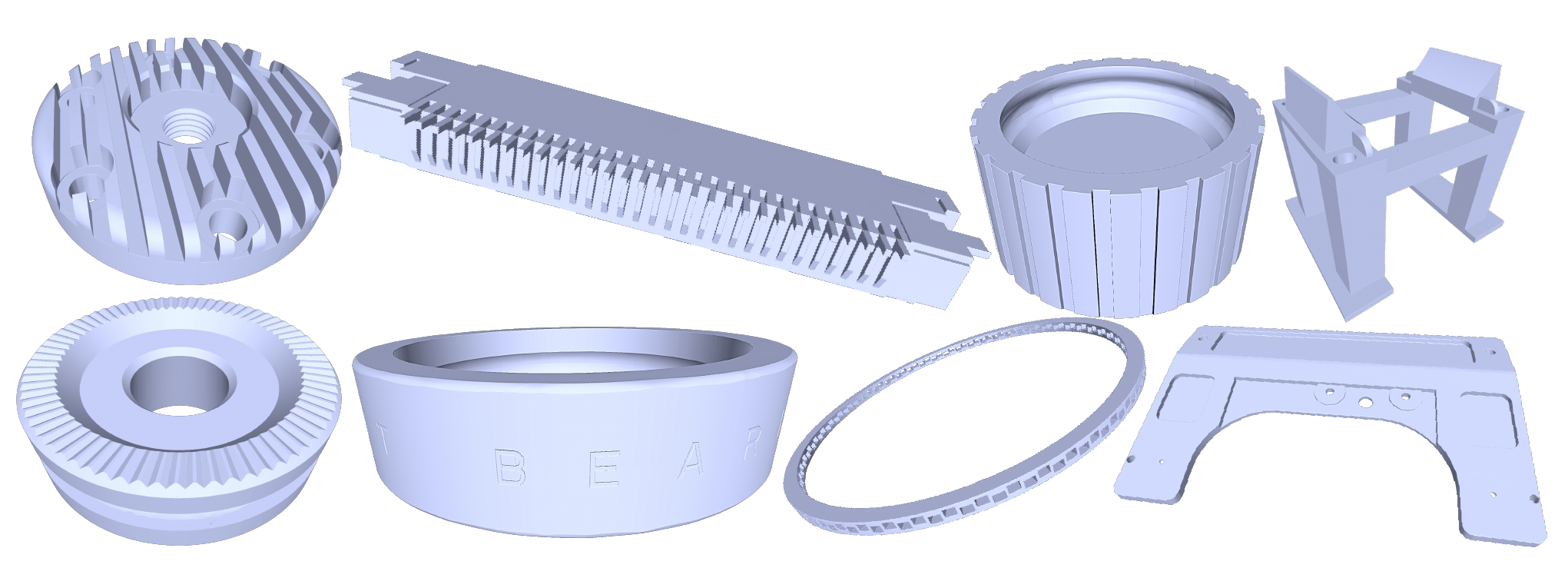}
    \caption{Examples of complex shapes represented by \textit{Patch-Grid}. The number of patches contained in each shape ranges from 105 to 670. These shapes were trained for 300 iterations, amounting to an average training time of 7.34 seconds.}
    \label{fig:large_patches}
\end{figure}

\begin{figure}
    \centering
    \includegraphics[width=1\linewidth]{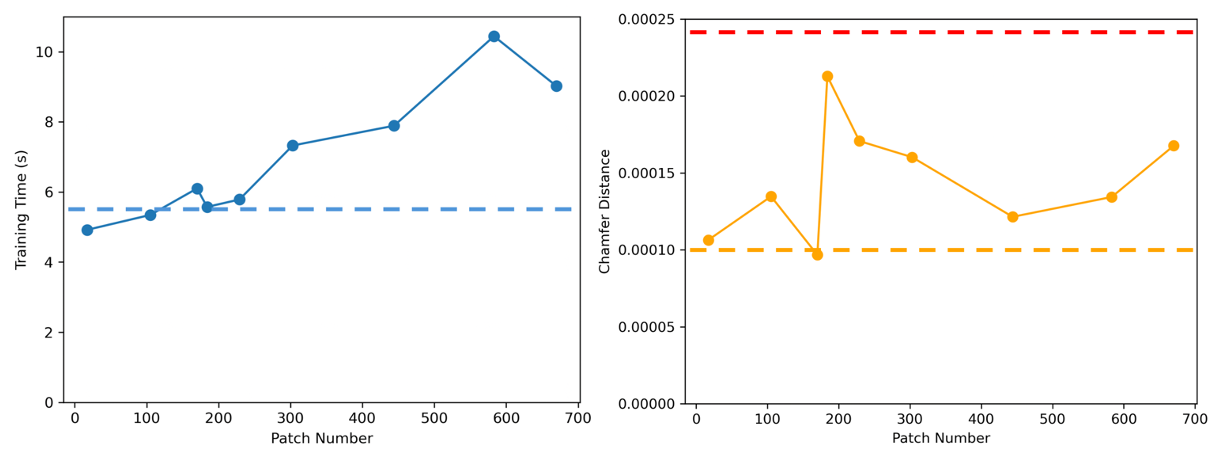}
    \put(-210,-8){(a) Training time}
    \put(-88,-8){(b) Fitting Error}
    \caption{Performance on complex shapes with over 100 patches. (a) shows the change in training time as the number of patches increases. (b) shows the change in fitting accuracy (in terms of Chamfer distance) as the number of patches increases. We have plotted the average time (blue dashed line) and average error (orange dashed line) tested on the pool of 100 shapes from the ABC dataset. The average error of \textit{InstantNGP} (obtained after training 7 seconds) on these complex shapes is plotted in the red dashed line, showing the robustness of the proposed method in modeling complex shapes. }
    \label{fig:large_patch_curve}
\end{figure}

\begin{figure}
    \centering
    \includegraphics[width=0.9\linewidth]{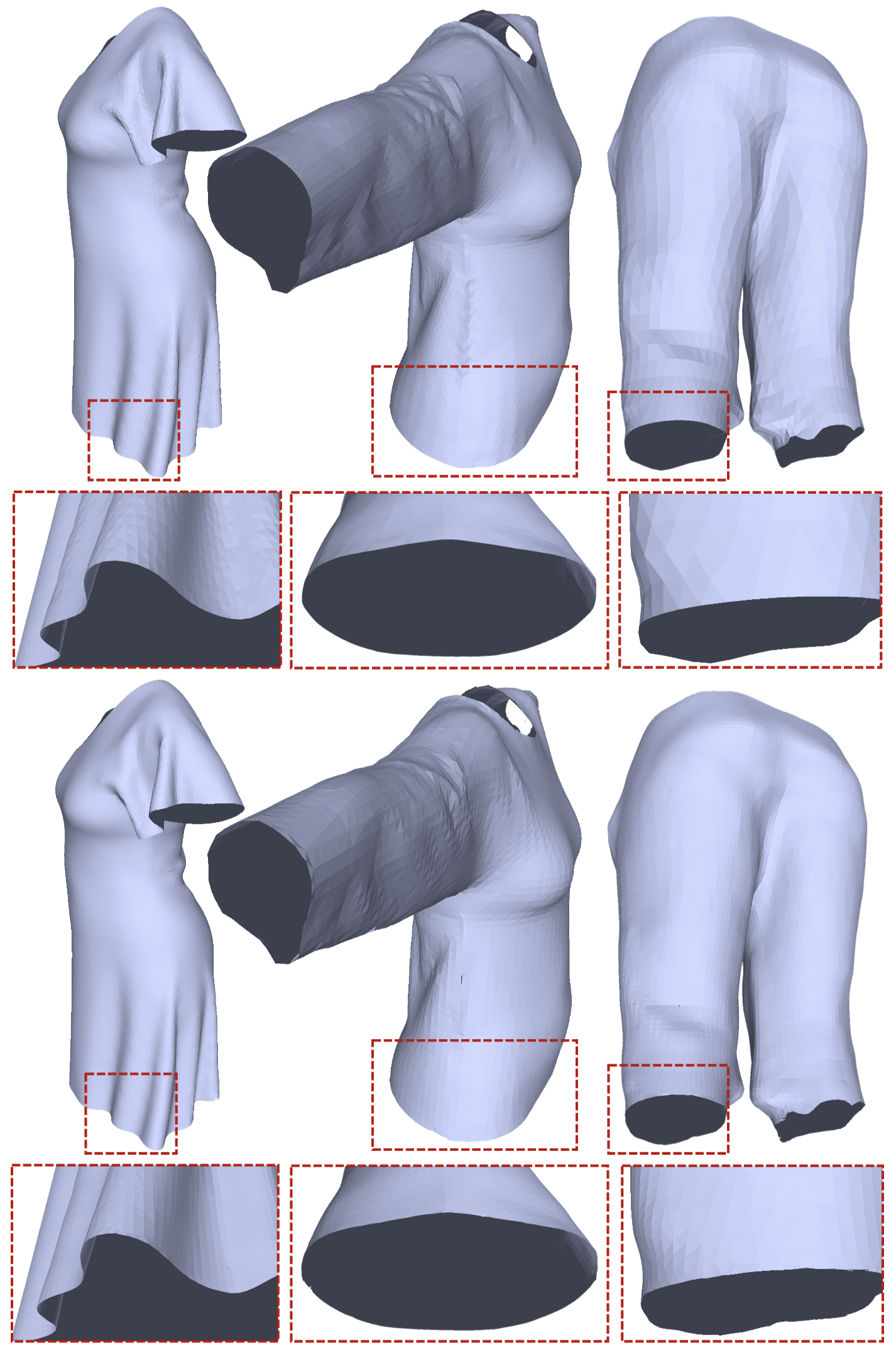}
    \put(-230,280){GT}
    \put(-243,100){Patch-Grid}
    \caption{Examples of open surfaces with the zoom-in views at the open surface boundaries. \textit{Patch-Grid}, modeling the open surface boundaries via a trimming process, cuts the boundaries in a precise and clean manner.}
    \label{fig:open_surfaces}
\end{figure}

\textbf{Robustness to highly-complex CAD models. }
To demonstrate the proposed method's ability to model complex shapes (over 100 patches), we present the fitting results of the \textit{Patch-Grid} method on 8 shapes as shown in Fig.~\ref{fig:large_patches}. Additionally, we present the accuracy and time performance curves of \textit{Patch-Grid} as the number of patches used to represent the shapes increases, shown in Fig.~\ref{fig:large_patch_curve}. From Fig.~\ref{fig:large_patch_curve} (a), we can see that as the number of patches increases, the training time increases but remains under 11 seconds, significantly lower than most of our baselines. In Fig.~\ref{fig:large_patch_curve} (b), we have annotated the average quantitative performance of our method on regular shapes using a yellow dashed line, and the average quantitative performance of \textit{InstantNGP}, the strongest baseline, on these complex shapes using a red dashed line. 

We observed that the fitting error for complex CAD models with more patches increased to around 1.5 times the fitting error on the large pool of 100 shapes randomly sampled from the ABC dataset. On the other hand, on this pool of more complex CAD models, the fitting error of our method is 40\% less than \textit{InstantNGP}. Results of our method are obtained with a training time averaged around 7 seconds. This experiment demonstrates that our method outperforms \textit{InstantNGP} on complex shapes as well.

\textbf{Garment shapes. }
In addition to the CAD models, we also show in Fig.~\ref{fig:open_surfaces} the reconstruction results of three garment models (i.e., a dress from VTO~\cite{santesteban2019virtualtryon} and a shirt and a pair of pants from MGN ~\cite{bhatnagar2019mgn}) with open surface boundaries to demonstrate the representational capability of the proposed method for modeling open surface boundaries. As shown, {\textit{Patch-Grid} accurately reproduces the open surface boundaries.

\begin{table}[t]
\centering
\small
\caption{Quantitative evaluation of the edited shapes. The accuracy of the edited shape is consistent with the accuracy of shape fitting. CD, HD, NC, and SPE are presented in units of $\times10^{-4}$, $\times10^{-3}$, $\times10^{-2}$, and $\times10^{-5}$, respectively. The training time is in the unit of seconds.}\label{tab:update}
\scalebox{0.9}{
    \begin{tabular}{c | c c c c c c c}
    \toprule
     & CD $\downarrow$ & HD $\downarrow$ & F-score $\uparrow$ & IoU $\uparrow$ & NC $\uparrow$ & SPE $\downarrow$ & Time $\downarrow$ \\ \hline
    \\[-1em]
    Patch-Grid & 1.59 & 5.35 & 99.23 & 99.17 & 99.86 & 19.8 & 1.52\\
    \bottomrule
    \end{tabular}}
    \vspace{-4mm}
\end{table}

\subsubsection{Shape update}
In addition to the shape fitting task, we have also provided the shape updating function. Specifically, when the shape is locally edited, we enable the proposed method to perform local updates to the learned neural representation. The shape update can be achieved in 1.52 seconds, halving the training time compared to \textit{Patch-Grid-TS}.

In addition to the updated shapes shown in Fig.~\ref{fig:teaser}(b), we provide more shape updating results in Fig.~\ref{fig:update}. As explained earlier, the process only involves updates of the feature volumes of the edited patches. 

We applied a variety of editing operations to the shape shown in Fig.~\ref{fig:update}, such as adding a trimming surface to form a chamfer to the central hole as in Fig.~\ref{fig:update}(e).
In each edited shape, we highlight the modified patches in orange. Additionally, the connected patches that are not directly edited but are affected by the update of the merge constraints are highlighted in green. Both the edited and involved patches are updated while the rest of the patches are kept fixed.

The fitting accuracy regarding these edited shapes (all results in Fig.~\ref{fig:teaser}(b) and Fig.~\ref{fig:update}) is reported in Tab.~\ref{tab:update}, which is similar to the performance obtained in the shape fitting task. Nevertheless, it takes an average of 1.52 seconds of training time to reach the fitting accuracy, demonstrating the feasibility of rapidly updating a learned neural implicit representation as its corresponding geometry is locally changed.

\begin{figure*}
\centering
  \includegraphics[width=0.9\linewidth]
  {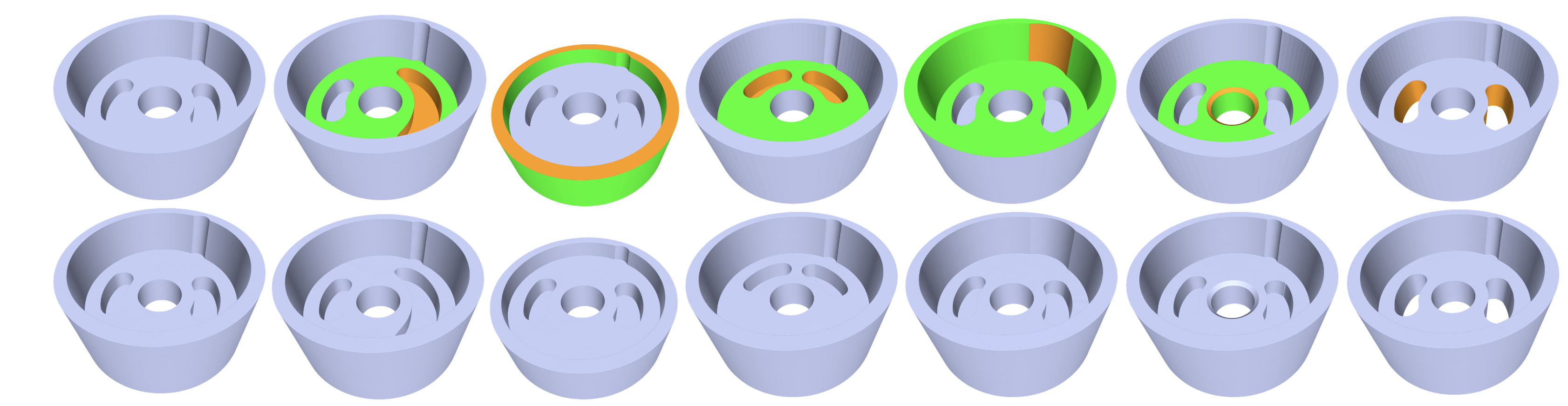}
    \put(-480,88){GT}
    \put(-490,28){Patch-Grid}
    \put(-419,-8){(a)}
    \put(-355,-8){(b)}
    \put(-294,-8){(c)}
    \put(-233,-8){(d)}
    \put(-166,-8){(e)}
    \put(-104,-8){(f)}
    \put(-41,-8){(g)}
  \caption{A collection of edited shapes (top) and the results (bottom) fitted by our shape updating strategy is presented. A variety of editing operations are applied to the original shape in (a).  (b): free-form deformation of the circular slot; (c): parametric editing of the height of the top; (d) rigid transformation of the pair of circular slots; (e): parametric editing of the radius of the cylindrical patch; (f): adding a new patch to create a chamfer to the central through-hole; (g): Extending several patches and removing two patches to obtain a pair of penetrated slots. 
  In each edited shape, the modified patches are highlighted in orange. Additionally, the connected patches that are not directly edited but are affected by the changes due to merge constraints are also highlighted in green. Patch feature volumes of both the directly edited and affected patches are updated while the rest of the patch feature volumes remain fixed.}
  \label{fig:update}
\end{figure*}

\subsection{Ablation study}

We conducted a series of ablation studies to validate the following aspects: 
1) The use of adaptive patch volumes, denoted as \textit{w/o Ada-PV}; 
2) The effect of each loss term;  
3) The weight assigned to each loss term; 
4) The effect of the two training strategies adopted in this paper;
5) The use of a 1-layer MLP; 
6) The use of the \textit{SoftPlus} activation function;

\begin{table}[t]
\centering
\caption{Ablation study on the adaptive patch feature volume. CD, HD, NC, and SPE are presented in units of $\times10^{-4}$, $\times10^{-3}$, $\times10^{-2}$, and $\times10^{-5}$ respectively.}\label{tab:ablation_adaptive}
\scalebox{0.95}{
\begin{tabular}{c|c c c c c c}
\toprule
 &  CD $\downarrow$ & HD $\downarrow$ & F-score $\uparrow$ & IoU $\uparrow$ & NC $\uparrow$ &  SFE $\downarrow$ \\ \hline
 w/o Ada-PV & \textbf{0.815} & 2.95 & 99.66 & 99.37 & 99.69 & 9.90 \\ 
 w/  Ada-PV & 0.934 & \textbf{2.70} &  \textbf{99.71}  & \textbf{99.54} & \textbf{99.70} & \textbf{8.87} \\ 
\bottomrule
\end{tabular}}
\end{table}

\begin{figure}[t]
    \centering
    \includegraphics[width=1.0\linewidth ]{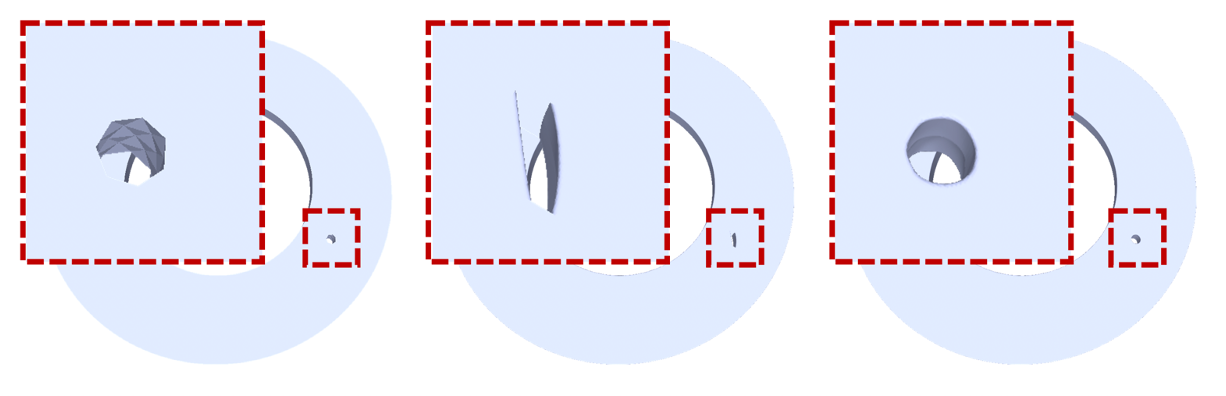}
    \put(-210,-3){(a) GT}
    \put(-150,-3){(b) w/o Ada-PV}
    \put(-70,-3){(c) Patch-Grid }
    \put(-58, -15){(w/ Ada-PV)}
    \caption{Adaptive patch volumes matter. (a) Without adapting the patch volume based on the shape diameter, the extracted mesh from the composed implicit field may fail to model thin tubes with smaller shape diameters due to a coarse resolution; (c) Using adaptive patch volumes can effectively resolve the issue, showing its effectiveness.}
    \label{fig:adaptive_grids}
\end{figure}

\begin{figure}
    \centering
    \includegraphics[width=1.0\linewidth ]{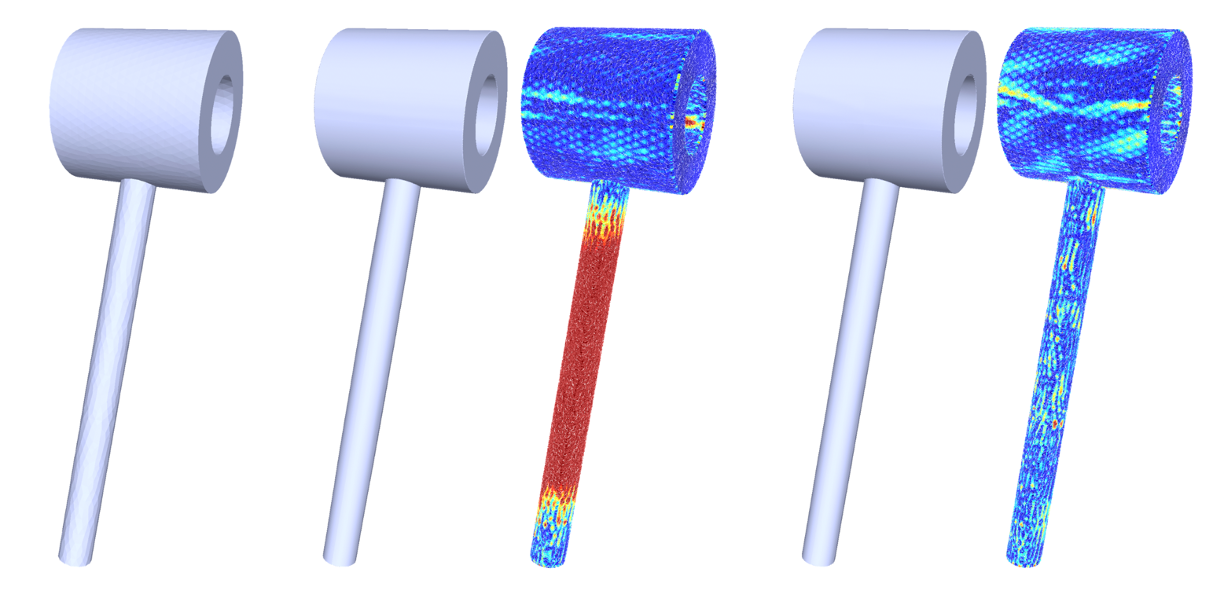}
    \put(-235,-6){(a) GT}
    \put(-170,-6){(b) w/o $\loss_{\mathrm{Surf}}$}
    \put(-75,-6){(c) w/ all }
    \caption{$\loss_{\mathrm{Surf}}$ is essential for fitting the zero-level set. In (b) and (c), the extracted meshes along their error maps are shown. Warmer colors indicate higher errors, with the error clipped at 0.001. Due to the presence of $\loss_{\mathrm{Merge}}$ and $\loss_{\mathrm{SDF}}$, the absence of $\loss_{\mathrm{Surf}}$ does not lead to severe surface artifacts, but the surface thus fit will have a large offset from the ground-truth surface, yielding large fitting error as shown in (b).}
    \label{fig:ablation_no_surf}
\end{figure}

\begin{figure}
    \centering
    \includegraphics[width=1.0\linewidth ]{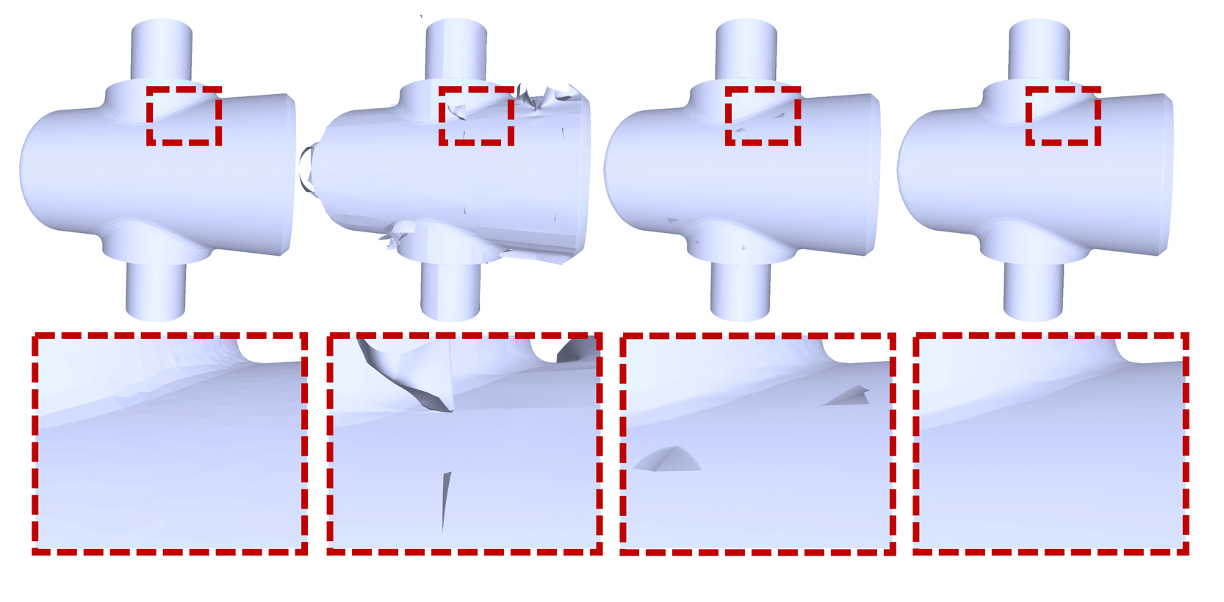}
    \put(-222,-6){(a) GT}
    \put(-118,-6){(c) w/o $\loss_{\mathrm{Merge}}$}    
    \put(-180,-6){(b) w/o $\loss_{\mathrm{Normal}}$}
    \put(-50,-6){(d) w/ all }

    \caption{Both $\loss_{\mathrm{Merge}}$ and $\loss_{\mathrm{Normal}}$ are crucial for shape fitting. Removing $\loss_{\mathrm{Merge}}$ leads to artifacts due to the lack of proper CSG operations between patches. Without $\loss_{\mathrm{Normal}}$, our method fails to fit the surface within a fixed timeframe, resulting in noticeable artifacts.}
    \label{fig:ablation_normal_merge}
\end{figure}

\begin{figure}
    \centering
    \includegraphics[width=1.0\linewidth ]{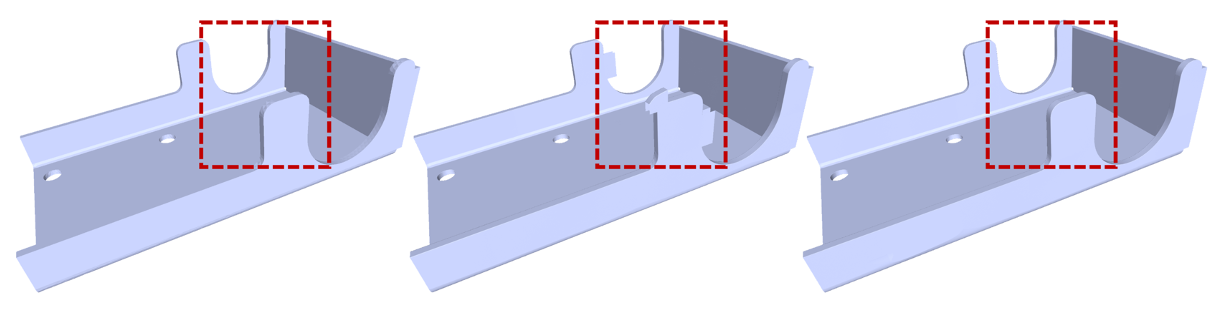}
    \put(-220,-6){(a) GT}  
    \put(-150,-6){(b) w/o $\loss_{\mathrm{SDF}}$}
    \put(-80,-6){(c) w/ all }

    \caption{$\loss_{\mathrm{SDF}}$ is beneficial to surface fitting. Without $\loss_{\mathrm{SDF}}$, artifacts can be observed due to the failure in learning a few individual patches.}
    \label{fig:ablation_jitter}
\end{figure}

\begin{figure}
    \centering
    \includegraphics[width=1.0\linewidth ]{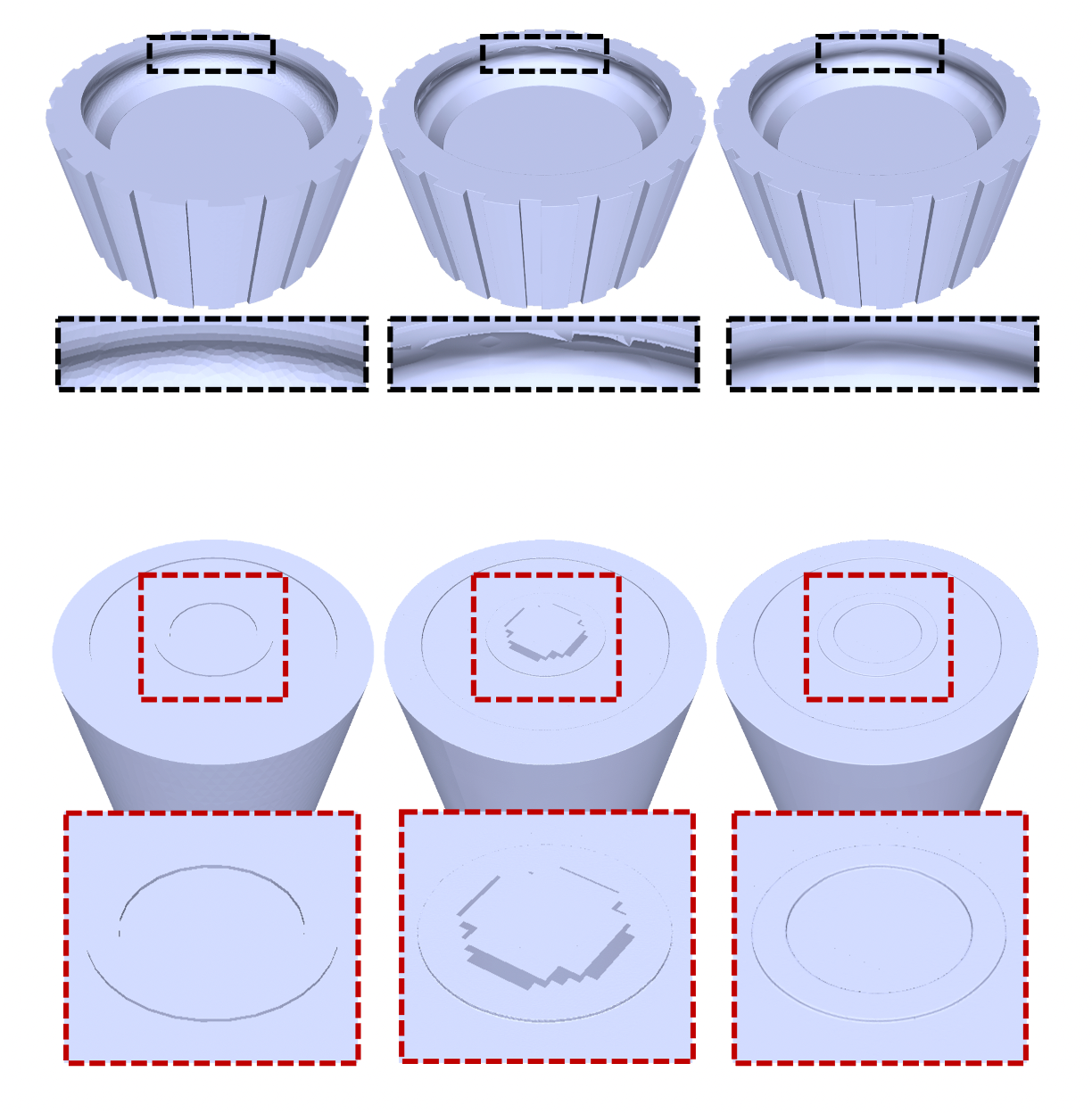}
    \put(-210,150){(a) GT}  
    \put(-210,0){(d) GT}  
    \put(-150,150){(b) $\lambda_{\mathrm{merge}}$=100}
    \put(-150,0){(e) $\lambda_{\mathrm{merge}}$=1200}
    \put(-75,0){(f) Patch-Grid}
    \put(-65,-12){($\lambda_{\mathrm{merge}}$=400)}
    \put(-75,150){(c) Patch-Grid}
    \put(-65,138){($\lambda_{\mathrm{merge}}$=400)}
    \caption{Influence of $\lambda_{\mathrm{Merge}}$ on shape fitting results. Inappropriate weights on $\loss_{\mathrm{Merge}}$ may lead to artifacts due to failure to handle the merging of multiple patches (b) or inaccurate fitting of certain patches (e).}

    \label{fig:ablation_merge_weight}
\end{figure}

\begin{figure}
    \centering
    \includegraphics[width=1.0\linewidth ]{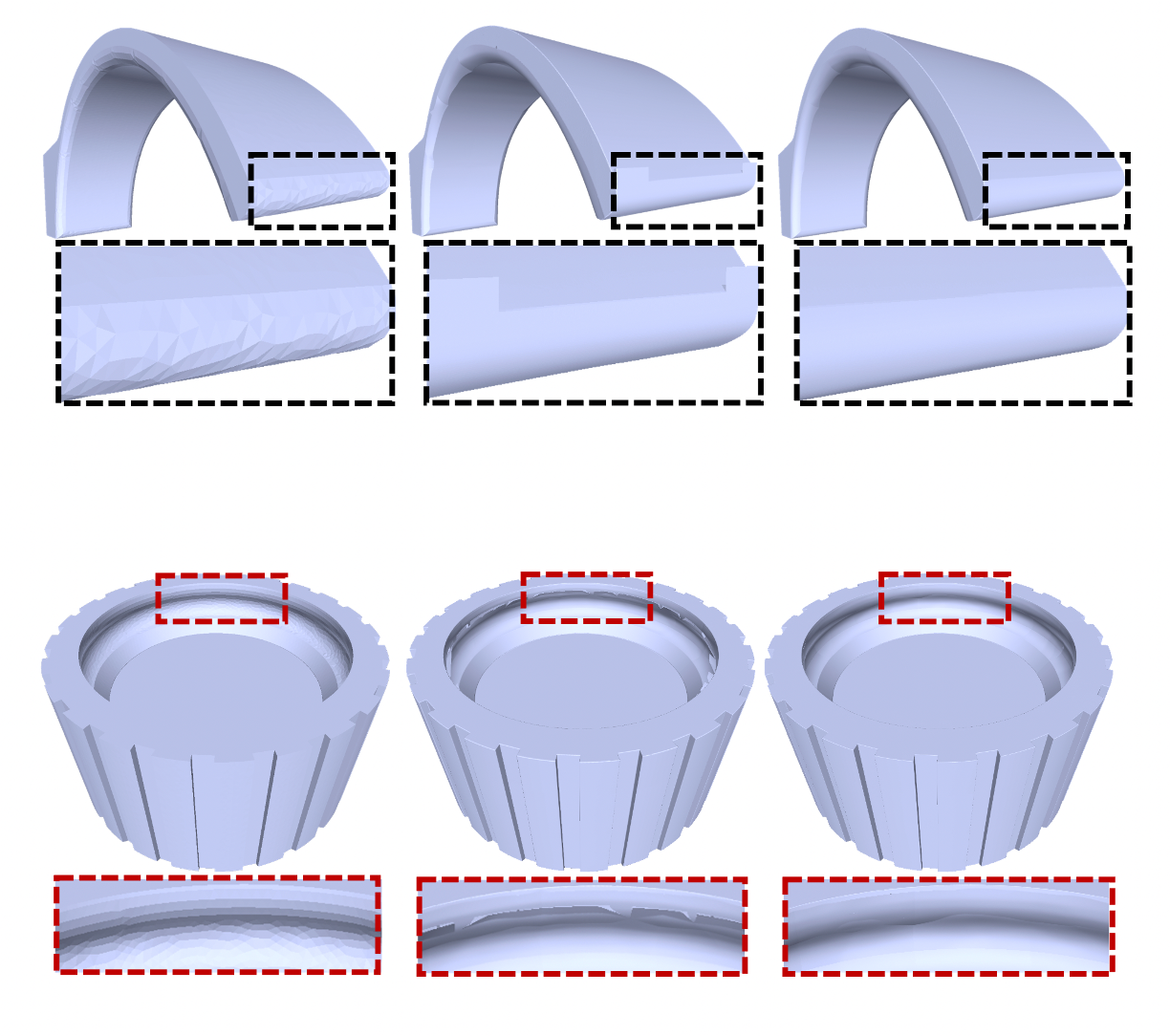}
    \put(-210,117){(a) GT}  
    \put(-210,0){(d) GT}  
    \put(-150,117){(b) $\lambda_{\mathrm{normal}}$=12}
    \put(-150,0){(e) $\lambda_{\mathrm{normal}}$=200}
    \put(-75,0){(f) Patch-Grid}
    \put(-65,-12){($\lambda_{\mathrm{normal}}$=50)}
    \put(-75,117){(c) Patch-Grid}
    \put(-65,105){($\lambda_{\mathrm{normal}}$=50)}
    \caption{
    Influence of $\lambda_{\mathrm{Normal}}$ on shape fitting results. Inappropriate weights on $\loss_{\mathrm{Normal}}$ may lead to artifacts due to failure to fit highly curved patches (b) or failure to handle the merging of multiple patches (e).
    }
    \label{fig:ablation_normal_weight}
\end{figure}

\begin{figure}
    \centering
    \includegraphics[width=\linewidth]{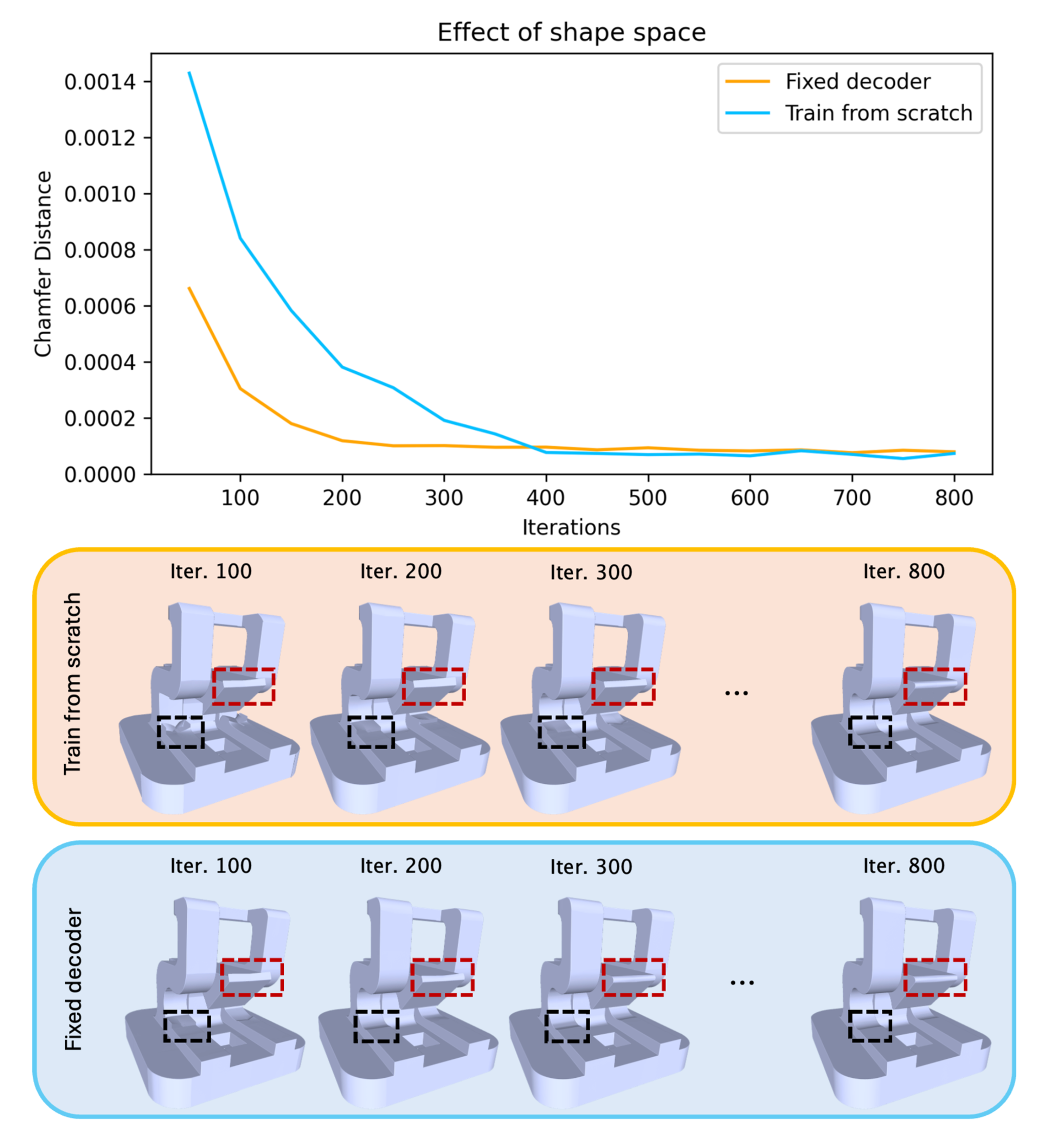}
    \caption{Pretrained fixed shape space matters. The ablation reveals that the use of a fixed, pretrained decoder significantly accelerates training as shown by the curves of CD value against the number of iterations (around 250 iterations). In contrast, without pretraining, our method takes 400 iterations to converge.
    The impact of the pretrained shape space is also evident in the visualization of the shapes, as demonstrated in the following figures.}
    \label{fig:ablation_shapespace}
\end{figure}

\textbf{Adaptive patch feature volume. } 
A patch feature volume's resolution is adaptively determined based on the shape diameter of this patch as introduced earlier. We show in Tab.~\ref{tab:ablation_adaptive} the necessity of this design choice to model surface patches of smaller size.
As shown in Tab.~\ref{tab:ablation_adaptive}, \textit{w/o Ada-PV} yields less satisfactory results across most quantitative metrics. A qualitative example is provided in Fig.~\ref{fig:adaptive_grids} to show the necessity of adaptive patch volumes for modeling slender geometry; otherwise, using patch volumes of uniform size fails to accurately fit thin tubes. 

\textbf{Loss terms. } We conducted an ablation study on the loss design. The quantitative results of removing each loss term are shown in Tab.~\ref{tab:ablation_term}. As can be seen from the table, the setting with the complete loss terms achieves the best accuracy in most quantitative metrics.
The two data terms, $\loss_{\mathrm{Surface}}$ and $\loss_{\mathrm{Normal}}$, are widely used to supervise surface fitting. Removing the former will degrade the surface reconstruction quality as shown in Fig.~\ref{fig:ablation_no_surf}, while removing the normal supervision will lead to severe artifacts as shown in Fig.~\ref{fig:ablation_no_surf}(b).

The merge loss term, $\loss_{\mathrm{Merge}}$, imposes an important constraint to coordinate the extended zero-level sets of patches to ensure the validity of the resulting neural field after applying the CSG operations. 

Due to the use of an octree-based merge grid, the merge loss is only imposed in the leaf nodes of the merge cells which account for a small portion of the shape. Hence, removing this loss term will not largely deteriorate the quantitative metrics in an average sense, but artifacts owing to the extraneous zero-level sets can be observed in Fig.~\ref{fig:ablation_normal_merge}(c). This qualitative observation can be reflected in the increased Hausdorff distance metric.

$\loss_{\mathrm{SDF}}$ applies a pseudo SDF supervision close to the geometric surface. Removing $\loss_{\mathrm{SDF}}$ would lead to artifacts such as obvious offsets in individual patches, as shown in Fig.~\ref{fig:ablation_jitter}(b).

Finally, $\loss_{\mathrm{Eikonal}}$ is designed to ensure that the SDF learned by the \textit{Patch-Grid} satisfies the distance field's property -- having a unit gradient magnitude~\cite{Gropp2020IGR}. 
In the absence of ground truth SDF supervision, it encourages the network to learn a more accurate signed distance field. An accurate distance field is critical for applications like offsetting, inverse optimization, and collision detection. As shown in Table \ref{tab:ablation_term}, removing $\loss_{\mathrm{Eikonal}}$ results in a very large \textit{Field Error}. 

\textbf{Loss weights. } We examine how the balancing weights of two important loss terms, $\loss_{\mathrm{Merge}}$ and $\loss_{\mathrm{Normal}}$, influence our results, validating that the weights are reasonably set. 
In our study, we multiplied the original weight in our setting by 1/4 and 4, respectively. We observe that a lower $\lambda_{\mathrm{Merge}}$ results in artifacts as depicted in Fig.~\ref{fig:ablation_merge_weight}(b), due to insufficient constraint on coordinating the extraneous zero-level sets learned by our method to satisfy the pre-defined CSG operations. On the other hand, a higher $\lambda_{\mathrm{Merge}}$ causes failure in accurately fitting small patches, resulting in holes on the surface, as shown in Fig.~\ref{fig:ablation_merge_weight}(e).
Regarding $\lambda_{\mathrm{Normal}}$
We found that a lower value could lead to less accurate fitting results, as illustrated in Fig.~\ref{fig:ablation_normal_weight}(b). Conversely, a higher $\lambda_{\mathrm{Normal}}$ can overwhelm the merge loss, resulting in similar artifacts as if the merge loss is set to a lower value. An example is shown in Fig.~\ref{fig:ablation_normal_weight}(e).
From the results, we can see the chosen weights can achieve satisfactory results both quantitatively and qualitatively.

\textbf{Training strategies. } 
Two training strategies, i.e.\ training from scratch (\textit{Patch-Grid-TS}) and training with a fixed, pretrained decoder (\textit{Patch-Grid}), are analyzed in depth.
As shown in Fig.~\ref{fig:ablation_shapespace}, we plot the curves of the Chamfer Distance metric and visualize the intermediate mesh results of the two training strategies. We can see that \textit{Patch-Grid} converges with $300$ training iterations, faster than \textit{Patch-Grid-TS} which requires around $500$ iterations to achieve similar performance. Note that $300$ and $500$ training iterations amount to 5 and 8 seconds, respectively.
From the qualitative results, one can see that \textit{Patch-Grid-TS} spends extra time in modeling the joinings of two or more patches (enclosed by red and black boxes in the lower part of Fig.~\ref{fig:ablation_shapespace}).

\begin{table}[t]
\centering
\small
\caption{Ablation study of the loss terms. CD, HD, NC, SPE and FE are presented in units of $\times10^{-4}$, $\times10^{-3}$, $\times10^{-2}$, $\times10^{-5}$ and $\times10^{-3}$, respectively.}\label{tab:ablation_term}
\scalebox{0.9}{
\begin{tabular}{c|c c c c c c c}
\toprule
 &  CD $\downarrow$ & HD $\downarrow$ & F-score $\uparrow$ & IoU $\uparrow$ & NC $\uparrow$ &  SFE $\downarrow$ & FE $\downarrow$  \\ \hline
 
w/o $\loss_{\mathrm{Normal}}$ & 7.88 & 58.3 & 91.82 & 91.33 & 98.12 & 28.8 & \textbf{2.14} \\
w/o $\loss_{\mathrm{SDF}}$ & 2.03 & 15.3 & 99.08 & 94.63 & 99.36 & 20.0 & 10.3  \\
w/o $\loss_{\mathrm{Eikonal}}$ & 0.886 & 2.53 & 9.77 & 99.48 & 99.70 & 9.56 & 24.5\\
w/o $\loss_{\mathrm{Merge}}$ & \textbf{0.775} & 4.72 & 99.75 & 99.35 & 99.70 & 8.70 & 8.99  \\
w/o $\loss_{\mathrm{Surf}}$ & 1.05 & 2.82 & 99.42 & 99.31 & 99.69 & 11.2 & 8.97 \\
w/ all & 1.00 & \textbf{2.47} &  \textbf{99.78}  & \textbf{99.61} & \textbf{99.83}& \textbf{8.08} & 9.35\\ 
\bottomrule
\end{tabular}}
\end{table}

\begin{table}[t]
\centering
\small
\caption{Ablation of the weights of the loss terms $\loss_{\mathrm{Normal}}$, and $\loss_{\mathrm{Merge}}$. The quantitative results validate our choice. CD, HD, NC, and SPE are presented in units of $\times10^{-4}$, $\times10^{-3}$, $\times10^{-2}$, and $\times10^{-5}$ respectively.}\label{tab:ablation_weight}
\scalebox{0.95}{
\begin{tabular}{c| c | c c c c c c}
\toprule
\multicolumn{2}{c|}{} & CD $\downarrow$ & HD $\downarrow$ & F-score $\uparrow$ & IoU $\uparrow$ & NC $\uparrow$ &  SFE $\downarrow$ \\ \hline
\multirow{3}{*}{$\loss_{\mathrm{Normal}}$}
&  12 & 1.52 & 3.88 & 99.15 & 99.371 & 99.80 & 14.1 \\
&  50 & 1.00 & \textbf{2.47} & \textbf{99.780} & \textbf{99.61} & \textbf{99.83} & 9.89\\
&  200 & \textbf{0.891} & 2.65 & 99.689 & 99.59 & 99.82 & \textbf{8.09} \\
\hline
\multirow{3}{*}{$\loss_{\mathrm{Merge}}$}
& 100 & \textbf{0.766} & 3.23 & \textbf{99.782} & 99.54 & 99.70 & \textbf{8.41} \\ 
&  400 & 1.00 & \textbf{2.47} & 99.780 & \textbf{99.61} & \textbf{99.83} & 9.89 \\
& 1200 & 1.20 & 3.13 & 99.523 & 99.30 & 99.69 & 13.1 \\
\bottomrule
\end{tabular}}
\end{table}

\begin{figure}
    \centering
    \includegraphics[width=1.0\linewidth ]{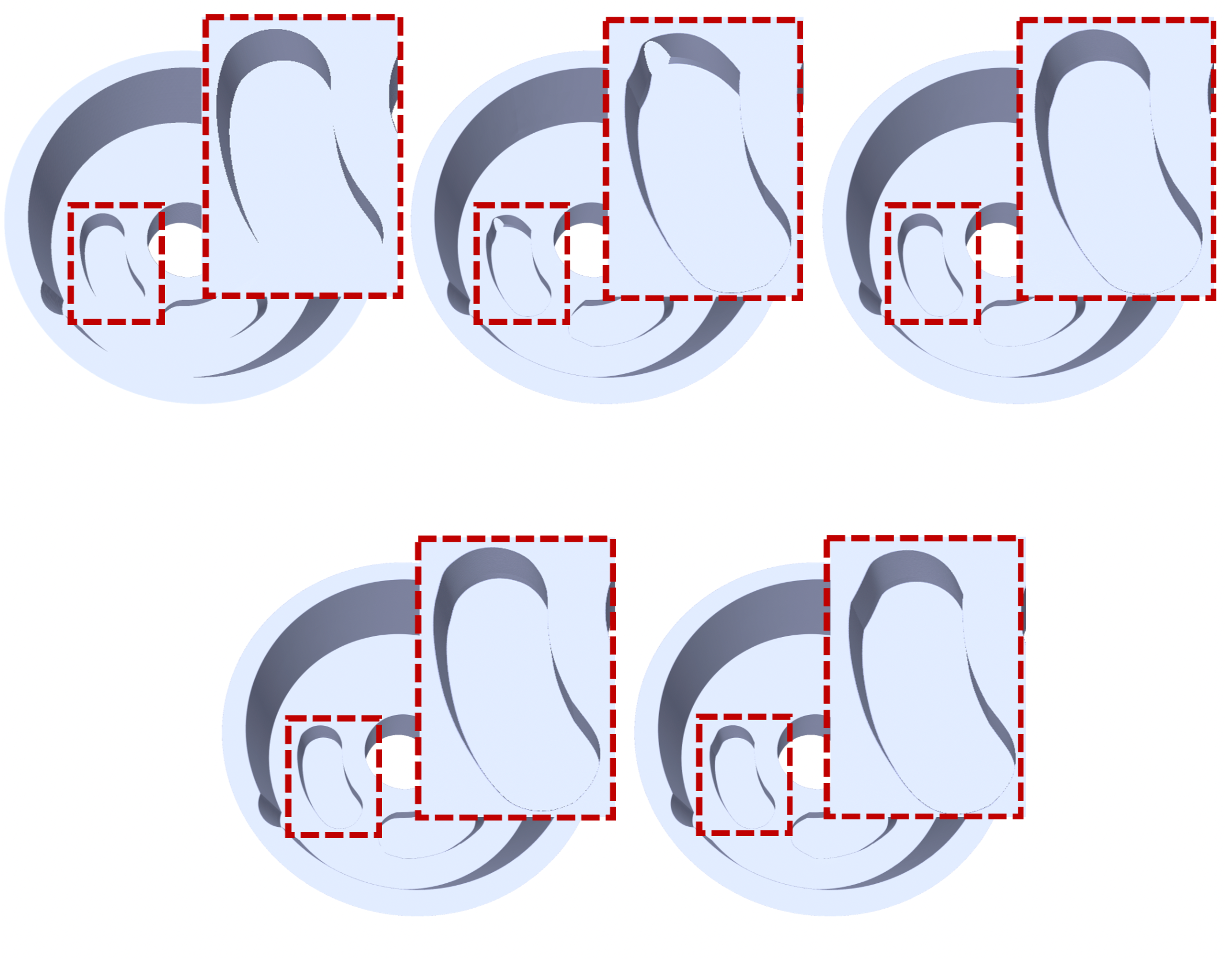}
    \put(-220,100){(a) GT}  
    \put(-145,100){(b) 1 layer}
    \put(-60,100){(c) 2 layers }
    \put(-180,-3){(d) 3 layers }
    \put(-100,-3){(e) 4 layers}
    \caption{Number of the decoder layers for the shape updating task. Decoders with 2 or 3 hidden layers tend to exhibit the best visual effects. }
    \label{fig:ablation_layer_update}
\end{figure}

\begin{figure}
    \centering
    \includegraphics[width=1.0\linewidth ]{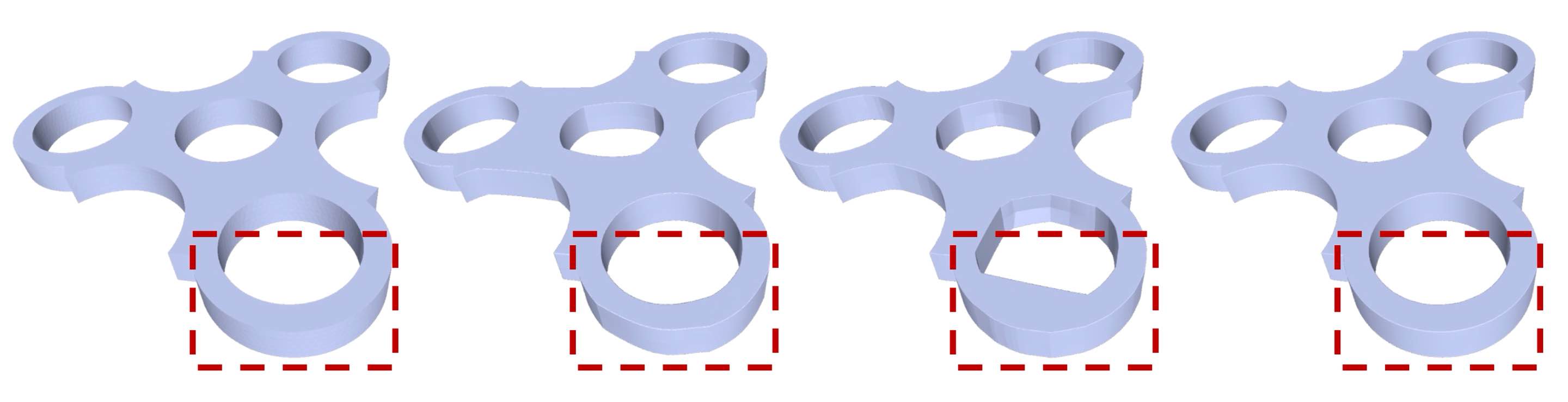}
    \put(-210,-6){(a) GT}  
    \put(-153,-6){(b) Sine}
    \put(-97,-6){(c) ReLU }
    \put(-46,-6){(d) SoftPlus }
    \caption{We evaluate our method with different activation functions, \textit{Sine} (b) and \textit{ReLU} (c), alternative to \textit{Softplus}. Only using \textit{Softplus} activation (d) can yield accurate fitting results.}
    \label{fig:activation}
\end{figure}

\begin{table}[htbp]
\centering
\small
\caption{Performance derived with different number of decoder layers. The upper and lower blocks are for the shape fitting and updating tasks, respectively. CD, HD, NC, and SPE are presented in units of $\times10^{-4}$, $\times10^{-3}$, $\times10^{-2}$, and $\times10^{-5}$ respectively. The unit for time is second.}\label{tab:ablation_layer}
\scalebox{0.95}{
\begin{tabular}{c|c c c c c c c}
\toprule
 &  CD $\downarrow$ & HD $\downarrow$ & F-score $\uparrow$ & IoU $\uparrow$ & NC $\uparrow$ &  SFE $\downarrow$ & Time $\downarrow$  \\ \hline
4 layers & \textbf{0.809} & \textbf{2.14} & \textbf{99.903} & \textbf{99.59}  & 99.700 & \textbf{8.24} & 5.64 \\
3 layers & 0.937 & 2.36 & 99.896 & 99.54 & 99.697 & 9.35 & 5.60 \\
2 layers & 0.989 & 2.17 & 99.901 & 99.47 & 99.701 & 9.39 & 5.61 \\
1 layer & 1.00 & 2.47 & 99.782 & 99.61 & \textbf{99.830} & 9.89 & \textbf{5.44}\\ 
\midrule
4 layers & 1.62 & 12.3 & 98.83 & 99.06 & 99.83 & 50.3 & 1.61 \\
3 layers & \textbf{1.36} & 6.44 & \textbf{99.40} & \textbf{99.24} & 99.85 & 19.8 & 1.53 \\
2 layers & 1.59 & \textbf{5.35} & 99.23 & 99.17 & \textbf{99.86} & \textbf{19.3} &  1.52\\
1 layer & 2.22 & 14.1 & 97.28 & 98.86 & 99.80 & 20.3 & \textbf{1.49} \\ 
\bottomrule
\end{tabular}}
\end{table}

\begin{table}[htbp]
\centering
\caption{Performance derived with different activation functions. CD, HD, NC, and SPE are presented in units of $\times10^{-4}$, $\times10^{-3}$, $\times10^{-2}$, and $\times10^{-5}$ respectively.}\label{tab:ablation_activation}
\begin{tabular}{c|c c c c c c }
\toprule
 &  CD $\downarrow$ & HD $\downarrow$ & F-score $\uparrow$ & IoU $\uparrow$ & NC $\uparrow$ &  SFE $\downarrow$ \\ \hline
 
Sine & 23.50 & 24.74 & 44.59 & 86.28 & 98.59 & 213.5 \\
ReLU & 3.98 & 10.75 & 93.59 & 98.37 & 99.43 & 36.2 \\
SoftPlus & \textbf{1.00} & \textbf{2.47} & \textbf{99.78} & \textbf{99.61} & \textbf{99.83} & \textbf{9.89}\\
\bottomrule
\end{tabular}
\end{table}

\textbf{Decoder depth. } The upper block of Tab.~\ref{tab:ablation_layer} demonstrates that an MLP with a single hidden layer is already sufficient for accurately fitting the shape. While further increasing the number of MLP layers can marginally improve the fitting quality, it leads to longer training times. Throughout our \textit{shape fitting} experiments, we report the results with a 1-layer MLP for simplicity. 

We also conducted another ablation study on the decoder's depth for the shape editing task, as shown in the lower block of Tab.~\ref{tab:ablation_layer} and Fig.~\ref{fig:ablation_layer_update}. From both the quantitative and qualitative results, we observed that 
Considering the trade-off between reconstruction quality and efficiency, we chose a 2-layer MLP as the decoder for the experiments of the shape updating task. We consider this detailed analysis helpful for users to decide on the decoder's depth based on their specific needs.

\textbf{Activation functions. } Tab.~\ref{tab:ablation_activation} validates our choice of the \textit{SoftPlus} activation function over \textit{ReLU} or the \textit{Sine} activation~\cite{Sitzmann2020SIREN}. The network structure based on \textit{SoftPlus} achieves the highest fitting accuracy. We also provide qualitative results for these three activation functions in Fig.~\ref{fig:activation} to support our design choice.
We hypothesize that the artifacts observed with \textit{ReLU} are due to the property of \textit{ReLU} as a piece-wise linear function, which may not be well-suited for our $\loss_{\mathrm{Normal}}$ supervision. Meanwhile, the artifacts with \textit{Sine} are likely caused by the mismatch between \textit{Sine} activation and our extremely compact MLP structure (i.e.\ 1-layer MLP).

%% file: sections/limitation.tex
\section{limitation and future works}\label{sec:limitation}

The amount of sampling points used to train the network scales roughly linearly to the number of patches and sharp features. In our current implementation, we have generated the sample points in one pass before training, which takes significant CPU memory storage for a large number of patches. In our stress test, we cannot process shapes with more than 670 patches due to the CPU memory bottleneck. Future work may consider generating the sample points online along side the training iterations.

\begin{figure}
    \centering
    \includegraphics[width=1.0\linewidth]{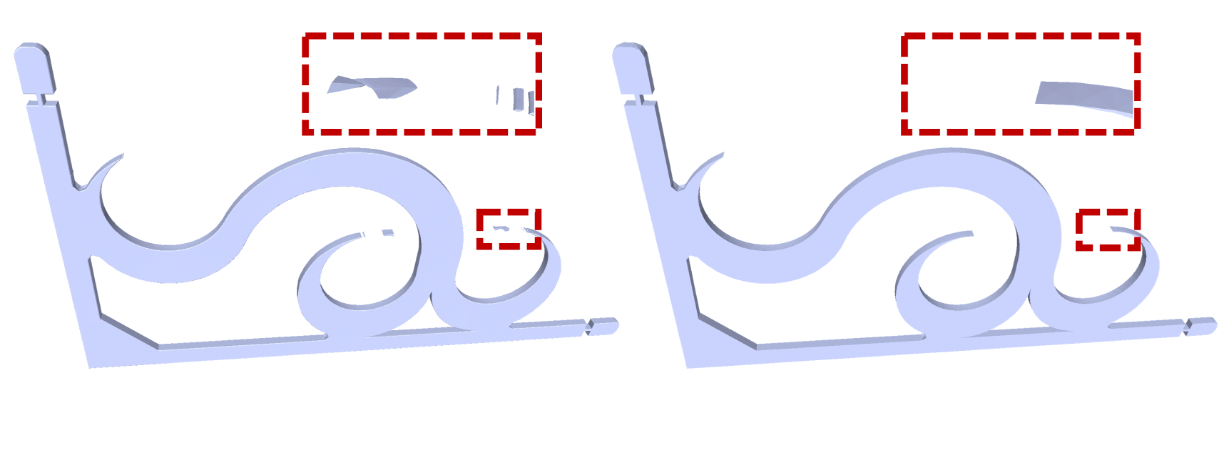}
    \put(-60,0){GT}
    \put(-175,0){Patch-Grid}
    \caption{Limitation of \textit{Patch-Grid} on fitting nearly parallel yet connected patches.}
    \label{fig:limitation_parallel}
\end{figure}

In addition, \textit{Patch-Grid} may fail to model thin features composed of two surface patches that are nearly parallel and spatially close at the same time, as shown in Fig.~\ref{fig:limitation_parallel}. This is because even small SDF fitting errors can result in inaccurate zero-level sets after CSG operations for nearly parallel patches, as illustrated in the zoomed-in region of Fig.\ref{fig:limitation_parallel}.

\begin{figure}
    \centering
    \includegraphics[width=1.0\linewidth]{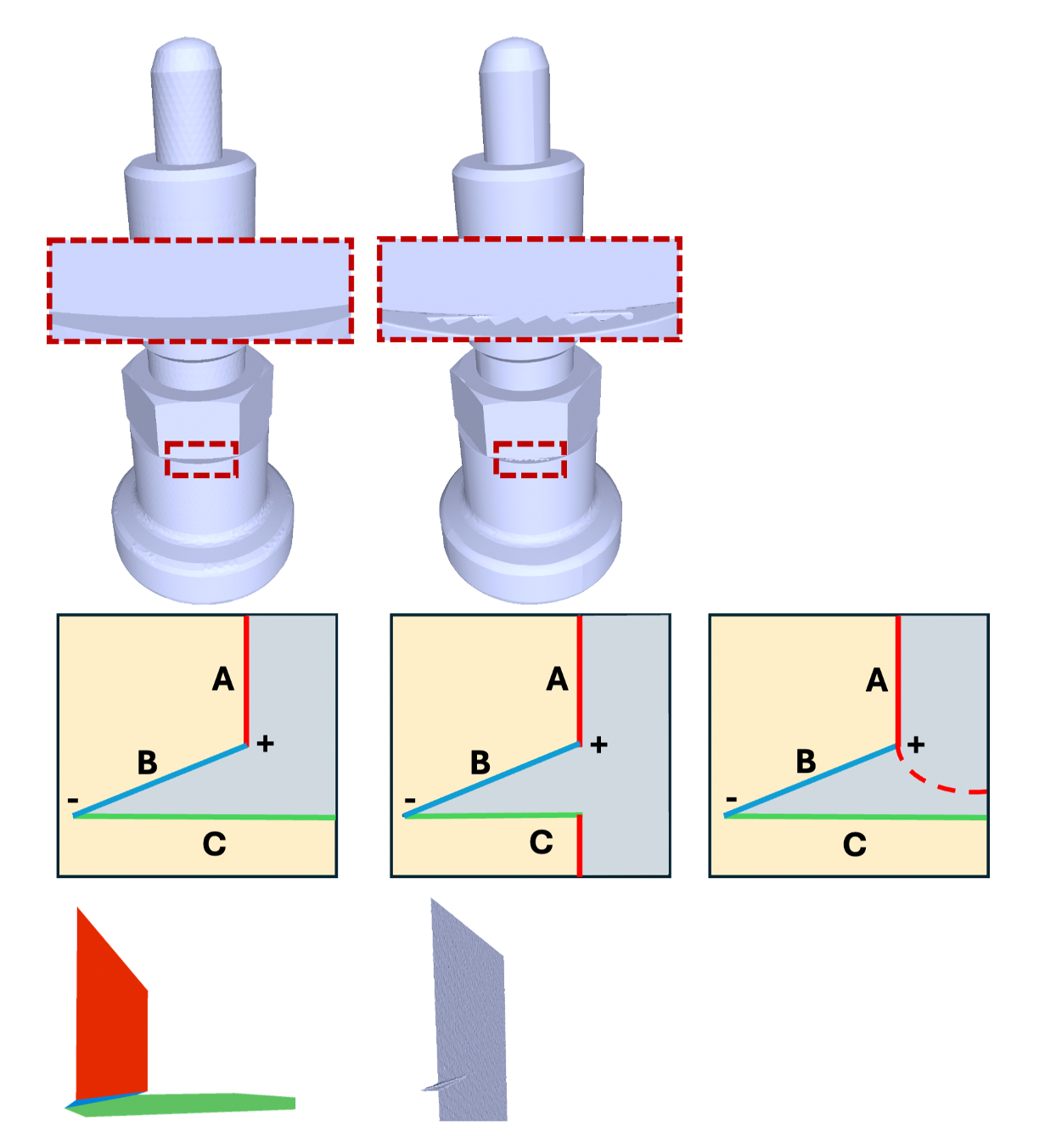}
    \put(-207,-5){(a) GT}
    \put(-144,-5){(b) Patch-Grid}
    \put(-70,-5){(c) Ideal result}
    \caption{Limitation of \textit{Patch-Grid} in the special case when a merge cell contains an incomplete adjacent graph. The first row shows the GT and extracted meshes with a zoom-in view of the artifact. 
    The second row exemplifies the cause of this artifact with a 2D example. Ideally, a sharp turn of the learned SDF of patch $A$ in its natural extension is expected to avoid intersecting patch $C$'s zero-level set in this confined region. However, due to the tiny sizes of patches $B$ and $C$, our method fails to model the detailed feature, yielding the observed artifact in (b). Note that this is statistically rare; this is the only case caused by this type of failure in the pool of 100 shapes.}
    \label{fig:limitation_fail_reason}
\end{figure}

Thirdly, \textit{Patch-Grid} is unable to avoid the failure cases observed in \textit{NH-Rep} in theory; similar failures may be observed in the merge cells that contain incomplete graphs at the maximum depth of the octree-based merge grid. However, we have examined all failure cases from the 100 shapes and found only one failure case due to this reason as shown in Fig.~\ref{fig:limitation_fail_reason}. Statistically, there are only 1.76\% merge cells (36,860 out of more than 2 million merge cells among the pool of 100 shapes) that contain incomplete connected graphs, and only one failure case is observed. We attribute this robustness to the significantly reduced CSG operations in each merge cell as a result of utilizing the merge grid.

Another limitation is that \textit{Patch-Grid} requires a longer training time to fit free-form geometries accurately. In our future work, we will integrate the hash table/hierarchical feature volumes to model the details in the free-form geometries efficiently.

Lastly, another future direction is to explore the potential of \textit{Patch-Grid} to support structure-level or semantic-level shape editing and shape manipulation due to its flexible representation with patch-wise latent spaces.  

%% file: sections/conclusions.tex
\section{Conclusion}\label{sec:conclusions}
We have presented a novel implicit neural surface using a patch-based representation with a merge grid to enable a local training and merging strategy.
\textit{Patch-Grid} performs CSG operations locally in merge grid cells to significantly improve the robustness of the patch-based representation, compared with the prior work~\cite{Guo2022NHRep} using a global approach. Moreover, \textit{Patch-Grid} can handle geometric features that previous methods fail to reproduce, such as thin geometries or open surfaces. We evaluate \textit{Patch-Grid} on various shapes to validate its superiority to existing works in terms of reconstruction quality, robustness, and efficiency.